\documentclass[prd,aps,preprint,epsfig,floats]{revtex4}
\usepackage{graphics}

\usepackage{graphicx}

\begin{document}
\input epsf.tex    
\newcommand{\be}{\begin{eqnarray}}
\newcommand{\ee}{\end{eqnarray}}

\title{\Large {Neutrino Mass and New Physics}\footnote{With permission 
from the 
Annual Review of Nuclear and Particle Science. Final
version of this material is scheduled to appear in the Annual Review of 
Nuclear and Particle Science Vol. 56, to be published in November 2006 by 
Annual Reviews
(www.annualreviews.org).
}}

\author{\bf R. N. Mohapatra$^1$ and A. Y. Smirnov$^{2,3}$}
\affiliation{$^1$Department of Physics, University of Maryland,
College
Park, MD-20742\\
$^2$Abdus Salam International Center for Theoretical Physics,
Trieste, Italy\\
$^3$Institute for Nuclear Research, RAS, Moscow, Russia}


\begin{abstract}
We review the present state and future outlook of our
understanding of neutrino masses and mixings. We discuss what we
think are the most important perspectives on the plausible and
natural scenarios for neutrinos and attempt
to throw light onto the flavor problem of quarks and leptons. This 
review focuses on the seesaw mechanism, which fits into a big picture 
of particle physics such as supersymmetry and grand unification
providing a unified approach to the flavor problem of quarks and
leptons. We argue that in combination with family symmetries, this
may be at the heart of a unified understanding of the flavor puzzle.
We also discuss other new physics ideas such as neutrinos in
models with extra dimensions and possible theoretical implications
of sterile neutrinos. We outline some tests for the various
schemes.
\end{abstract}

\maketitle

\tableofcontents

\section{INTRODUCTION}
The Standard Model  (SM) of electroweak (EW) and strong interactions is an
overwhelmingly successful theory for particles and forces. It
probes physics below a 100 GeV's  (in some cases up to several
TeVs ) and has met the challenge of many high-precision
experiments. There are, however,  strong reasons from considerations
of both particle physics as well as cosmology to suspect that
there is a good deal of new physics beyond the standard model.
In cosmology, the SM does not explain  
dark matter, inflation and dark energy.  
On the side of particle physics, the recent discovery of
flavor conversion of solar, atmospheric, reactor, and accelerator
neutrinos has established conclusively  that neutrinos have
nonzero mass and they mix among themselves much like  quarks,
thereby providing the first evidence of new physics beyond the
standard model. This discovery has triggered enormous theoretical
activity attempting to uncover the nature of this new physics.
These attempts include further developments of the already existing
mechanisms and theories such as the grand unified theories (GUT's) 
and the appearance of new  ideas 
and approaches. Various aspects of these developments have to some
extent been covered in several recent reviews \cite{threv}. The
present review  is an update that  focuses on what we think are
the most important perspectives on the most plausible and natural
scenarios  of physics beyond the standard model.

The main points stressed in this review are as follows:

1). After all recent developments, the seesaw mechanism with a large
scale of $B-L$ (baryon minus lepton number) violations still appears to be the most 
appealing
and natural mechanism of neutrino mass generation. 
However, it is not excluded that some more complicated version of this
mechanism is realized.

2). Grand unification  (plus supersymmetry in some form) still
appears to be  the most preferable (plausible) scenario of physics
that naturally embeds the seesaw.

3). The seesaw - GUT scenario does not
provide a complete understanding of the masses and mixing of neutrinos, 
nor the  masses and mixing of other fermions (in other words,
the flavor structure of the mass matrices). Some new physics 
in addition to  this scenario seems essential. 
In this connection two issues are of great importance:

- possible existence of new symmetries that   show up mainly
(only?) in the lepton sector; and 

- understanding the relation between quarks and leptons.

To uncover this new additional element(s) of theory, the bottom-up
approach is extremely important.

4). Alternative mechanisms and pictures are not
excluded, although they have not reached the same level of
sophistication as the GUT approach and appear less plausible.
Nonetheless, it is quite possible that they play a  role in
the complete picture, e.g., as the source of new neutrino states,
as some element in addition to the ``seesaw - GUT'' scenario, or as
some element of physics at a  level deeper than  or above the GUT
scale. Flavor structure can appear as a result of 
compactification of extra dimensions.

This review  presents a  detailed discussion of these statements
and current arguments in favor of this possibility. 
In sect. 2 we introduce the main
notions on neutrino mass and mixing and summarize available
experimental results. We proceed with the bottom-up approach in
the following three  sections. In sect 3.  we analyze results on
neutrino mixings and the mass matrix.
We address the question of particular neutrino
symmetries in sect 4.,  and in sect. 5 we consider possible
relations between quarks and leptons. In the rest of the review the
top-down approach is presented. We study properties of the seesaw
mechanism in sect 6. Neutrino masses in the grand unified theories
are discussed in sect. 7. We consider  achievements and
limitations of this picture. The need for flavor symmetries in
addition to GUTs is described.  We also discuss alternative
mechanisms  of neutrino mass generation and the alternative
scenarios for the  big picture.
In sec. 8, we discuss sterile neutrinos,  their implications and
possible theories. Sec. 9 contains the conclusion.


\section{ Masses,  flavors  and mixing}

\subsection{Neutrinos and the Standard model}

According to the standard model (SM), left-handed neutrinos form
the EW doublets $L$ with charged leptons,
have zero electric charge and no color. The right-handed
components, $\nu_R$, are not included  by choice.

The masslessness of the neutrinos at the tree level in this model
owes its origin to the fact that there are no right-handed
neutrinos. This result holds not only to all orders in
perturbation theory but also when nonperturbative effects are
taken into account, owing  to the existence of an exact B-L 
symmetry in the model even though B+L is
violated by weak sphaleron configurations. It would therefore
appear that nonzero neutrino masses must somehow be connected to
the existence of right-handed neutrinos and/or to the breaking of B-L
symmetry both of which imply new physics beyond the standard
model.

Vanishing  conserved charges (color and electric) distinguish neutrinos
from other fermions of the standard model.  This difference leads
to several new possibilities for neutrino masses all of which
involve new physics:

\noindent(i)   The neutrino masses could be the Majorana type
thereby breaking $L$ by two units.

\noindent (ii) Neutrinos have the  possibility to mix with
singlets of the SM symmetry group, in particular,  singlet fermions in
extra dimensions.


The main question is whether these features are   enough to
explain all the salient properties  of the neutrino masses and mixing
observed in experiment. \\



One way that the neutrino mass can be generated even if
the SM particles are the only light degrees of freedom 
requires that the condition of explicit
renormalizability of the theory should be abandoned. Indeed, the non-renormalizable
operator~\cite{eff1}
\begin{equation}
\frac{\lambda_{ij}}{M} (L_i H)^T(L_j H), ~~~ i,j = e, \mu, \tau ,
\label{nonren}
\end{equation}
where $H$ is the Higgs doublet $\lambda_{ij}$  are the
dimensionless couplings and ${M}$ is the cutoff scale, after the
EW symmetry breaking generates the Majorana neutrino
masses
\begin{equation}
m_{ij} = \frac{\lambda_{ij} \langle H \rangle^2}{M}.
\end{equation}
The operator breaks $L$ and $B-L$ quantum numbers. One may think
that this operator is generated by some gravitational  Planck
scale effects, so that $M \sim M_{Pl}$ and $\lambda_{ij} \sim 1$
\cite{eff}. In this case however, $m_{ij} \sim 10^{-5}$ eV are too
small to explain the observed masses (although such a contribution
can still produce some  subleading features \cite{BerV}). Therefore,
new scales of physics below $M_{Pl}$ must exist to give the
desired mass to neutrinos. The operator in eq.(\ref{nonren}) can
appear after integrating out some new heavy degrees of freedom
with masses $M \ll M_{Pl}$.


Another important conclusion from this consideration is that the
neutrinos can get relevant contributions to masses from all
possible energy/mass scales $M$ from  $\sim  1$ eV
to the Planck scale. If there are 
two or more differentsubstantial contributions 
(from different scales and
different physics)  to the mass, the 
interpretation of results can be extremely difficult.
In this case one can argue that the ``coincidence'' of 
sizes of contributions looks
unnatural. However, this will be not the only place where we face
the coincidence problem.

\subsection{Right-handed neutrinos, neutrino mass and seesaw.}

Let us consider possible extensions of the standard model which
can lead to non-zero neutrino masses.

1). If  right-handed neutrinos exist (we  consider the
conceptual implications of their existence in sect. 6.2) one can
introduce the Yukawa coupling
\be
Y_\nu\bar{L}H \nu_R + h.c.
\label{dyukawa}
\ee
which after the electroweak (EW) symmetry
breaking leads to the Dirac neutrino mass
\be m_D = Y_\nu \langle H \rangle.
\label{dmass}
\ee

The observed neutrino masses would require $Y_{\nu} \leq 10^{-13}
- 10^{-12}$. If $\nu_R$ is the same type of field as right-handed components
of other fermions, such smallness looks rather unnatural.
However, if the Dirac mass is formed by coupling with some new
singlet fermion, $S$, beyond the usual fermion family structure,
possible new symmetries associated to $S$ and/or $\nu_L$ can
suppress $Y_\nu$. In this case  $Y_\nu$  appears as the effective
coupling: $Y_\nu \sim  (v_S/M)^n$, where $v_S \ll M$  are the
scales of some new interactions and new symmetry breaking, and $n$
is some integer determined by the charges of the fields.

2). The right handed neutrinos are allowed to have Majorana masses
\be
M_R \nu^T_RC^{-1} \nu_R + h.c. ,
\label{mmass}
\ee
where $C$
is the Dirac charge-conjugation matrix. Because $\nu_R$ are singlets
under the SM gauge group the $M_R$ can appear as a
bare mass term in the Lagrangian or be generated by interactions
with singlet scalar field $\sigma$ (so that $M_R \rightarrow
f_\sigma \sigma$ in eq.(\ref{mmass})):
\be
M_R = f_{\sigma} \langle \sigma \rangle,
\ee
where $\langle \sigma \rangle$ is the vacuum expectation value (VEV) of $\sigma$. 
The latter possibility is realized if, {\it e.g.},   $\nu_R$ is a component of a
multiplet of an extended gauge group.

3). The left handed neutrinos can also acquire the Majorana masses
$m_L$. The corresponding mass terms have the weak isospin $I = 1$
and violate lepton number by two units. Thus,  they can be generated
either
 via the non-renormalizable operators in eq.(\ref{nonren}) with two Higgs
 doublets  and/or by coupling with the Higgs triplet $\Delta$:
\be f_{\Delta} L^T  L \Delta + h.c.. \ee The non-zero VEV of
$\Delta$  then gives $m_L =  f_{\Delta} \langle \Delta \rangle$.

In the case of three neutrino species the mass parameters $m_D$, $M_R$
and $m_L$ should be considered as $3 \times 3$ 
(non-diagonal) matrices. In general all these mass terms are
present. By introducing the charge-conjugate left handed component
$N_L \equiv (\nu_R)^C$ the general mass matrix in the
basis $(\nu_L, N_L)$, can be written as
\begin{eqnarray}
{ M}_\nu~=~\left(\begin{array}{cc}m_L & m_D^T\\
m_D & M_R
\end{array}
\right).
\label{mgeneral}
\end{eqnarray}
The eigenstates of this matrix are the Majorana neutrinos with
different Majorana masses. The Dirac mass term mixes the active
neutrinos with the sterile singlet states $N_L$.

The matrix has several important limits.

Suppose $m_L = 0$. The Yukawa couplings $Y_\nu$
eq.(\ref{dyukawa}) are expected to be of the same order as the
charged-fermion couplings.
Because the $N_L$'s are singlets under the SM gauge
group, their Majorana masses unlike the masses of the charged
fermions, are not constrained by the gauge symmetry and can therefore be
arbitrarily large, {\it i.e.} $M_R \gg  m_D$.
In this case the diagonalization of the mass matrix in eq.(\ref{mgeneral}),
leads to an approximate form for the mass matrix for the light neutrinos
$m_\nu$:
\begin{eqnarray}
{\cal M}_\nu~=~-m^T_DM^{-1}_R m_D. \label{seesaw}
\end{eqnarray}
As  noted, because  $M_R$ can be much larger than $m_D$
one finds  that $m_\nu \ll m_{e,u,d}$ very naturally, as is
observed. This is known as the seesaw (type I)
mechanism~\cite{seesaw} and it provides a natural explanation of
why neutrino masses are small.

If elements of the matrix $m_L$ are non-zero but  much smaller
than the other elements of $M_\nu$,  we can write the resulting
light neutrino mass matrix in the form
\begin{eqnarray}
{\cal M}_\nu~=~m_L-m^T_D M^{-1}_R m_D. \label{seesaw2}
\end{eqnarray}
We refer to this as  mixed seesaw \cite{seesaw2,valle} and
when the first term dominates, we refer to  it as a type II seesaw.

In the matrix eq.(\ref{mgeneral}), 
elements of both $m_L$ and $M_R$ may have magnitudes much
smaller than those of $m_D$. In this case, the neutrinos 
are predominantly Dirac type,  with a  small admixture of Majorana mass.
We refer to this case the pseudo-Dirac~\cite{pseudo}.

The main question  we discuss subsequently is whether
the see-saw mechanism  can  explain all the observed features of
neutrino mass and mixing.

%

\subsection{Flavors and mixing}

The electron, muon and tau neutrinos, $\nu_e$, $\nu_{\mu}$, $\nu_{\tau}$ 
 - particles produced in
association with definite charged leptons: electron, muon and tau
correspondingly are called the flavor neutrinos. For 
example, the neutrinos emitted in weak processes such as the
beta decay or pion decay with electron or muon  are termed the
electron or muon neutrinos. In the detection process, it is the
flavor neutrino that is  picked out because the detectors 
are sensitive to the charged lepton flavors such as $(e,\mu,\tau)$. In
the SM neutrinos  $\nu_e$, $\nu_\mu$, and  
$\nu_{\tau}$  are described by the fields 
that  form the weak doublets (or weak charged currents) with
charged lepton fields of definite mass:
\be
J^{\mu} = \bar{l} \gamma^{\mu} (1 - \gamma_5)\nu_l, ~~~ l = e,
\mu, \tau. \label{cc} \ee
Note that phenomenologically defined
flavor states $|\nu_f \rangle$, as states produced in certain 
weak processes, may 
not coincide precisely with theoretical flavor states - the states
produced by the fields from certain weak doublets. The difference can 
appear because  of 
neutrino mixing with heavy neutral leptons that  cannot be
produced in low-energy weak processes owing to kinematics. In
fact, this situation is realized in the seesaw mechanism. However, 
in  standard situation, 
the admixture and difference of the states is negligible.

Flavor mixing means that the flavor neutrinos $\nu_{\alpha}$
($\alpha = e, \mu, \tau$) do not coincide with neutrinos of definite  mass
$\nu_i$ ($i = 1,2,3$). 
The electron, muon
and tau neutrino states have no definite masses but turn out to be 
coherent combinations of the mass states.
The weak charged current processes mix neutrino mass states.  

The relationship between the  flavor fields, $\nu_f^T \equiv (\nu_e, 
\nu_\mu, \nu_\tau)$  and the mass fields $\nu^T \equiv (\nu_1, \nu_2,
\nu_3)$ can be written as 
\be \nu_f = U_{PMNS} \nu, \label{pmns}
\ee 
where $U_{PMNS}$ is $3 \times 3$ unitary matrix termed the
Pontecorvo - Maki - Nakagawa - Sakata lepton mixing
matrix~\cite{pontosc,mns}. 
The fields $\nu_f^T \equiv (\nu_e, \nu_{\mu}, \nu_\tau)$ form the
flavor basis. Inserting
eq.~(\ref{pmns}) into  eq.(\ref{cc}) we can write the  weak
charged currents as \be J^{\mu} = \bar{l} \gamma^{\mu} (1 -
\gamma_5) U_{PMNS} \nu. \label{ccmix} 
\ee 
Thus, the lepton mixing
matrix connects the neutrino mass fields and charged lepton
fields with definite mass in the weak charged currents.
Neutrino mass states are the eigenstates of the Hamiltonian in a 
vacuum,  and we refer to the  mixing in eq.(\ref{pmns}) as vacuum mixing.
Vacuum mixing is  generated by the non-diagonal mass matrices.

In addition to  the flavor $\nu_f$ and mass bases let us introduce the
{\it ``symmetry'' basis} $(\tilde{\nu}, \tilde{l})$ - the basis in which
the underlying  theory of the fermion masses is presumably
formulated. This can be some flavor symmetry, or GUT, or  some
dynamical principle, or selection rule originating from  string
theory. We do not know this basis a priori, and in fact, its
identification is one of the key problems in the bottom-up
approach.

In the symmetry basis, both the charged lepton and the
neutrino mass matrices are in general nondiagonal (although  models exist
in which the symmetry basis coincides with the flavor basis).
The  mass terms of the Lagrangian can then be written as
\begin{eqnarray}
{\cal L}_m~=~ \tilde{\nu}^T_LC^{-1} {\cal M}_\nu \tilde{\nu}_L +
\bar{\tilde{l}}_L M_l \tilde{l}_R + h.c. . \label{massterm}
\end{eqnarray}
We have assumed that neutrinos are Majorana particles. (Notice
that in the flavor basis $(\nu_l, l)$ the mass matrix of charged
leptons is diagonal, so the  existence of mixing implies that
the mass matrix of neutrinos should be non-diagonal in this basis.)
We diagonalize the matrices  in eq.(\ref{massterm}) as
\be 
U^T_\nu {\cal M}_\nu U_\nu = M^d_\nu, ~~~~~~  U_l M_l
V^{\dagger}_l = M^d_l ,
\label{diagmat} 
\ee
where  $M^d_\nu \equiv diag(m_1,m_2,m_3)$ and
$M^d_\ell \equiv diag(m_e, m_{\mu}, m_{\tau})$ are the diagonal matrices
and the rotation matrices, $U_{\nu}, U_\ell, V_\ell$,
connect the symmetry fields with the mass fields:
$\tilde{\nu} = U_\nu \nu$, $\tilde{l}_L = U_l l_L$, $\tilde{l}_R =
V_l l_R$. Plugging these relations into the charged current we
obtain
\be 
J^{\mu} = \bar{ \tilde{l}} \gamma^{\mu} (1 - \gamma_5)
\tilde{\nu} = \bar{l} \gamma^{\mu} (1 - \gamma_5) U_l^{\dagger}
U_{\nu} \nu. \label{ccmixgen} 
\ee 
Thus, the physical neutrino mixing
matrix is given by
\begin{eqnarray}
 U_{PMNS} ~=~U^{\dagger}_l U_\nu \label{ell}.
\end{eqnarray}


It is convenient to parameterize the mixing matrix as
\be
U_{PMNS}
= U_{23} (\theta_{23}) U_{13} (\theta_{13}, \delta) U_{12}
(\theta_{12}) I_{\phi},
\ee
where the  $U_{ij}$  are matrices of rotations
in the $ij$ plane by angle $\theta_{ij}$,  $\delta$ is the Dirac
CP-violating phase attached to 1-3 rotation.
In the case of Majorana neutrinos
sometimes the mixing matrix is defined as
$U_{PMNS}' =  U_{PMNS}I_{\phi}$, where
$I_{\phi} \equiv diag (1, e^{i\phi_1},
e^{i\phi_2})$ is the diagonal matrix of the Majorana CP-violating
phases. The latter can be incorporated into the mass eigenvalues which
can then be considered  as  complex parameters.

Note that in the first approximation 
the above parameterization  allows us to connect
immediately the rotation angles with physical observables 
$\theta_{23} = \theta_{atm}$, is the angle
measured in the atmospheric neutrino oscillations; $\theta_{12} =
\theta_{sol}$ is the angle determined from solar neutrino studies,
and $\theta_{13} = \theta_{CHOOZ}$ is the angle restricted by the
reactor experiment CHOOZ.


\subsection{Experimental results and global fits}

We use the results of global analysis of the neutrino
data published through the middle of 2006. The 
analysis assumed that (i)  there are only three mixed, active neutrinos;
(ii) CPT is conserved, so that masses and mixing angles in the
neutrino and antineutrino channel coincide; and (iii)  neutrino masses
and mixings have pure ``vacuum origin'': that is, due to
the interaction
with Higgs field(s) that develop a  VEV at a scale  
much  larger than neutrino mass. We comment below on
possible changes when some of these assumptions are abandoned.

The parameter space includes the oscillation parameters: 
mass-squared differences $\Delta m^2_{ij} \equiv m_i^2 - m_j^2$, mixing
angles $\theta_{ij}$ as well as the Dirac  CP-violating phase $\delta$. 
Non-oscillation parameters are the absolute mass scale which can
be identified with the mass of the heaviest neutrino
and two  Majorana CP-violating phases.

The experimental results used in the analysis can be split into
three sectors:

1). Solar neutrinos~\cite{homestake} and
the reactor experiment KamLAND~\cite{kamland} are sensitive
to mainly $\Delta m^2_{21}$ and $\theta_{12}$ (solar sector). The 1-3
mixing, if not zero, may give subleading effect.

2). Atmospheric neutrino studies~\cite{atm},  K2K~\cite{k2k} and MINOS~\cite{minos}    
accelerator
experiments  are sensitive to $\Delta m^2_{23}$ and
$\theta_{23}$ (atmospheric sector). The solar parameters  $\Delta
m^2_{21}$ and $\theta_{12}$ as well as $\theta_{13}$ 
give small subleading effects. 

3). The CHOOZ experiment~\cite{CHOOZ} gives a bound  $\theta_{13}$
as a function of $\Delta m^2_{31}$.

The physical effects involved in the interpretation are

- vacuum oscillations \cite{pontosc,mns,pontprob} (atmospheric
neutrinos - main mode, K2K, MINOS and CHOOZ);

- the Mikheyev-Smirnov-Wolfenstein (MSW) effect - the adiabatic conversion~\cite{w1,ms1} 
(conversion
of  solar neutrinos in the matter of the Sun; at low energies
solar neutrinos undergo the averaged vacuum oscillation with small
matter effect); and 

- oscillations in matter  ~\cite{w1,ms1} (oscillations of solar and atmospheric
neutrinos in the matter of Earth). These oscillations produce
sub-leading effects  and have not yet been  established  at the
statistically significant level.


Let us consider the results of global analysis from \cite{sv,bari}.



The mass splitting responsible for the dominant mode of the
atmospheric neutrino oscillations equals 
\be 
|\Delta m_{32}^2| =
(2.4 \pm 0.3) \cdot 10^{-3} ~ {\rm eV}^{2}, ~~(1\sigma).
\label{atmspl} 
\ee 
The sign of the mass split determines the type
of mass hierarchy:  normal, $\Delta m_{32}^2 > 0$,   or inverted, 
$\Delta m_{32}^2 <0 $, and it  is not identified yet.
The result of eq.(\ref{atmspl}) allows one to get a lower bound on the heaviest
neutrino mass:
\be m_h  \geq \sqrt{|\Delta m^2_{13}|} > 0.04~ {\rm
eV},  ~(2\sigma), \label{eq:ab1}
\ee
where $m_h = m_3$ for the
normal mass hierarchy,  and $m_h = m_1 \approx m_2$ for the
inverted hierarchy.

A much smaller mass-squared splitting drives the solar neutrino
conversion and oscillations detected by KamLAND \be \Delta
m_{12}^2 = (7.9 \pm  0.4) \cdot 10^{-5}~~ {\rm eV}^{2},
~~(1\sigma). \ee In the case of the hierarchical mass spectrum
that would correspond to $m_2 \sim 0.009$ eV. The ratio of the
solar and atmospheric neutrino mass scales,
\be r_{\Delta} \equiv \frac{\Delta m_{21}^2}{\Delta m_{31}^2} =
0.033 \pm 0.004, 
\label{r-delta}
\ee
gives a lower bound on the normal  mass hierarchy 
\be
\frac{m_2}{m_3} \geq \sqrt{r_{\Delta}} =  0.18 \pm 0.01. 
\label{rmass} 
\ee 


\begin{figure}[!tbp]
\begin{center}
\epsfxsize9cm\epsffile{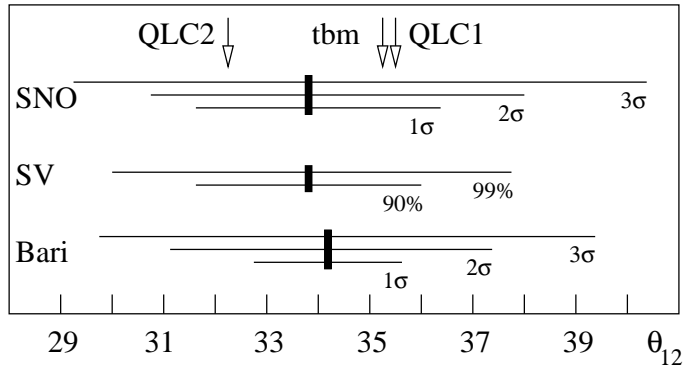}
\end{center}
    \caption{The best-fit values and
the allowed regions of lepton mixing angle $\theta_{12}$ at
different confidence levels determined by different groups. From
SNO~(from ref.\cite{homestake}), SV~\cite{sv} and Bari~\cite{bari}.
Shown are
predictions from quark-lepton complementarity (QLC) and tri-bimaximal 
mixing (see secs. 4,5).}
  \label{12mix}
\end{figure}

\begin{figure}[!tbp]
\begin{center}
\epsfxsize8cm\epsffile{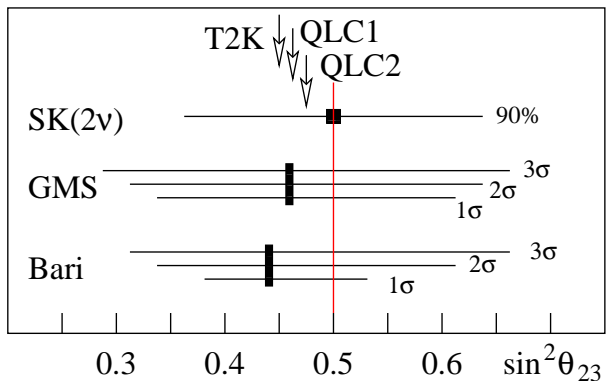}
\end{center}
\caption{The best-fit values and the allowed regions of
$\sin^2 \theta_{23}$ at different confidence levels
determined by different groups: SK~\cite{atm}, GMS~\cite{concha},
Bari~\cite{bari}. Shown are expectations from QLC (sec. 5) and sensitivity
limit of T2K experiment~\cite{T2K}.}
  \label{23mix}
\end{figure}

\begin{figure}[!tbp]
\begin{center}
\epsfxsize10cm\epsffile{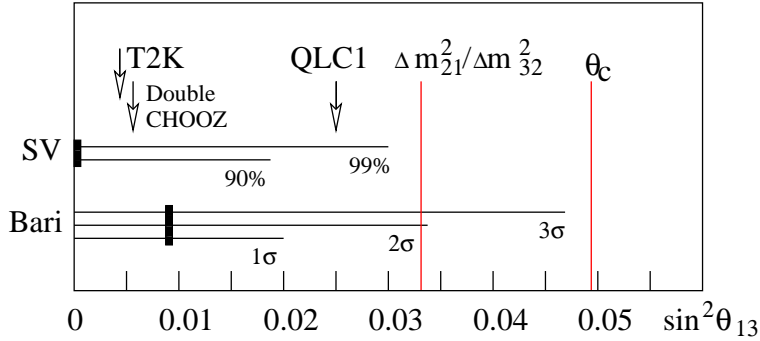}
\end{center}
\caption{The best-fit values and allowed regions of
$\sin^2\theta_{13}$ at different confidence levels
determined by different groups: SV~\cite{sv} and Bari~\cite{bari}.
Shown are also some theoretical predictions and  sensitivity limits of
Double CHOOZ~\cite{DC} and T2K~\cite{T2K}.} \label{13mix}
\end{figure}

In figs. \ref{12mix}, \ref{23mix}, \ref{13mix} we summarize
results of determination of the mixing angles obtained by
different groups. We also show some theoretical benchmarks that
will be discussed below. 

The best-fit value of the 1-2 mixing angle and $1\sigma$ error equal 
\be 
\theta_{12} =
33.9^{\circ} \pm 1.6^{\circ}, 
\ee 
or $\sin^2 \theta_{12} = 0.315^{0.028}_{-0.025}$ (see fig.~\ref{12mix}). 
It deviates from maximal mixing,
$\sin^2 \theta_{12} = 0.5$,
by approximately  $(6 - 7) \sigma$. 

The 2-3 mixing is in agreement with  maximal mixing,  $\theta_{23}
= \pi/4$ (fig.~\ref{23mix}).  A shift of the best-fit point from
$\pi/4$ to smaller angles appears when the effect of the 1-2 mass
splitting and mixing is included in the analysis \cite{concha,bari}.
According to \cite{concha}, $\sin^2\theta_{23} = 0.47$, and  a slightly
larger shift, $\sin^2\theta_{23} = 0.44$,  follows from 
analysis \cite{bari}. Thus, the deviation from maximal mixing
can be quantified as
\be D_{23} \equiv 0.5 - \sin^2\theta_{23} \sim 0.03 - 0.06. \ee
The shift is related to an  excess of the so-called $e$-like
atmospheric neutrino events in the sub-GeV range. The excess can
be explained  by the oscillations driven by the ``solar'' oscillation parameters
and it is proportional to the deviation $D_{23}$ \cite{orl}.
The experimental  errors  (the lower bound $\sin^2 2\theta_{23} > 0.92$) still allow 
substantial deviations from
maximal mixing:
\be D_{23}/\sin^2\theta_{23} \sim 0.4 ~~~(2\sigma). \ee

Results on the 1-3 mixing are consistent with zero $\theta_{13}$ (fig.
\ref{13mix}). A small, nonzero best-fit value of $\sin^2
\theta_{13}$ from the analysis \cite{bari} is related to the
angular distribution  of the multi-GeV $e$-like events measured by
Super-Kamiokande \cite{atm}. The most conservative $3\sigma$ bound
is \cite{bari} \be \sin^2 \theta_{13} < 0.048, ~~~  (3 \sigma).
\ee

%

In the first approximation, the pattern of lepton mixing has been
established. There are two large mixings: The 2-3 mixing is consistent with
maximal mixing, 1-2 mixing is large but not maximal, and 1-3  is small and
consistent with zero. 
Further precise measurements of the mixing angles and,  in
particular, searches for the deviations of 1-3 mixing from zero
and of 2-3 mixing - from maximal,  are crucial
to  understanding the underlying physics.
The figures show the accuracy required to make important
theoretical conclusions and the potential  of
next generation experiments. \\

\begin{figure}[!tbp]
\begin{center}
\hspace{-0.1cm} \epsfxsize8cm\epsffile{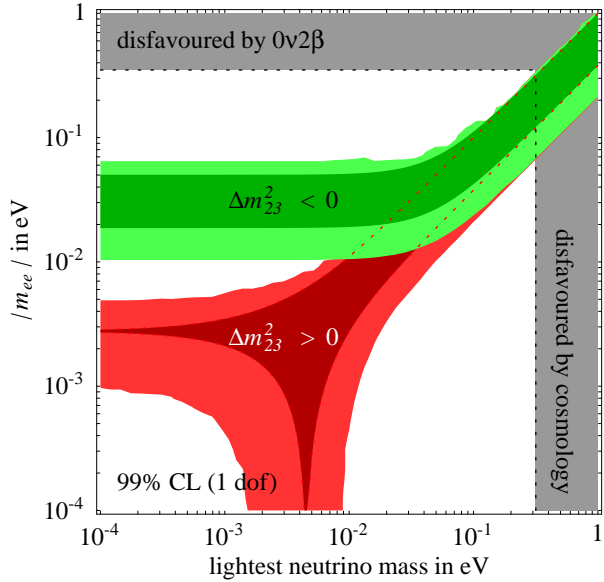}
\caption{The 99\% C.L. range for $m_{ee}$ as a function of the
lightest neutrino mass for the normal ($\Delta m_{23}^2 > 0$) and
inverted ($\Delta m_{23}^2 < 0$) mass hierarchies. The
darker regions show the allowed range for the present best-fit
values of the parameters with negligible errors; from \cite{sv}. }
\label{bb}
\end{center}
\end{figure}

Information on non-oscillation parameters  has been  obtained from
the direct kinematical measurements, neutrinoless double beta
decay and cosmology. 
The effective Majorana  mass of the electron neutrino - the
$ee$-element of the neutrino Majorana mass matrix, $m_{ee}$, determines the
life time of the neutrinoless double beta decay: $T(2\beta 0\nu) \propto 
m_{ee}^{-2}$. In terms of masses and mixing parameters,  it can be
written as
\begin{equation}
m_{ee} = |\sum_k U_{ek}^2 m_k e^{i\phi(k)}| = \left| \sum_k
U_{ek}^2   \sqrt{m_L^2 + \Delta m_{kL}^2 } e^{i\phi(k)} \right|,
\label{eq:it}
\end{equation}
where $\phi(k)$ is the phase of the $k$ eigenvalue,  $U_{ek}$ its
admixture in $\nu_e$, and $m_L$ -  the lightest  neutrino mass.
Fig.~\ref{bb}  from \cite{sv} summarizes the present knowledge
of the absolute mass scale. The regions allowed by
oscillation results in the plane of $m_{ee}$
and $m_L$ - the mass of the lightest neutrino  probed by the  direct
kinematical methods and cosmology are shown. The two bends correspond to the
normal and inverted mass hierarchies. For a given $m_1$, the range
of $m_{ee}$ is determined by variations of the Majorana phases
$\phi$ and uncertainties in the oscillation parameters.

The  Heidelberg-Moscow experiment gives the 
best present bound on $m_{ee}$:  $m_{ee} < (0.35 - 0.50)$ eV 
\cite{HM-neg}. Part of the collaboration claims an evidence for a positive signal 
\cite{HM-pos} which would correspond to $m_{ee} \sim 0.4$ eV. If this positive signal is 
confirmed and if it is due to the exchange of light Majorana neutrinos, the neutrino 
mass  spectrum is  strongly degenerate: $m_{1} \approx m_2 \approx m_3 \equiv m_0$ 
\cite{deg}.

The cosmological observations put a bound on the  sum of neutrino
masses\\ $ \sum_{i = 1}^3 m_i  < 0.42~ {\rm eV}$ $(95\%~ C.L.)$
\cite{cos} (see also \cite{goobar}) which corresponds to $m_0 < 0.13$ eV  
in the case of a degenerate spectrum.  An even stronger bound,
$ \sum_{i = 1}^3 m_i  < 0.17~ {\rm eV}$ ($95\%$ C.L.) \cite{Seljak06}
was established after publication of WMAP3 results.
This limit disfavors a strongly degenerate mass spectrum.
Combining the cosmological and  oscillation Eq. (\ref{eq:ab1})
bounds,  we conclude that at least one neutrino mass should be in
the interval
\be
m \sim  (0.04 - 0.10) ~ {\rm eV} ~~~(95\% ~~ {\rm C.L.}).
\ee

Direct kinematic measurements give the weaker limit  $m < (2.0 -2.2)$ eV
\cite{troitsk}. The planned experiment KATRIN \cite{katrin} is
expected to improve this limit down to $\sim 0.2$ eV.

How robust are these results? Can we expect some substantial
change in this picture in future? There are three types of effects
(in fact, related to lifting of assumptions made in the analysis)
that can influence the interpretation of the present neutrino
results:

1). Possible existence of new neutrino states - sterile neutrinos.
If these states are light, they can directly (dynamically)
influence the observed effects used to determine neutrino
parameters.

2). Possible presence of non-standard neutrino interactions
can change values of the extracted neutrino parameters.

3). Interactions with light scalar fields \cite{scalar} that can produce
the soft neutrino masses that  depend on properties of the medium.
These masses may also change with time and be related to dark
energy in the universe \cite{mavan}.

At present, however there are  no well-established results that 
indicate deviation from the ``standard'' $3\nu$ mixing and
standard matter interactions.

There are various tests of validity of the standard picture and of the 
theory of neutrino conversion. Different sets of data confirm
each other. Consistent interpretation of whole bulk of various results in
terms of the vacuum masses and mixing provides 
further confidence.
The fit of the data is not improved substantially with inclusion of new states and 
non-standard interactions and, if  exist, they may produce sub-leading
effects only. One can perform, {\it e.g.}, a  global fit of
neutrino data considering the normalization of the matter potential as
free parameter. According to \cite{bari} the best-fit value of the
potential is close to the standard one. In this way one tests not
only the validity of the refraction theory for
neutrinos, but also the consistency of the whole picture.\\

\subsection{Mass and flavor spectrum}

\begin{figure}[!tbp]
\begin{center}
\hspace{-0.1cm} \epsfxsize9cm\epsffile{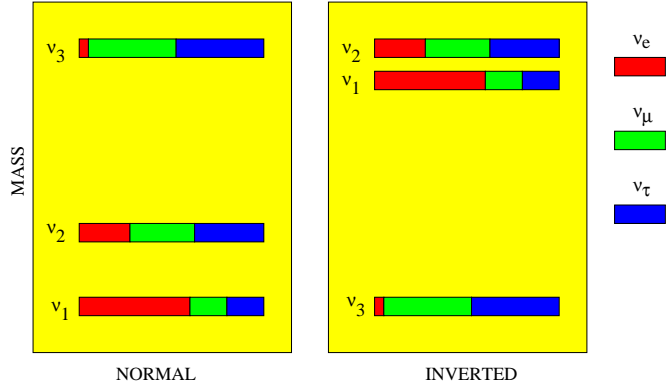} \caption{Neutrino
mass and flavor spectra for the normal (left) and inverted (right)
mass hierarchies. The distribution of flavors (colored parts of
boxes) in the mass eigenstates corresponds to the best-fit values
of mixing parameters  and  $\sin^2 \theta_{13} = 0.05$. }
\label{sp}
\end{center}
\end{figure}

Information obtained from the oscillation experiments allows us to
partially reconstruct the neutrino mass and flavor spectrum
(Fig.~\ref{sp}).
There are four unknowns: 
 (i) the admixture of $\nu_e$ in $\nu_3$
described by  $U_{e3}$; (ii) the type of mass spectrum -  hierarchical;
non-hierarchical with a certain ordering; or degenerate, which is
related to the value of the absolute mass scale, $m_1$; (iii) the type
of mass hierarchy (ordering): normal, or inverted; and (iv) the  CP-violating
phase $\delta$.  Determining these unknowns is the goal of future
phenomenological and experimental studies.


There are some weak indications in favor of a normal mass hierarchy
from supernova SN1987A data.
However in view of small statistics and uncertainties in the
original fluxes it is not possible to make a firm statement.

As is clear from the fig. \ref{bb}, future high-sensitivity
measurements of the effective mass $m_{ee}$ can allow one to
establish the hierarchy.  The bound $m_{ee} < 0.01$ eV will exclude
the inverted mass hierarchy and also degenerate mass spectrum if neutrinos are Majorana 
particles. 
Future detection of a  galactic supernova can also contribute to
the determination of the type of mass hierarchy and 1-3 mixing
\cite{DS}.

\subsection{Toward the underlying physics}

What is behind all these observations?
To uncover the underlying physics two general strategies are
invoked:

\noindent 
1.  The  Bottom-Up approach 
is essentially  an attempt to uncover the underlying physics
starting from observations. The strategy is to (i) reconstruct the
neutrino mass matrix in the flavor basis using the 
information available on masses and mixings ($\Delta m^2_{ij}$,
$\theta_{ij}$, $m_{ee}$); (ii)  take into account the
RGE effect and obtain the mass matrix at the
scale of new physics, (iii) search for the symmetry basis in
which flavor or some other  symmetry is realized; and (iv) identify
the symmetry and  mechanism of symmetry violation, if needed, as
well as the  underlying dynamics. This is our  standard  way of 
understanding  things, but it does  not preclude that the explanation will
require something new.

\noindent
2.  Top-Down approach  starts with a
general unified theory  framework, be it a GUT,
TeV scale theory or extra dimension theory which has motivation
outside neutrino physics - and uses it to make predictions for
neutrinos, i.e.,  to go from the big picture to the observed properties
of neutrinos.

At some point these two approaches should merge. Both approaches
are needed. It seems difficult to uncover the underlying picture
by just moving from observations, and some {\it a priori} ideas (or
context) are needed to relate neutrinos with other physics. This
opens the  possibility of using results from other areas ({\it e.g.}, collider
experiments). Thus, we need the big picture. However,
working solely within the big picture, one can miss some
important elements, which is where the bottom-up approach can help. 
The results of the  bottom-up approach are considered in sect.  3 
- 5. The top-down approach is developed in sect. 6 - 7.

\section{Mixing and mass matrix of neutrinos}

Analyzing  results on neutrino mixing and mass matres one can  (i)
search for some particular features in the data, such as  empirical
relations,  equalities, hierarchies of elements,  zeros, {\it
etc.}; (ii) identify possible dominant structures in the mixing
and mass matrices (the idea being that matrices can have a
structure as ``lowest order plus small corrections'' which in turn
can correspond to some dominant mechanism plus  sub-dominant
effects); (iii) search for simple parameterizations in terms of
small number of parameters; and (iv) present matrices in terms of
powers of some small quantity, {\it etc.}. All these may hint at the
underlying physics.

\subsection{Properties of neutrino mixing matrix}

Let us first to analyze the mixing matrix.
Maximal 2-3 mixing, large 1-2 mixing and small 1-3 mixing indicate
that the  following matrices can play some important role
({\it e.g.},  be the lowest-order mixing matrices).

1). The bimaximal mixing matrix \cite{bim}: 
\be 
U_{bm} = U_{23}^m
U_{12}^m, 
\ee 
where $U_{23}^m$ and  $U_{12}^m$ are the matrices of
the maximal ($\pi/4$) rotations in the 2-3 and 1-2 subspaces
respectively. Explicitly,  we have
\be U_{bm} = \frac{1}{2} \left(\begin{array}{ccc}
\sqrt{2} & \sqrt{2} & 0\\
-1 & 1 & \sqrt{2}\\
1 & - 1 & \sqrt{2}
\end{array}
\right). \label{bimax} 
\ee 
The identification $U_{PMNS} = U_{bm}$ is
not possible owing to strong  deviation of the 1-2
mixing from maximal. However, $U_{bm}$ can play the role of a
dominant structure. In the latter case, the correction can
originate from the charged lepton sector (mass matrix), so that
$U_{PMNS} = U'U_{bm}$.  Suppose $U' \approx U_{12}(\alpha)$, where 
$\alpha \sim O(\theta_C)$ in
analogy to  quark mixing. 
Then $U'$  simultaneously generates deviations of the 1-2 and 2-3 mixing
from maximal as well as  non-zero 1-3 mixing which are related as
$$
\sin \theta_{13} =  \frac{\sin\alpha}{\sqrt{2}}
$$
and
$$
\sin \theta_{23} \simeq \frac{1}{\sqrt{2}}\left(1 - 
\sin^2\theta_{13} \right), 
$$
or $D_{23} \approx  \sin^2\theta_{13}$. Confirmation of this
equality will be very suggestive.

2). The tri-bimaximal mixing matrix~\cite{tbm} \be U_{tbm} =
U_{23}^m U_{12}(\theta_{12}), ~~~ \sin^2 \theta_{12} = 1/3, \ee or
explicitly, \be U_{tbm} = \frac{1}{\sqrt{6}}
\left(\begin{array}{ccc}
2 & \sqrt{2} & 0\\
-1 & \sqrt{2} & \sqrt{3}\\
 1 & - \sqrt{2} & \sqrt{3}
\end{array}
\right) \label{bimax1}
\ee
is in good agreement with data
including 1-2 mixing. Here $\nu_2$ is trimaximally mixed: In the
second column the three flavors mix maximally, whereas $\nu_3$ is
bi-maximally mixed. Mixing parameters turn out to be simple
numbers such as  0, 1/3, and 1/2 and can appear as the Clebsch-Gordon
coefficients.

The bi-maximal and tri-bimaximal  matrices can be considered as
matrices in the  lowest order of  some approximation. Then one can
introduce  parameters that describe deviations of the true matrix
from these lowest-order structures \cite{xing}.
The matrices $U_{bm}$ and $U_{tbm}$ reveal certain symmetries,
and the deviation parameters may describe effects of the  violation of
these symmetries.

\subsection{Reconstructing neutrino mass matrix}

The mass matrix is probably more fundamental than mass eigenvalues and
mixing angles because it combines information about masses and
mixings. Dynamics and symmetries can be realized in terms of mass
matrices and not their eigenstates and eigenvalues. However, it is
possible (and models of this type exist, see below) that symmetry 
immediately determines  the mixings but  not the  masses which are 
left
as free parameters.

As mentioned above, the first step in the bottom-up approach is
the reconstruction of the mass matrix in flavor basis. Note that
in the case of Majorana neutrinos, the elements of this mass
matrix are physical parameters.  They can be  measured directly,
{\it e.g.}, in  neutrinoless double beta decay and, in principle, in
other similar processes.


The answer to the question what is more fundamental:
mass matrices or observables ($\Delta m^2$, $\theta$), 
may depend on the type
of mass spectrum. In the case of a 
hierarchical spectrum, the observables are  imprinted visibly into
the structure of the mass matrix. In contrast, for the
quasi-degenerate spectrum they are just very small perturbations
of the dominant structure  determined by the non-oscillatory
parameters: the absolute mass scale and the Majorana CP-violating
phases. The oscillation parameters can originate from some small,
in particular radiative, corrections.

In the flavor basis the mass matrix of the charged leptons is
diagonal and therefore the  neutrino mass matrix is diagonalized
by $U_{\nu} = U_{PMNS}$. Consequently, according to
eq.(\ref{diagmat}) the neutrino  mass matrix in the flavor basis
can be written as
\begin{equation}
{\cal M}_{\nu} = U^*_{PMNS} {\cal M}^{d}_{\nu} U^{\dagger}_{PMNS},
\label{eq:mass}
\end{equation}
where
\begin{equation}
{\cal M}^{d}_{\nu} \equiv diag (m_1, ~ m_2e^{-2i\phi_2} ,~ m_3
e^{-2i\phi_3}). \label{eq:mass1}
\end{equation}
Here $\phi_i$ are the Majorana phases
and we take here $\phi_1 = 0$. Results of reconstruction
show \cite{Frige} that a large
variety
of different structures of mass matrices is possible, depending
strongly on the unknown $m_1$, type of mass hierarchy and Majorana
phases. The dependence on $\sin \theta_{13}$ and $\delta$ is
relatively weak. This means  huge degeneracy of mass matrices now
and perhaps even in the far future because,  in reality, it is not
possible to measure all the parameters including CP-violating
phases. Variations of one Majorana phase (even if all other
parameters are known) can lead to a 
 strong change in the structure. Nature would have to be very 
``collaborative''
for us to determine the mass matrix. Or we may uncover some
principle that  allows us to predict the mass matrix,
which we can then check by certain precision measurements.

\subsection{Extreme cases}

To give some idea about various possibilities, we present
simple parameterizations of the neutrino mass matrix in the
flavor basis for three extreme cases: normal mass hierarchy,
inverted mass hierarchy
and degenerate spectrum. \\

1). Normal mass hierarchy ($m_1 \ll m_2 \ll m_3$).
The mass matrix indicated by data can be parameterized as
\begin{eqnarray}
{\cal M}_\nu~=~\frac{\sqrt{\Delta m^2_{31}}}{2}\pmatrix{d\epsilon &
b\epsilon & a\epsilon\cr b\epsilon & 1+c\epsilon & -1\cr a\epsilon
& -1 & 1+\epsilon}, \label{normhier}
\end{eqnarray}
where $a,b,c,d$ are complex parameters of the order one, and
$\epsilon$ is essentially the ratio of the solar and the
atmospheric mass hierarchies: $\epsilon =  2\sqrt{r_{\Delta}} F(a,
c,b,d)$, and   $F(a, c,b,d) \sim O(1)$. A salient feature of this matrix
is the dominant $\mu - \tau$ block. With the present
accuracy of measurements of the parameters, 
a sharp difference between dominant and subdominant  elements may 
not exist and some moderate hierarchy of elements with the unique
expansion parameter 0.6 - 0.7 is realized \cite{Frige}.

An important property of the above mass matrix  is that in the
limit of $a= b$ and $c=1$, it is symmetric under $\mu-\tau$
interchange and one gets maximal 2-3 mixing and zero 1-3
mixing, {\it i.e.} $\theta_{23}=\frac{\pi}{4}$ and
$\theta_{13}=0$ \cite{mutau}.
This symmetry should  of course be approximate one, as the masses of
muon and tau leptons are different. Any resulting $\mu-\tau$
breaking must therefore reflect itself in
nonzero $\theta_{13}$ and $D_{23}$ with the two connected to each
other \cite{mutaubr}. For example, if $a=b$ and $c\neq 1$, i.e.,
the symmetry is broken in the dominant block, the induced
$\theta_{13}$ and $D_{23}$ are  given by
\be \sin \theta_{13} = -
b\epsilon^2 (c - 1)/4, ~~~~~ D_{23} = \epsilon (c - 1)/4 ,
\ee
and therefore are related as 
\begin{equation}
\tan \theta_{13} = b\epsilon D_{23}.
\end{equation}
Furthermore, $\theta_{13} \ll D_{23}$. In contrast, if the
symmetry is broken in the sub-dominant block only: $a \neq b$ but $c =
1$,  the situation is opposite: $\theta_{13} \gg D_{23}$, {\it i.e.}
\begin{eqnarray}
\sin \theta_{13}\simeq \frac{a-b}{\sqrt{2}}\epsilon,  ~~~~~
D_{23} = \frac{b^2-a^2}{8}\epsilon^2.
\end{eqnarray}
Thus,  measurements of  $\theta_{13}$ and
$D_{23}$ will provide an important probe of the mass matrix
structure. 
Note that when $a=b=d$ and $c=1$, we get the tri-bi-maximal
mixing pattern.\\

2). Inverted mass hierarchy ($m_1 \approx m_2 \gg m_3$).
The structure of the mass matrix in this case depends strongly on
the Majorana CP-violation phase. An approximate form of the mass matrix in
the case of opposite CP-parities of $\nu_1$ and $\nu_2$ is 
\begin{eqnarray}
{\cal M}_\nu=\sqrt{\Delta m^2_A}~\left(\begin{array}{ccc}
z\epsilon & c & s\\ c & y\epsilon& d\epsilon\\
s & d\epsilon & x\epsilon
\end{array}
\right), 
\label{invma}
\end{eqnarray}
where $\epsilon \ll 1$. In the limit $\epsilon\rightarrow 0$,
this mass matrix has the symmetry $L_e-L_\mu-L_\tau$
\cite{emutau}. In the symmetry limit one has $\Delta m^2_{12} = 0$ and
$\theta_{12}=\pi/4$. Furthermore, if an additional $\mu-\tau$
exchange symmetry is imposed on this mass matrix, the atmospheric
mixing angle also becomes maximal.

The breaking of $L_e-L_\mu-L_\tau$ symmetry leads to nonzero
$\Delta m^2_{12}$ and  deviation of $\theta_{12}$ from maximality. 
It has, however, been difficult in general, although not
impossible, to accommodate the observed ``large'' departure from
maximality of $\theta_{12}$ using ``small'' breakings of
$L_e-L_\mu-L_\tau$ symmetry. One needs as much as 40\% breaking to
fit data.

In the case of the same CP-parities of mass eigenstates the mass
matrix has completely different form with interchange of dominant
- sub-dominant elements in eq.(\ref{invma}). This illustrates
strong dependence of the matrix structure on the unknown CP
violating phases.
\\

3). Degenerate spectrum: $m_1 \approx m_2 \approx m_3 = m_0$.

Here also the structure of mass matrix depends strongly on the
CP-violating phases. Two possibilities are particularly
interesting:

(i) If the relative phases between the mass eigenvalues are zero
($2\pi k$),  the mass matrix is close to the unit matrix 
\be {\cal
M}_\nu = m_0 I + \delta M, 
\label{unitmat} 
\ee 
where $\delta M \ll
m_0$ is the matrix of small corrections.

(ii) In the case of opposite CP-parities of $\nu_2$ and $\nu_3$
\be {\cal M}_\nu = m_0~\left(\begin{array}{ccc}
1 & 0 & 0\\
0 & 0  & 1\\
0 & 1 & 0
\end{array}\right) + \delta M.
\label{trianglem} \ee

Both matrices give $m_{ee} = m_0$ and can  explain the
Heidelberg-Moscow positive result. 
Theoretical understanding of such a situation would  require
underlying symmetries that guarantee mass
degeneracy. The basic strategy here is to consider symmetries that
have a three-dimensional representation to which the three lepton
doublets of the standard model can be assigned and then design a
Higgs sector that will lead to a quasi-degenerate neutrino
spectrum. A list of such symmetries includes $S_4$~\cite{s4},
$SO(3)$~\cite{so3}, and $A_4$~\cite{a4}.

In this connection an interesting possibility is a theory with
mixed seesaw, where the type II contribution gives the dominant
quasi-degenerate term  $m_0 I$ or the first term in eq.(\ref{trianglem}) 
with the conventional type I contribution giving
mass splittings  and mixings: $ \delta M = - m^T_DM^{-1}_R m_D$. A
very generic way to see how these models could explain
observations is as follows:
In a quark-lepton unified theory such as an SO(10) model, we would
expect the Dirac mass term for the neutrinos to have a
hierarchical pattern for its eigenvalues so that roughly speaking,
the atmospheric and solar mass differences will be given by
$$
\Delta m^2_{13}\sim \frac{m_0 m^2_{D,33}}{M_{3}}, ~~~~
 \Delta m^2_{12} \sim
\frac{m_0 m^2_{D,22}}{M_{2}}
$$
respectively, roughly similar to observations.

The matrices of eqs.(\ref{unitmat},\ref{trianglem}) offer various
possibilities that relate the degeneracy of the spectrum with large
or maximal mixing. The matrix of eq.(\ref{trianglem}) 
leads immediately to maximal 2-3 mixing. Nearly maximal mixings are
generated by small off-diagonal elements in eq.(\ref{unitmat}).

\subsection{Mass matrices with texture zeros}

Another approach in analyzing possible mass matrices is to see if
some elements can be exactly zero or equal each other. This may 
also uncover dominant structures and certain underlying
symmetries. This approach allows one  to reduce the number of
free parameters and therefore can lead to certain predictions.
Recall that the Majorana mass matrix for three neutrinos has six 
independent elements.

Mass matrices with different  numbers of zeros and with zeros in
various places of matrix have been considered.
Two of the cases discussed widely in the literature are
 textures with  three zeroes and  two zeros.
It is easy to be convince that the three zeros cannot be
along the off-diagonal entries nor  be in any of the
$2\times 2$ submatrices and still give a fit to already known data.
In the former case all mixings vanish and in the latter case, one
cannot satisfy the requirement from observation that $\Delta
m^2_{12} \ll \Delta m^2_{23}$ if $\theta_{23}$ and  $\theta_{12}$
are large, as observed. 
The case when all zeros are along the diagonal~\cite{zee}
 (or two along the diagonal and the third is off diagonal)
 is more subtle because 
 now one can satisfy the requirements of
 large solar and atmospheric mixings as well as $\Delta
m^2_{12} \ll \Delta m^2_{23}$. However,  in this case, there are
only three (real)  parameters  which can be determined from
$\Delta m^2_{12}$  $\Delta m^2_{23}$ and $\theta_{23}$.
One then  predicts a value for the solar mixing angle, $\sin^2
2\theta_{12} = 1- r_{\Delta}/16$, which is incompatible with
observations. 

As far as textures with two zeros are concerned, they have five
free parameters: four real parameters and a complex phase and are
therefore interesting candidates for neutrino mass
matrices~\cite{frampton}. These have been analyzed to give their
characteristic predictions. There are seven different
(out of fifteen)  possibilities  that are
currently in accord with data and make predictions for various
parameters such as neutrinoless double beta decay and
$\theta_{13}$. As an interesting example, consider the matrix
\be
{\cal M}_\nu = \pmatrix{0 & 0 & X\cr 0 & X & X \cr X & X & X},
\ee
where $X$ indicate non-zero entries.
This leads to a hierarchical mass matrix with the prediction 
\be
\sin^2\theta_{13}\sim
\frac{r_\Delta}{\tan^2\theta_{12}-\cot^2\theta_{12}}\sim 0.01
\ee and
zero amplitude for neutrinoless double beta decay.

Another possible texture is
\be
{\cal M}_\nu = \pmatrix{X & X &
0\cr X & 0 & X \cr 0 & X & X}.
\ee
This produces a degenerate mass
spectrum with an effective mass in neutrinoless double beta decay
exceeding 0.1 eV.

Such an approach should be taken with some caution: (i) for
instance, it is not clear why the zeros appear in the flavor basis?
(ii) there is the possibility that interesting symmetries
 may not correspond to zeros in the flavor basis,  and (iii) finally,
 the zeros may not be
exactly zeros, in which case we are unlikely to learn much from
these exercises. In any case, such situations should be considered
on equal footing with cases in which  various elements of the neutrino
mass matrices are related or simply equal,
as in the case of $\mu-\tau$ symmetric models discussed in sec. 4.




\subsection{Anarchy approach}

Although the limiting examples of mass matrices described above may
contain interesting hints of symmetries, it is quite possible that
one is far from these cases. In a large part of the parameter
space, the allowed mass matrix has no clear
structure: All the elements are of the same order and could be
taken as random numbers. This fact, and the existence of two large
mixings, motivates the anarchy approach~\cite{anarchy} where all
parameters of the neutrino mass matrix are allowed to take random
values. One then calculates the probability that
 $\theta_{12}$, $\theta_{23}$ and
$\theta_{13}$ satisfy the experimental results. There can be
variations to this approach depending on whether the anarchy is
assumed to be in the high-scale sector of the theory (such as in
the right handed neutrino mass matrix in the seesaw models 
described below) or in low-scale sector. One feature of these
models is that in general they predict ``large'' $\theta_{13}$
(closer to $0.1$ or higher) for a large domain of parameters.
One can also consider partial ``anarchy'' -  randomness of some 
parameters on the top of a dominant structure determined by a  certain
symmetry. 
Questions for this approach include: which kind of physics leads to
the anarchy and  why does anarchy manifest itself 
for lepton mixings and neutrino
masses only? To some extend, studying  anarchy can be considered as
a test of complexity behind neutrino masses and mixings. In fact,
several comparable contributions to
the neutrino mass matrix may exist, each having a simple structure and
obeying a certain symmetry, but the totality of it having   the
anarchy effect.

\subsection{Renormalization group equation  effects}

If the underlying theory is formulated at some high-energy scale
$M$, {\it e.g.}, much above the electroweak scale, one needs to
use the renormalization group equations (RGE) to extrapolate the mass matrix 
from the low-energy scale, where it has been reconstructed, to the
high-scale $M$. In general, renormalization can change the
structure of the Yukawa coupling matrix. Thus, to uncover the
mechanism of mass generation one needs to calculate RGE
corrections~\cite{rad}. In fact, some features of the mass matrix and
observables at low scale can be due to RGE 
effects.

Suppose  neutrinos are Majorana particles and their masses are
generated at the EW scale  by the operator eq. (\ref{nonren})
(For renormalization of masses of the Dirac neutrinos
see Ref. \cite{manfD}.) Then, in general, two RGE effects should be taken
into account:

1).   renormalization of the operator in eq.(\ref{nonren}) from
the low scale up to the scale at which it is formed, and

2). renormalization between and above the scale of the operator
formation \cite{manfth}. In general, different terms of the
operator in eq.(\ref{nonren}) are formed  at different scales
separated  by many orders of magnitude rather than at the unique
scale. This happens, {\it e.g.}, in the case of the type I seesaw 
mechanism with strongly hierarchical masses of the right handed
neutrinos: $M_1 \ll M_2 \ll M_3$. The underlying physics is
formulated at  $M \geq M_3$. In this case one should take into
account threshold effects - a different renormalization group
running between the masses and also above the seesaw scales (see sect. 6).

In this section we  consider the RGE effects below the scale
of operator eq.(\ref{nonren}) formation. In the seesaw
version with a hierarchy  of the RH neutrino masses, that would correspond to
$\mu < M_1$,  where $\mu$ is the running scale.

In the SM as well as Minimal Supersymmetric Standard Model (MSSM), 
treatment of the RGE  effects on the
mass matrix in the flavor basis  is
simpler than RGE effects on observables - angles or
masses. The observables can be found after
renormalization by diagonalizing  the obtained mass matrix (matrix
of the Yukawa couplings).

The RG equation  for the effective mass matrix has a very transparent structure
\cite{rad,rge-eq}
\be
\frac{d {\cal M}_{\nu}}{dt} = C_l Y_l^{\dagger} Y_l  {\cal M}_{\nu} +  {\cal M}_{\nu}
 C_l Y_l^{\dagger} Y_l +  \alpha {\cal M}_{\nu},
\label{rge-eq} \ee where  $t \equiv (1/16\pi^2) \log (\mu/\mu_0)$,
$ C_l = -3/2 $ in the SM and $C_l = 1$ in MSSM. The first two
flavor dependent terms correspond to the neutrino wave function
renormalization  due to Yukawa couplings of the charged leptons,
the last term is the flavor independent renormalization due to
gauge couplings and also Yukawa coupling renormalization of the
Higgs field wave function.

\subsubsection{Renormalization of the neutrino mass matrix.}

In  lowest order the gauge couplings produce  the overall
renormalization of the mass matrix only and do not change its flavor
structure. (This is not true for threshold corrections because 
couplings of  different right-handed  neutrinos are flavor dependent
\cite{manfth}.) In contrast, the Yukawa interactions modify the
flavor structure of the mass matrices. In the flavor basis, Yukawa
corrections do not generate the off-diagonal elements of the
charged lepton mass matrix in the SM and MSSM. This matrix remains
diagonal and therefore the correction does not change the flavor
basis. On the contrary, RGE corrections change structure of the
non-diagonal neutrino mass matrix.

To understand the RGE effects we consider the one-loop
corrections and neglect all the Yukawa couplings except the tau
lepton one,  $Y_{\tau} = m_{\tau}/v_d$, where $v_d$ is the VEV
that generates masses of down fermions. Essentially, the RGE
effects are reduced to the wave-function renormalization and can be
written as
\be {\cal M}_{\nu}(\mu) = I_C(\mu) R(\mu){\cal M}_{\nu}(m_Z)
R(\mu).
\ee
Here  $I_C$ is flavor
independent renormalization factor
and
\be
R \approx diag(1,
1, Z_{\tau}(\mu)), ~~~~ Z_{\tau} - 1 = C_l \frac{Y_{\tau}^2}{16\pi^2}\log
\frac{\mu}{m_Z},
\ee
The size of the effect is different in the SM and MSSM: In the SM $v_d
= v  = 265$ GeV and $Y_\tau = 9.5\times 10^{-3}$ at the EW scale,  so the
corrections are very small: For $\mu = 10^{10}$ GeV we obtain
$(Z_{\tau} - 1)_{SM} \simeq 10^{-5}$. The effect is strongly
enhanced in MSSM with large $\tan\beta$, where $v_d \sim  v/\tan
\beta$, so that $(Z_{\tau} - 1)_{MSSM} \approx (Z_{\tau} - 1)_{SM} \tan^2 \beta$.
For $\tan \beta = 50$, we obtain $Z_{\tau} - 1 \sim 0.03$.

Corrections appear as factors multiplying the bare values of the matrix
elements. Thus, the zero elements will remain zeros \cite{massmatrrg}.
This allows us to draw important conclusion:
The RGE effects (at least in the SM and MSSM) do not change
the structure of the mass matrix significantly. Therefore the
mass matrix reconstructed at low energies will have nearly the
same form at high scales (before threshold corrections are
included). This is not true if some new interactions
exist above the EW scale (see {\it e.g.} \cite{frigren}).

In contrast, effect of corrections on the observables - angle
and mass differences - can be strong for particular forms of the
zero order mass matrix.

\subsubsection{Renormalization of observables.}

The strongest effect on the observables is in the case of the
quasi-degenerate mass spectrum \cite{degen,rad1,small}. Indeed,
the corrections are proportional to the absolute mass scale  $m_0$:
$\delta m  = m_0 (Z_{\tau} - 1)$.  It generates  the mass squared
difference  $\Delta m^2 \approx  2 m_0  \delta m = 2 m_0^2
(Z_{\tau} - 1)$. Effect increases as square of the overall mass
scale. For $m_0 = 0.3$ eV  and $(Z_{\tau} - 1) \sim 10^{-3}$ it
can give the solar mass split \cite{small}. Furthermore, the
mixing angles depend on mass differences whereas corrections are
proportional to the absolute values of the elements so that the
relative corrections to the mixing get enhanced \cite{rad1} by
$\Delta \theta \propto m/ \Delta m    \propto m_0^2/ \Delta
m^2 $. Essentially, the  enhancement of mixing occurs when
neutrinos become even more degenerate at low energies. 
In the case of normal hierarchy, however, the effect of the RGE's
is small.

Let us summarize possible effects \cite{degen,small,manf13,renphases,rad1,bmrad}.

1). The angle $\theta_{12}$ can undergo the strongest renormalization
since it is associated to the smallest mass split.
Some important dependences and results
can be traced from the approximate analytical expression for running
\cite{manf13}:
\be
\dot{\theta}_{12} \approx - A \sin 2\theta_{12}\sin^2\theta_{23}
\frac{|m_1 + m_2 e^{\phi_{12}}|^2}{\Delta m^2_{21}} +
{\cal O}(\theta_{13}),
\label{12mixren}
\ee
where $A \equiv  C_l Y_{\tau}^2/32\pi^2$ and
$\phi_{12} \equiv \phi_2 - \phi_1$. Notice that
other parameters in this formula, and especially $\Delta m^2_{12}$,
also run and  their dependence
on renormalization scale $\mu$ should be taken into account.
As a result the dependence of $\theta_{12}$ on the
$log \mu$ turns out to be  nonlinear.

According to eq.(\ref{12mixren}) for $\theta_{13} = 0$,
with increase of $\mu$ the angle
$\theta_{12}$ decreases in the MSSM and increases in SM.
For the degenerate spectrum
the enhancement factor can reach  $4 m_0^2/\Delta m^2_{12}$.
Maximal enhancement corresponds to zero relative phase,
$\phi_{12} = 0$, and
running is suppressed for the  opposite phases.

For the degenerate spectrum  the
angle $\theta_{12}$ can run practically to zero.
This means that the large 1-2 mixing at low energies may have
the  radiative origin being small at, {\it e.g.},  the  GUT scale
\cite{degen,rad1}.
Another interesting possibility is that at $M_{GUT}$
equality of  the quark and leptonic 1-2 mixings, $\theta_{12} = \theta_C$,  
is realized.
So, the difference of quark and lepton mixings
is related via the RGE running to degeneracy of the neutrino  spectrum.

In MSSM the angle $\theta_{12}$  can increase with $\mu$
due to effect of non-zero $\theta_{13}$ provided that  $\phi_{12} = \pi$, which
equality ensures that the effect of the main term in eq. (\ref{12mixren}) is 
suppressed.

2). Evolution of 1-3 mixing  associated to the larger
mass split is   weaker and nearly linear in  $\log \mu$.
It can be approximated as
\begin{eqnarray}
\dot{\theta}_{13} \approx  A \sin 2\theta_{12}\sin 2\theta_{23}
\frac{m_3}{\Delta m^2_{13}}
[m_1 \cos(\phi_1 -\delta) -  m_2 \cos(\phi_2 -\delta)
\nonumber\\
- r_{\Delta} m_3 \cos\delta] +
{\cal O}(\theta_{13}).
\label{13mixren}
\end{eqnarray}
The enhancement factor for the degenerate spectrum is
$m_0^2/\Delta m^2_{13}$ and the strongest evolution is when
phases $\phi_1$ and $\phi_2$  are different. For equal phases
running is suppressed by an additional factor  $r_{\Delta}$.
In the case of normal mass hierarchy $\theta_{13}$ decreases with
$\log \mu$.

The main term in eq.(\ref{13mixren}) does not depend on
$\theta_{13}$ at all, and therefore it evolves to nonzero value even if  $\theta_{13}
= 0$ at some high energy scale. Consequently at low scales  non-zero
$\theta_{13}$ may have purely radiative origin. For instance, one
can get $\theta_{13} \sim 8^{\circ}$, i.e. at the level of present
experimental bound,  if it is zero at $M_{GUT}$ \cite{manf13}.

If the mass hierarchy is inverted  $\theta_{13}$ increases with $\mu$
and at the GUT scale it could be larger than $\theta_C$.
The RGE effect is strongly suppresses when  $m_3$ is small.

3). Running of the 2-3 mixing can be described approximately by \cite{manf13}
\begin{eqnarray}
\dot{\theta}_{23} \approx - A \sin 2\theta_{23}
\frac{1}{\Delta m^2_{23}}
[\sin^2\theta_{12} |m_1 e^{\phi_1} + m_3|^2
\nonumber\\
+ \cos^2\theta_{12} |m_2 e^{\phi_2} + m_3|^2]
+ {\cal O}(\theta_{13}).
\label{23mixren}
\end{eqnarray}
As in the previous case the enhancement factor
for the degenerate spectrum is $m_0^2/\Delta m^2_{23}$ and  running is
suppressed if $\phi_{1} = \phi_{2} = \pi$.
In MSSM with increase of $\log \mu$ the angle $\theta_{23}$
decreases and can be as small as $(20 - 30)^{\circ}$ at the GUT scale.
So, one can obtain  the {\it  radiative enhancement} of the mixing
\cite{rad,degen,rad1}:
$\theta_{23}$  is small  (similar
to $\theta_C$) at high energies and it reaches
$\sim 45^{\circ}$ at low energies.
Here again the large lepton mixing is related to the neutrino mass
degeneracy.

The RGE should  lead to deviation of the 2-3 mixing from maximal
when running to small scales if it is maximal at high scale: 
$D_{23} = \frac{1}{2}(Z_{\tau} - 1)$ \cite{bmrad}, though  for
normal hierarchy the deviation is below $1^{\circ}$.

Finally let us consider the renormalization of the 1-2 mass split:
\be \Delta \dot{m}^2_{12} \approx \alpha \Delta m^2_{12} - 4 A
[2\sin^2\theta_{23} (m_2^2  \cos^2\theta_{12} - m_1^2
\sin^2\theta_{12}) + {\cal O}(\theta_{13})], \label{23mixren1} \ee
where the first term is the overall renormalization of all masses
due to the gauge radiative corrections and also renormalization of
the Higgs boson wave function. The second term can dominate for
the degenerate spectrum. Depending on parameters the
renormalization can enhance splitting up to the atmospheric one or
suppress it down to zero. So,  zero split at some high scales and
radiative origin of the 1-2 split at low scales can be realized
\cite{small}. Another possibility is that at high scales all the
mass splits are of the same order and the hierarchy of  splits at
low scales is produced by radiative corrections in the case of
the degenerate spectrum.

As is clear from our discussion the  role of the RGE effects
depends on a number of unknowns: possible extensions of the SM
like two higgs doublet model,  MSSM, (ii) on value of $\tan
\beta$, (iii) on type of spectrum (degenerate, hierarchical), (iv)
on CP violating phases. Depending on these unknowns the effects
can vary from negligible to dominant, thus explaining main
features of the neutrino mass spectrum and mixing.  Apparently
many uncertainties related to RGE effects will disappear if it
turns out that the neutrino mass spectrum  is hierarchical. Even
in this case the corrections can be larger than accuracy of future
measurements of the neutrino parameters. 
Strong effects are expected also from the ``threshold'' corrections
\cite{manfth}.

\subsection{Searching for the symmetry basis}

The structure of the mass matrix and its symmetries depend on the
basis. In general, symmetry basis can differ substantially from
the flavor basis considered so far. Therefore identification of
the symmetry basis is crucial for uncovering the underlying
physics. 
Unfortunately,  there is no clear guideline on how to look for
this basis. One can perform a continuous change of the basis, 
searching for situations where both neutrino and charged lepton mass
matrices have certain common symmetries. Some hints can be
obtained from explicit models. 

It is not impossible for  the symmetry basis to coincide with the flavor
basis, and models where  they do coincide exist. One can expect that in this
case symmetries associated to  some combinations of $L_e, L_{\mu},
L_{\tau}$ play an important role.

The symmetry basis may not coincide with but may be close to,  the flavor
basis. They can differ by rotation through an  angle of the order of
the Cabibbo angle, $\theta_C \sim \sqrt{m_\mu/m_{\tau}}$. The strong
hierarchy of masses of charged leptons favors this
possibility. 
However,  the symmetry basis can strongly differ from the
flavor basis. In some  models, as a consequence of symmetry, the
neutrino mass matrix is diagonal and maximal mixing comes from
diagonalization of the charged lepton mass matrix. Further studies
in this direction are necessary.

\section{Neutrinos and new symmetries of nature}

The most striking and unexpected outcome of the bottom-up
approach is an indication of  particular symmetries in the neutrino
sector, namely, symmetries of the lepton mixing matrix and the 
neutrino mass matrix in the flavor basis. It is strange that
symmetry is associated somehow with neutrinos and does not
show up in other sectors of theory. Several observations  testify
to such a  ``neutrino'' symmetries:

- Maximal or close to maximal 2-3 mixing;

- Zero or very small 1-3 mixing;

- Special values of 1-2 mixing;

- Hierarchy of mass-squared differences;

- Quasi-degenerate mass spectrum, if established.

As far as the last item is concerned, independent confirmation of the
Heidelberg-Moscow positive result is needed. Note that
in physics, large or maximal mixing is often related to degeneracy.
THus ,  possibility of the degenerate spectrum does not seem 
implausible. 
Some of the features  listed above can
originate from the same underlying symmetry.\\

\subsection{$\nu_{\mu} - \nu_{\tau}$ symmetry}

Both maximal 2-3 mixing and zero 1-3 mixing indicate the same
underlying symmetry and therefore deserve special attention. They
are consequences of the $\nu_{\mu} - \nu_{\tau}$ permutation
symmetry of the neutrino mass matrix in the flavor
basis~\cite{mutau}. The general form of such a matrix is \be M =
\left(\begin{array}{ccc}
A & B & B\\
B & C & D\\
B & D & C
\end{array}
\right). \label{23sym} \ee  This symmetry can be a part of
discrete $S_3$, $A_4$ or  $D_4$  groups and also can be embedded into
certain continuous symmetries. The   matrices in
eq.(\ref{normhier}) for  $c=1$ and $a = b$, and in eq.(\ref{invma})
for $x = y$ and $c = s$  are special cases of the general matrix in 
eq.(\ref{23sym}).

At first, this  
might seen  problematic because this
symmetry cannot be extended to the  charged lepton sector. To see
this, note that in the flavor basis  for the charged leptons we
have zero off diagonal elements, $D_l = B_l = 0$, and therefore the symmetry implies 
$M_l = diag(A_l, C_l, C_l)$ which  contradicts  $m_\mu \ll
m_\tau$. If $D_l \neq 0$, implying that 
the symmetry basis does not coincide with the flavor basis, 
one can get the required mass hierarchy.
However, in this case the charged lepton mass matrix also produces
maximal mixing rotation. Neutrino and charged lepton rotations
cancel leading to zero lepton mixing. 

Apparently the $\nu_{\mu} - \nu_{\tau}$ symmetry should be broken.  
One way to resolve the problem is to couple  the  Higgs bosons, which
violate $\mu - \tau$ symmetry spontaneously,  weakly with neutrinos,
but strongly - with the charged leptons. Examples where this is
achieved using simple auxiliary symmetries such as  $Z_2$, have been
discussed in the literature~\cite{grimus}. A difference between 
charged leptons and neutrinos appears because the right handed
components $l_R$ and $\nu_R$ have different transformation
properties under $Z_2$. This is not unnatural in the context of
supersymmetric (SUSY) theories in which  charged leptons get mass from the
down Higgs, whereas the neutrinos get mass from the up Higgs
doublet. Such models can be embedded into SUSY SU(5)
GUTs \cite{yu}.

Other possibilities are to introduce the symmetry basis which
differs from the flavor basis (in this case the symmetry will be
the 2-3 permutation symmetry),  or to use other (approximate)
symmetries which in the flavor basis are reduced to $\nu_{\mu} -
\nu_{\tau}$ permutation. 
Small breakings of these symmetries would manifest themselves in the
appearance of a small but nonzero $\theta_{13}$. The question of
course is what is small? It is  reasonable to assume 
that  $\theta_{13}\sim r_{\Delta}$ (or smaller values of 1-3 mixing) 
indicates 
an underlying $\nu_\mu - \nu_\tau$ symmetry. However,  if $\theta_{13}\sim
\sqrt{r_{\Delta}}$, no conclusion about this symmetry can be drawn
because there are many examples in which  larger values of $\theta_{13}$
are possible.

Can this symmetry be extended to the quark sector?
Ref. \cite{joship} argues that in fact smallness of the 2-3
quark mixing, $V_{cb} \ll \sqrt{m_s/m_b}$, can also be a consequence
of this symmetry.

There are two shortcomings of the discussed symmetry:

1). It does not determine masses: Symmetry fixes  general form of
the mass matrix (equalities of certain matrix elements) and not
masses which are given by the values of the  elements.

2). Symmetry does not determine the  1-2 sector.
Thus, one needs to use some more extended symmetries which involve
all three generations.
One such symmetry that is  widely studied is the $A_4$ symmetry 
\cite{a4}.
Other possibilities are: $S_3$\cite{kubo},  $Z_4$~\cite{Z4},
$D_4$~\cite{D4}.


\subsection{ A symmetry example,  $A_4$}

An interesting class of models is based on the $A_4$ symmetry
group of even permutations of four elements \cite{a4,a4a,a4b}. It is
the symmetry of the tetrahedron and has the irreducible
representations {\bf 3}, {\bf 1}, ${\bf 1'}$ and ${\bf 1''}$. The
products of representations ${\bf 3} \times {\bf 3} = {\bf 3} +
{\bf 3}  + {\bf 1} + {\bf 1'} + {\bf 1''}$ and also ${\bf
1'} \times {\bf 1''} \sim {\bf 1}$ both contain the invariant ${\bf
1}$. This allows one to introduce the Yukawa couplings with
special flavor structure. Furthermore, it is the existence of
three different singlet representations that leads to substantial
freedom to reproduce the observed pattern of masses and mixings.
Notice also that $A_4$ is subgroup of $SO(3)$.

In all the models proposed so far, three lepton doublets form the
triplet of $A_4$: $L_i = (\nu_i, l_i)  \sim {\bf  3}$, where $i = 1, 2,
3$. The right handed components of the charged leptons, $l^c_i$,
neutrinos and Higgs doublets transform 
in different ways as either ${\bf
3}$, or as ${\bf 1},~ {\bf 1'},~ {\bf 1''}$,  depending on the model. 
Essentially, the large
(maximal)  mixing originates from the fact that $l_R$ and $\nu_R$
have different $A_4$ transformation properties. 
A disadvantage of the model is that it also requires the
introduction of new Higgs multiplets,
and often  new heavy leptons as well as quarks
which are generic features of most symmetry
approaches.

In $A_4$ models one should introduce Higgs fields with non-trivial
$A_4$ transformation properties, that is in  representations {\bf
3} and ${\bf 1},~ {\bf 1'},~ {\bf 1''}$. One can ascribe  these
properties to the $SU_2$ Higgs doublets, in which case  6 such
doublets are required. Alternatively, one can  keep SM Higgs
doublet to be  singlet of $A_4$  but introduce new SU(2) singlet scalar 
fields which form non-trivial representations of $A_4$ (in the
spirit of Froggatt-Nielsen approach). The latter however requires
introduction of non-renormalizable operators (see {\it e.g.}
\cite{AF06}) or explicitly new heavy leptons and quarks
\cite{a4a}.

There are different versions of the $A_4$ models. By appropriate
choice of the Higgs fields and their VEVs and/or right handed
neutrino couplings, one can  obtain the tri-bimaximal mixing. The
models constructed are based on the fact that tri-bimaximal mixing
is given by the product of the trimaximal (``magic'') rotation
$U_{tm}$ and maximal 1-3 rotation: \be U_{tbm} = U_{tm} U_{13}^m.
\ee
Here
\be  U_{tm} \equiv \left(\begin{array}{ccc}
1 & 1 & 1\\
1 & \omega & \omega^2\\
1 & \omega^2 & \omega
\end{array}
\right), ~~~~  \omega \equiv e^{-2i\pi/3}. \ee

\begin{figure}[!tbp]
\centerline{\hfil\epsfxsize=9.5cm\epsfbox{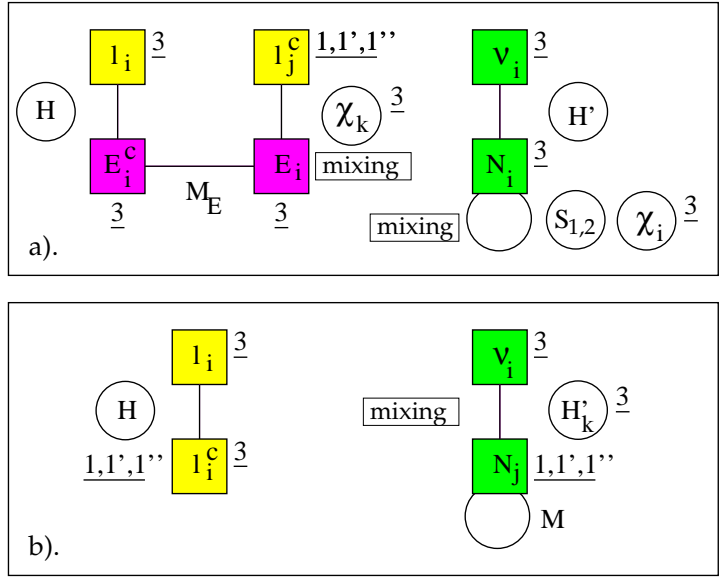}\hfil}
\caption{Generation of the
lepton masses and mixing in two models based on $A_4$:
a). model \cite{babu06} and b). model \cite{ma06}.
Lepton and Higgs multiplets are in shadowed boxes and circles.
Numbers at the boxes and circles indicate the $A_4$ representations of
the corresponding particles. The lines show the
Yukawa couplings or bare mass terms in the models. We indicate also
places where mixing is generated.}
\label{schemes}
\end{figure}

As an illustration, in fig.~\ref{schemes} we show schemes of generation of
the neutrino and charge lepton masses which lead to tri-bimaximal mixing
in two models based on $A_4$.

In the model a) \cite{babu06}  which is certain modification of
the early proposal \cite{a4,a4a}, the two Higgs doublets $H$ and
$H'$ are invariant under  $A_4$, and the SM singlets $\chi$ form
$A_4$ triplet. New heavy leptons $E$, $E^c$ have to be
introduced. The model content and  transformation properties of
the fields have been arranged in such a way that in the symmetry
basis the charged lepton mass matrix produces the $U_{tm}$
rotation, whereas the neutrino mass matrix produces maximal 1-3
mixing \cite{a4b}. The latter requires also certain VEV alignment.

In the model b) \cite{ma06} the $SU_2$ Higgs doublets  form the
triplet and singlets of $A_4$.

This illustrates generic problems and complexity of realization of
``neutrino symmetries''.

\subsection{Real or accidental}

The main question in this approach  is whether
the symmetries hinted at by  observations are simply accidental or
backed by real physics,  especially in view of the fact
that the price one pays in terms of the number of Higgs fields and/or
extra leptons for constructing a realistic model can be very
heavy. Thus, a fruitful approach is to look for possible
deviations from symmetries in data ({\it e.g.}, small $\theta_{13}$ along
with correlation between $\theta_{13}$ and $\theta_{23}-\pi/4$)
and explore models that may give other tests, because
if symmetries are not accidental, they have consequences of
fundamental importance: new structures are predicted, unification
path may differ substantially from what we are using now,  {\it etc.}. 
The symmetries may hold the key to understanding the flavor problem.



Another possibility is that the symmetries are not accidental but
the underlying theory has not been found yet and observed symmetry
relations are hints of a new sector in physics, {\it e.g.} 
flavor-universal mixing with new singlets may produce a symmetric
contribution to the active neutrino mass matrix (sec. 8).

The only way to establish that symmetry is not accidental is to
find new consequences of the symmetry, {\it i.e.} to make predictions in
the context of a certain model and to test the predictions in
experiment. It is important to find not just one but  several
predictions (see discussion in sec. 5.E).

Finally let us stress that the observational basis for the
existence of symmetries (real or accidental) is not yet well
established. As we described in sec. 2.D, still significant
deviation from maximal 2-3 mixing is possible and 1-3 mixing could
be relatively large. So, further experimental measurements will be
decisive.



\section{Leptons and Quarks} 

Joint consideration of quarks and leptons and searches for
possible relations between quark and lepton parameters are of
fundamental importance, because  this may (i)  provide a unified clue
for understanding  fermion masses and mixings and (ii) give more
insight into the unification of particles and forces in nature.


Below  we compare quark and lepton masses and mixings
and discuss various ideas about possible relations of quarks
and leptons such as (i) quark-lepton symmetry; (ii) quark-lepton
unification; (iii) quark-lepton universality; and (iv) quark-lepton
complementarity (QLC).


\subsection{Comparing leptons and quarks}

\begin{figure}[!tbp]
\centerline{\hfil\epsfxsize=10cm\epsfbox{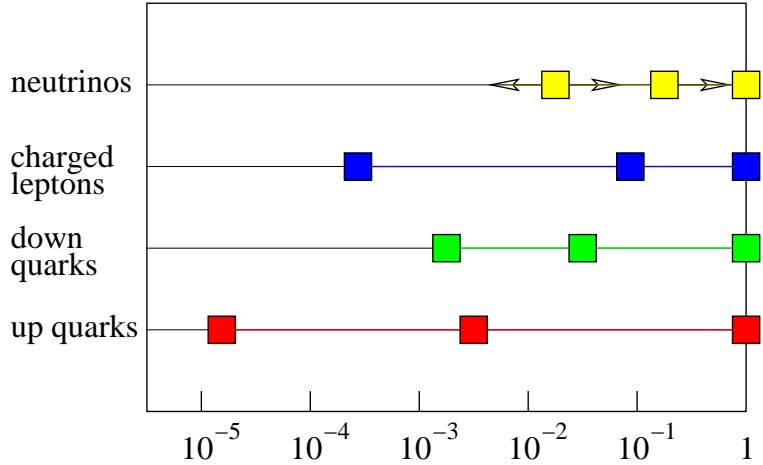}\hfil}
\caption{Mass hierarchies of quarks and leptons. The mass of
the heaviest fermion of a given type is taken to be one.}
  \label{ratios}
\end{figure}

There is a strong difference between the masses and
mixings in the quark and lepton sectors (fig. \ref{ratios}). The
ratios of masses of neutrinos and the corresponding upper quarks
are $m_{2} / m_c < 10^{-10}$, $m_3 / m_t < 10^{-12}$. Lepton
mixings are large,  quark mixings are small. The 1-2 mixing is the
largest for quarks, whereas 2-3 mixing is the largest lepton mixing. The
only common feature is that the 1-3 mixing (mixing between the
remote generations) is small in both cases.

More careful consideration, however,  reveals some interesting
features: It seems the 1-2 as well as 2-3 mixing angles in the
quark and lepton sectors are complementary in the sense that they
sum up to maximal mixing angle~\cite{qlc,raidal,qlc-ms}:
\be
\theta_{12} + \theta_C \approx \frac{\pi}{4},
\label{qlc12}
\ee
\be \theta_{23} + V_{cb} \approx \frac{\pi}{4}. \label{qlc23} \ee
Although  for various reasons it is difficult to  expect exact
equality in the above relations, 
one can say, qualitatively, that
there is a certain correlation:   2-3 mixing in the lepton sector
is close to maximal mixing because the corresponding quark mixing is
small, the 1-2 mixing deviates from maximal mixing substantially because
1-2 (Cabibbo) quark mixing is relatively large. For the 1-3 angles
we do not see a simple connection, and apparently the quark relation
$\theta_{13} \sim \theta_{12} \times \theta_{23}$ does not work in
the lepton sector. Below we explore the implications  of the above
equalities known in the literature as QLC relations~\cite{qlc}.

Comparing the ratio of neutrino masses in eq.(\ref{rmass}) 
with ratios for charged leptons and quarks (at $m_Z$ scale), 
\be \frac{m_\mu}{m_\tau} =  0.06, ~~ \frac{m_s}{m_b} = 0.02 -
0.03,~~~ \frac{m_c}{m_t} = 0.005 \ee
one concludes that the neutrino hierarchy (if it exists at all) is the
weakest one. This is consistent with possible mass-mixing
correlation,  so that large mixings are associated with a weak
hierarchy: $\sqrt{m_i/m_j}
\sim \theta_{ij}$. 

It is also intriguing that 
\be 
\sqrt{\frac{m_\mu}{m_\tau}}
\approx \sin \theta_{12}^q  \approx \sin \theta_C.
\label{univtheta} 
\ee 
Does this perhaps  indicate that the Cabibbo
angle is the universal flavor parameter for both quark and lepton
physics? In a class of grand unified models to be discussed below, 
the Cabibbo angle, indeed, becomes the key parameter of the neutrino
mass matrix and describes the neutrino masses as well as mixings.

In fig. \ref{ratios} we show the mass ratios for  three
generations. The strongest hierarchy and geometric relation $m_u
\times m_t \sim m_c^2$ exist for the upper quarks. Apart from that
no simple relations show up.
Furthermore, it looks like the observed pattern is an interplay of some regularities
- flavor alignment and randomness - ``anarchy''.
Below we explore the possible meaning behind this picture.

\subsection{Quark-lepton symmetry}

There are good reasons to suspect that quarks and leptons may be two
different manifestations of the same form of matter. The first
hint arises from the observed similarities between weak-interaction 
properties of quarks and leptons. Each quark has its
own counterpart in the leptonic sector, which has the same weak
isospin properties: $u_L$ corresponds to $\nu_L$, $d_L$ - $e_L$,
{\it etc.}. This is generally known as quark-lepton symmetry, and
although it manifests itself only in the left-handed helicity
sector of quarks and leptons, it is often considered as a hint of
further unification among these two very different kinds of
matter. It can be extended to the right-handed sector, when leptons are
treated as the fourth color~\cite{pati} following Pati-Salam
$SU(4)_C$ unification symmetry.

The second hint comes from the attractive hypothesis of grand
unification of matter and forces which argues that at very short
distances, all forces and all matter unify. In GUTs, quarks 
and leptons form multiplets of the
extended gauge group. The most appealing such group is SO(10)
where all known components of quarks and leptons including the right-handed 
neutrinos fit into the unique 16-plet
spinor multiplet \cite{so10}.
It is difficult to believe that all these features are
accidental.



However, at first sight different patterns of  masses and mixing 
strongly break the quark-lepton symmetry. Furthermore, if
particular symmetries are found to exist only in the leptonic
sector, they may indicate that quarks and leptons are
fundamentally different.

\subsection{Quark-lepton universality}

In spite  of the strong difference between the  masses and mixings  of quarks and
leptons we still can speak about the approximate quark-lepton
symmetry or even universality. The universality is realized in
terms of mass matrices  or matrices of the Yukawa couplings  and
not in terms of observables - mass ratios and mixing angles.
The point is that very similar mass matrices can lead to
substantially different mixing angles and masses (eigenvalues) if
the matrices are nearly singular (approximately equal matrices of
rank-1) \cite{dors}. The singular matrices are ``unstable'' in the
sense that small perturbations can lead to strong variations of
mass ratios and (in the context of the seesaw) mixing angles. 
The well-known examples of singular matrices  are the ``democratic'' mass
matrix~\cite{sing} and the matrix with only one non-zero element
$m_{33}$.

Let us assume that in the zeroth order of some approximation all
fermion (quark and lepton) mass matrices are equal 
to the same universal singular matrix $Y_0$: 
\be 
Y_u = Y_d = Y_D = Y_M =  Y_l = Y_0.
\label{matequ} 
\ee
Here  $Y_D$ is the Dirac-type
neutrino Yukawa matrix.   The Majorana type matrix for the RH
neutrinos, $Y_M$, can in general differ from the others being, {\it
e.g.},  as $Y_M \approx Y_0^2$.

As an important
example we can take \be Y_0 = \left(\begin{array}{ccc}
\lambda^4 & \lambda^3 & \lambda^2\\
\lambda^3 & \lambda^2 & \lambda\\
\lambda^2 & \lambda & 1
\end{array}
\right), ~~~~ \lambda \sim \sin \theta_C \sim 0.2 - 0.3. 
\ee
This matrix has only one non-zero eigenvalue and no physical
mixing appears because  matrices of all fermions are diagonalized
by the same rotation.
In this respect all singular matrices are equivalent (up to basis
definition) until corrections are introduced - corrections break
the equivalence.

Let us introduce perturbations of matrix structure, $\epsilon$, in
the following form \be Y^f_{ij} = Y^0_{ij} (1 + \epsilon_{ij}^f),
~~~ f = u, d, e, \nu, N , \ee where $Y^0_{ij}$ is the element of
the original singular matrix. This form can be justified, {\it
e.g.}, in the context of the Froggatt-Nielsen mechanism~\cite{FN}.
It turns out that small perturbations $\epsilon \leq 0.25$ 
are enough
to reproduce all large observed differences in mass hierarchies and mixings of
quarks and leptons \cite{dors}.

The smallness of neutrino masses is explained by the seesaw mechanism.
A nearly singular matrix of the right-handed neutrinos, which appears in the
denominator of the seesaw formula leads to enhancement of lepton
mixing and can flip the sign of the mixing angle which comes from
diagonalization of the neutrino mass matrix. Thus,the  angles from
the charged leptons, $U_\ell$,  and neutrinos, $U_\nu$,  sum up,
whereas in quark sector,  mixing angles from up- and down-quark mass
matrices subtract leading to small quark mixing.

Note that the different mass hierarchies of the up and down quarks
(as well as charged leptons) may testify that two different
universal matrices should be introduced for fermions with the
third projection of weak isospin 1/2 and -1/2.  This can also be
related to existence of two different Higgs doublets giving masses to
those components (as in SUSY theories). 

Instead of mass matrices one can  consider the universality of
rotation (mixing) matrices $V_{U}$ which diagonalize mass matrices of
all fermions in certain basis. The model can be arranged in such a way
that in the lowest order $U_{CKM} = V_{U}^{\dagger} V_U = I$, whereas
$U_{PMNS} = V^{T}_U V_U$  contains large mixings \cite{JS05}.

In connection to the quark-lepton universality one can consider the following
working hypothesis:

1). No particular ``neutrino'' symmetry exists, and in general one
expects some deviation of the 2-3 mixing from maximal as well as
non-zero 1-3 mixing. Nearly maximal 2 -3 mixing would be
accidental in this case. Instead, some family symmetry is realized, 
which ensures universality of mass matrices and their particular
structure.

2). The seesaw mechanism with the scale of riht-handed neutrino masses $M \sim
(10^7 - 10^{15})$ GeV explains the smallness of neutrino mass.

3). The quark-lepton unification or grand unification are realized
in some form, {\it e.g.}, $SO(10)$.

4). The quark-lepton symmetry is weakly broken with still some
observable consequences such as  $m_b \approx m_\tau$.

5). Large lepton mixing is a consequence of the seesaw mechanism
 - seesaw enhancement of lepton mixing
(special structure of the right-handed neutrino mass matrix), and/or due to
contribution from the type II seesaw (which we  consider in
sect. 7.1).
 Flavor symmetry and/or physics of extra dimensions could determine
this special structure.

\subsection{Quark-lepton complementarity}

As noted in eqs. (\ref{qlc12})  and (\ref{qlc23}), the latest 
determination of solar mixing
angle gives 
\be \theta_{12} + \theta_C = 46.7^{\circ} \pm 2.4^{\circ}
~~~(1\sigma) \ee which is consistent with maximal mixing angle
within $1\sigma$ (fig. \ref{12mix}). 
The fact that the approximate complementarity is also fulfilled 
for the 2-3 mixings hints at  more serious  reasons than just
numerical coincidence~\footnote{Notice that due to smallness of the 1-3
mixings in quark and lepton sectors we have $\sin\theta_{12} \approx
V_{12}$ and $\sin\theta_C \approx V_{us}$. The elements of the mixing
matrices $V_{12}$ and $V_{us}$ are the physical parameters and therefore
the QLC relation is essentially parameterization independent in contrast
with statement in \cite{jarlskog}.}.  A possibility that the lepton mixing
responcible for solar neutrino conversion equals maximal mixing minus $\theta_C$ was 
first proposed in \cite{petcov},  and
corrections of  the bimaximal mixing by the CKM type rotations
discussed in \cite{parametr}.

If not accidental,  the quark-lepton complementarity would require
certain modification of the picture described in the previous section
\cite{raidal,qlc-ms,qlc-fm}.
It implies the existence of some
additional structure  in the leptonic (or quark?) sector, which
generates bi-maximal mixing. In this sense it might indicate a
fundamental difference between leptons and quarks. However, 
there should also be the  quark- lepton unification or symmetry,  which
communicates the quark mixing to the lepton sector. A general
scheme could be  that~\cite{raidal,qlc-ms}
\be ``{\rm lepton~ mixing} =  {\rm bimaximal~mixing} - {\rm
CKM}''.
\label{implic}
\ee
(Another option is  ``CKM = bimaximal - PMNS'', which may have different
implications).

There are  a number of non-trivial conditions  that must be met for the {\it exact} QLC
relation to be realized~\cite{qlc-ms}.

1). The matrices $U_{12}^m$ and
$U_{12}^{CKM \dagger}$ should be multiplied in the following
order:
 \be U_{PMNS} \equiv U_l^{\dagger}U_{\nu} =
...U_{23}^m ... U_{12}^m U_{12}^{CKM \dagger} \label{order} \ee
(last two matrices can be permuted). Different orders of rotations lead to
corrections to the exact QLC relation.

2). A matrix with CP violating phases should not appear between
$U_{12}^{CKM \dagger}$ and $U_{12}^m$ or the CP violating phase in
this matrix should be small \cite{qlc-ms,qlc-cp}.

3). The RGE  effects should be small because 
presumably the quark-lepton symmetry which leads to the QLC
relation is realized at high mass scales.

Let us first describe two possible (to some extent extreme) scenarios
of eq.(\ref{implic})
that  differ by the origin of the bi-maximal mixing and lead to
different predictions.

(1). QLC-1: In the symmetry basis maximal mixing is generated by
the neutrino mass matrix: $U_{\nu} = U_{bm}$;
it can be produced by the seesaw
mechanism. The charged lepton mass matrix gives  the CKM rotation
$U_{\ell} = U_{CKM}^{\dagger}$,
as a consequence of the quark-lepton symmetry: $m_l = m_d$. In this case
the order of matrices in eq.(\ref{order}) is not realized; 
$U_{12}^{CKM}$ should be permuted with $U_{23}^m$  and
consequently the QLC relation is modified:
$\sin \theta_{12} =  (1/ \sqrt{2}) \cos \theta_C - 0.5 \sin \theta_C$ or
\be
\sin \theta_{12}
\approx \sin (\pi/4 -\theta_C) + 0.5\sin \theta_C (\sqrt{2} -1).
\label{qlc1} 
\ee 
Numerically we find $\sin^2\theta_{12} = 0.331$
which is practically indistinguishable from the tri-bimaximal
mixing value. The predicted  1-3 mixing, $\sin \theta_{13} =
\sin\theta_C/\sqrt{2}, $ is close to the upper experimental bound
(fig. \ref{13mix}).
Combining this with expression for 1-2 mixing we get an
interesting relation $\theta_{12} \approx \pi/4 - \theta_{13}$ \cite{qlc-cp}.

2). QLC-2: In the symmetry basis maximal 1-2 mixing originates from the
charged lepton mass matrix, $U_{\ell} = U_{bm}$,
and the CKM, $U_{\nu} = U_{CKM}^{\dagger}$, appears from the neutrino
mass matrix owing  to the quark-lepton symmetry: $m_D \sim m_u$ (assuming also
that in the context of the see-saw the right-handed neutrino mass matrix does not
influence mixing, {\it e.g.},  owing  to factorization). In this
case  the QLC relation is satisfied precisely: $\sin \theta_{12} =
\sin (\pi/4 -\theta_C)$. The 1-3 mixing is very small - of the
order $\sin^2 \theta_{12}|V_{cb}|$ .

According to fig. \ref{12mix} the best fit experimental value
of $\theta_{12}$  is in between the QLC-1 and QLC-2
predictions and further measurements of the angle
with accuracy $\Delta \theta_{12} \sim 1^{\circ}$ are required
to disentangle the scenarios.

Other possibilities exist as well. For instance, one maximal mixing may
come from the neutrino mass matrix in the symmetry basis and another 
from the charged lepton mass matrix.


There are two main issues related to the QLC relation:

- origin of the bi-maximal mixing;

- mechanism of propagation  of the CKM mixing from the quark to
lepton sector.

The main challenge here is that the required quark-lepton symmetry
is broken. In particular,
the  leptonic mass ratio $m_e/m_\mu = 0.0047$
is much smaller than the quark ratio  $m_d/m_s = 0.04 - 0.06$; also the 
masses of the muon and $s$-quark are very different at the GUT scale.
Precise  QLC relation may imply that

- the q-l symmetry is actually weakly broken as  we
discussed in sec. 5.3;

- the q-l symmetry is very weakly broken for up quarks and neutrinos
in a sense that for Dirac matrices $M_u \approx M_D$. Then
CKM propagates via the up-sector;

- the breaking affects  mainly the masses and mass ratios
  but not mixings.

Anyway, the mass matrices are different for quarks and leptons and
propagation of the CKM mixing leads to corrections to the QLC
relation at least of the order $\Delta \theta_{12} \sim \theta_C m_d/m_s
\sim 0.5 - 1.0^{\circ}$ \cite{qlc-ms}.

Consider the case of QLC-1 (bimaximal mixing from neutrinos),
where deviation of quark mixing from zero and lepton mixing from
maximal follow from the down quarks and charged leptons.
If the leptonic mass matrix has
similar structure to the d-quark mass matrix  with Gatto-Sartori-Tonin (GST)
relation one would expect
$\theta_l \sim \sqrt{m_e/m_\mu} \approx \theta_C/3$ and
deviation from maximal mixing
$\theta_l/\sqrt{2} = 1/3\sqrt{2} \theta_C$ turns out to be
too small \cite{shift}.
There are several proposals to enhance the shift angle.
In particular, the neutrino mass matrix can be modified as 
${\cal M}_\nu  = {\cal M}_{bm} + \delta {\cal M}$,
where ${\cal M}_{bm}$ produces the bi-maximal mixing and
$\delta \delta {\cal M}$ leads to deviation \cite{shift}. 
$\delta {\cal M}$  can be due to  the seesaw type-II contribution 
\cite{falcone}.
However in this case  connection to
quark mixing  is lost and  the relation eq.(\ref{qlc12}) is simply
accidental. Notice that the ratio of the mass squared differences,
$r_{\Delta} \sim \sin\theta_C$, so that  the shift, $\theta_l$,
can be related simply with generation of the solar mass split and
therefore be of purely leptonic origin.

In the context of  quark-lepton symmetric models, the enhancement
may have the group theoretical origin.
In \cite{qlc-km} for certain operators generating fermion masses the
relation $\theta_l = 3 \theta_C/2$ has been found, where  factor
3/2 is the ratio of  Clebsh-Gordon coefficients.

The renormalization group effects on 1-2 mixing are in general
small and furthermore they lead to  increase of the angle
$\theta_{12}$ at low scales. The negative shift can be obtained
from renormalization group effects in presence of the non-zero 1-3
mixing \cite{qlc-ms}. Also the threshold corrections due to some
intermediate scale physics like low scale supersymmetry can
produce the negative shift thus enhancing the deviation from
maximal mixing \cite{qlc-ren}. Strong shift can also be obtained
from RGE effects between and above the seesaw scales related to
the RH neutrino masses (see sec. 6.3) \cite{manfth}.

To avoid the additional 1/3 suppression of $\theta_l$ one can abandon
the GST-type relation for charged leptons.
Then $\theta_l \sim \theta_C$ would  imply nearly
singular character of the 1-2 leptonic submatrix.

As remarked before, the quark-lepton symmetry can propagate $\theta_C$
to lepton sector exactly if the neutrino mass matrix is the source
of both bi-maximal mixing and the CKM rotations. The charged
lepton and down quark mass matrices should be  diagonal, and as a
consequence of the quark-lepton symmetry,   $m_u = m_D$. The left rotations
for these matrices give $U_{CKM}$ and the rest of the seesaw
structure  generates the  bimaximal mixing. In this case, however,
the GST-relation in the quark sector becomes accidental. If the
bi-maximal mixing is generated by the charge leptons  (lopsided
scenario, see sec. 6.4.4) the QLC relation becomes precise \cite{raidal}.

The role of CP-phases can be important in the q-l relations \cite{qlc-ms}.
CP violating phase in $U_{CKM}$
produces very small effect on QLC due to smallness of $V_{ub}$.
Also  in this scenario the leptonic CP phase is very small.
On the other hand appearance of the phase matrices
in between $U_{12}^{CKM}$ and $U_{12}^m$ will both modify
the QLC and the leptonic Dirac phase, $\delta$. Apparently,  the
relation between these two modifications should appear.
In the QLC-1 scenario insertion of the  phase matrix
$I_{phase} \equiv diag(e^{i\alpha}, e^{i\beta}, 1)$, between two 1-2 rotations:
$U^{CKM \dagger} I_{phase} U_{bm}$, 
leads to the following change of the QLC relation \cite{qlc-cp}:
\be
\theta_{12} \approx \frac{\pi}{4}  - \frac{\theta_C}{\sqrt{2}}\cos(\delta
- \pi).
\ee
So, the phase diminishes the shift, thus destroying the
relation. Maximal shift required by  QLC  implies
$\delta \approx \pi$, that is,  suppressed CP violation phase.

There are few attempts to construct consistent
quark-lepton model which reproduces the QLC relation. 
The simplest possibility is the
$SU(2)_L \times SU(2)_R \times SU(4)_C$ model
which  implements the quark-lepton symmetry in the most straightforward
way \cite{qlc-fm,qlc-km}.
The strategy is to obtain (using an additional flavor symmetry)
the neutrino mass matrix with inverted mass hierarchy
which leads naturally to the bimaximal mixing
(QLC-1 realization). The quark-lepton symmetry
provides equality of the mass matrices $M_l = M_d$,
and consequently the same CKM type mixing in both sectors. The perturbations
to the matrices,  $\delta M_l$ and  $\delta M_d$, should be  introduced which
break the q-l symmetry and correct the masses.
(In \cite{qlc-fm} they are due to the non-renormalizable operators
with new Higgs fields which transform  as $\bf 15$ of  $SU(4)_C$).
These corrections however modify relation between 
$\theta_l$ and $\theta_C$, and  their equality is matter of
tuning of  continuous parameters.
Another possibility \cite{qlc-km} is to introduce the non-renormalizable
operators which include couplings with Higgs in $\bf 4$ of $SU(4)_c$
as well as singlet flavon fields {\it a la} Froggatt-Nielsen.
Selecting particular type of operators
one can get inequality of matrices $M_l$ and  $M_d$ already in the lowest order
and  enhance  the leptonic angle: {\it e.g.} like
$\theta_l = 3\theta_C/2$, as we have marked previously.
The enhancement allows to reproduce the QLC
relation eq.(\ref{qlc12}) almost precisely.

Different approach to resolve the problem
of decoupling of masses and mixing
is to use non-abelian  flavor symmetries \cite{raidal}.
Via minimization of the potential
the symmetries lead  to zero or to maximal (bi-maximal) mixing
independently of the mass eigenvalues.

The Cabibbo mixing can be transmitted to the lepton sector in a more
complicated way (than via the quark-lepton symmetry). In fact, $\sin
\theta_C$ may turn out to be a generic parameter of the theory of
fermion masses - the ``quantum'' of flavor physics,  and therefore to
appear in various places: mass ratios, mixing angles. The relation 
in eq.(\ref{univtheta})
favors of this possibility. 
However, the same relation in eq.(\ref{univtheta}) may
suggest that the QLC relation is accidental. Indeed, it can be
written and interpreted as  pure leptonic relation 
\be \theta_{12}
+ \theta_{\mu \tau} = \frac{\pi}{4},   ~~~ \tan \theta_{\mu \tau}
\equiv \sqrt{\frac{m_{\mu}}{m_{\tau}}}. \ee 
This relation may be even more difficult to realize in models.

Following an  idea that $\lambda \approx \sin \theta_C$ is the ``quantum''
of the flavor physics one can consider
the Cabibbo angle as an expansion parameter
for mixing matrices.
In zero approximation the quarks have unit mixing matrix:
 $U_{CKM}^0 =  I$,
whereas leptons have
$U_{PMNS}^0 = U_{bm}$ or bi-large mixing matrix.
The $\lambda$ size corrections can
be included as
$U_{PMNS} = U_{\lambda}^{\dagger} U_{bm}$
or $U_{PMNS} = U_{bm} U_{\lambda}^{\dagger}$.
Interesting possibility (in a spirit of the QLC relation)
is that  $U_{\lambda} = U_{CKM}(\lambda)$
in the Wolfenstein parameterization \cite{parametr,parametr2}.
In this case one gets universal description (parameterization) of quark
and lepton mixing matrices. This apparently
reduces the number of free parameters in the problem and also
establishes various relations between mixing angles.

In general on can take   $U_{\lambda}$ as a matrix with
all three $\lambda$ size rotations and
study properties of the PMNS matrix obtained by insertion of the
$U_{\lambda}$ in various places of the zero order
structure \cite{par-gen}, that is,  $U_{\lambda}^{\dagger} U_{bm}$,
$ U_{bm}  U_{\lambda}^{\dagger}$ or
$U_{23}^m  U_{\lambda}^{\dagger} U_{12}^m$, {\it etc.}.


\subsection{Empirical relations}




Establishing  empirical relations
between masses and mixings of fermions may give a clue to the
underlying physics. The tri-bimaximal mixing scheme and QLC
equality are examples of relations ``between mixings without
masses''. One should note, of course, the GST relation  $\sin
\theta_C \approx V_{us} \approx \sqrt{m_d/m_s}$  \cite{GST}, and
$m_d/m_s = \sqrt{m_u/m_c}$ which determine substantially the form
of quark mass matrices, {\it etc}.

A particularly intriguing such a relation is the Koide relation
\cite{koide,koide1,koide2} between 
the pole masses of
charge leptons 
\be Q_l \equiv \frac{m_e +
m_{\mu} + m_{\tau}} {(\sqrt{m_e} + \sqrt{m_{\mu}} +
\sqrt{m_{\tau}})^2} = \frac{2}{3} \label{koi-1} 
\ee 
which is satisfied
with accuracy $10^{-5}$: \be Q_l^{(pole)} = 2/3~ ^{+
0.00002}_{-0.00001}. \label{precision} 
\ee
The Koide formula eq.(\ref{koi-1}) is interesting not only because
of precision but also because it was obtained in the context
of certain model and not as empirical relation. In fact, it was obtained 
in attempt to explain the relation (also rather precise) 
between the Cabibbo angle and lepton masses 
in  the composite model of the quarks and leptons \cite{koide1}. 
It allowed to predict precise value of the tau-lepton mass.





There are several properties of the relation eq.(\ref{koi-1})
which could have interesting implications \cite{koide05}.

(i)  Varying masses one finds that the  minimal value, $Q_{min} =
1/3$,  corresponds to the degenerate spectrum and the maximal one,
$Q_{max} = 1$, to the strongly hierarchical spectrum.  So, the
quantity $Q_l$ is a good measure of degeneracy of spectrum. The
experimental value 2/3 is exactly  in between  the two extremes.

(ii) The relation involves 3 generations explicitly.
The mass of electron can not be neglected and therefore  in the
underlying theory $m_e$  can not be considered as perturbation.
In fact, the value 2/3 may be interpreted as
$2/N_{f}$,  where $N_f = 3$ is the number
of flavors.

(iii) The formula  is invariant under interchange of flavors
$e \leftrightarrow \mu$, {\it  etc.},
and therefore implies $S_3$ (or wider) underlying symmetry.
The value 2/3 may have certain group theoretical origin.

(iv) Essentially eq.(\ref{koi-1}) gives relation between the
two mass hierarchies $r_e \equiv m_e/m_{\tau}$,
$r_\mu \equiv m_\mu/m_{\tau}$, and does not depend on the absolute
scale of masses:
\be
Q_l = \frac{1 + r_e + r_{\mu}}{(1 + \sqrt{r_e} + \sqrt{r_\mu})^2} =
\frac{2}{3}.
\ee
So, it can be realised for different sets of hierarchies.

(v) The formula may have certain geometrical
origin \cite{foot,esposito}.
Introducing vectors $\vec{M} = (1,1,1)$ and
$\vec{L} = (\sqrt{m_e}, \sqrt{m_{\mu}}, \sqrt{m_{\tau}})$
we can rewrite it as
\be
Q_l =  \frac{1}{3\cos^2\theta_{ML} },~~~~
\cos\theta_{ML} \equiv \frac{\vec{L} \cdot {\vec{M}}}{|\vec{L}||\vec{M}|}.
\ee
Apparently the experimental result corresponds to
$\theta_{ML} = 45^{\circ}$.

(vi) The relation has bilinear structure in $\sqrt{m}$
which may imply that masses are bilinear of some other
physical quantities: coupling constants or VEV's.
In fact, the Koide relation is reproduced if
\be
m_i = m_0(z_i + z_0)^2,
\label{prop}
\ee
where $z_0 = \sqrt{\sum_i z_i^2/3} $ and
$\sum_i z_i = 0$.
Such a situation can be realized in the case of the radiative
mechanism of mass generation: in one loop $m \propto Y^2$,
or in the seesaw mechanism $m \propto \mu \mu' M^{-1}$.

(vii) The quantity $Q_l$ is not invariant under RGE running.
At the $Z^0$ - mass  $Q_l(m_Z) \approx 1.002Q_l^{(pole)}$
\cite{bo05,xing05}.
Above  $m_Z$ the renormalization is negligible
in the SM, and it can lead to further increase of $Q_l$ by about
$0.7\%$ in MSSM at $M_R \sim 10^{14}$ GeV and for
$\tan \beta = 50$ \cite{xing05}. So, the renormalization effect
is much larger than the  error bars in eq.(\ref{precision}) and therefore
$Q_l$ deviates from 2/3 at high scales (already at the EW scale).
This may indicate various things: the  relation is accidental;
the accuracy for the pole masses is accidental;
physics responsible for the relation, and therefore the lepton masses,
is at low scales.

(viii) The relation eq.(\ref{koi-1}) is not universal:
it can still be valid for the down quarks: $Q_d \sim 0.7$
at $m_Z$,  but it is certainly violated for the up quarks:
$Q_u \sim 0.9$.

(ix) Important aspect is that the mass relation does not depend on
mixing. That is,  physics of mass generation and that of mixing
should decouple. Mixing can be included if the relation $Q = 2/3$
is considered, {\it e.g.}, for the ``pseudo-masses'' introduced as
$\tilde{m}_{\alpha} \equiv \sum_i U_{L \alpha i} m_i$, where
$U_{L}$ is the matrix of rotation of the LH components which
diagonalizes the mass matrix in the ``symmetry'' basis
\cite{gerard}. For charge leptons one should take $U_{L}^l = I$.
For quarks one can select the matrices so that universality $Q^{u}
= Q^{d} = 2/3$ is restored.

Till now no realistic and consistent model
for the Koide relation is constructed
(see \cite{koide05} for review).
Among interesting proposals one should mention
the radiative (one loop) mechanisms of charged leptons masses
generation \cite{koide,koide1}; the seesaw mechanism \cite{koide2,koidess};
mechanism based on the democratic mass matrices
and $S_3$ symmetry \cite{koidedem}.
An interesting possibility is that lepton masses are
generated by bi-linear of VEV's of new scalar fields:
$
m_i \propto \langle \bar{\phi}_i \rangle \langle {\phi}_i \rangle.
$
Then as a consequence of symmetry of the scalar potential
($S_3$ and $SU(3)$ symmetries have been considered),
the VEV's have the property
$\langle \bar{\phi}_i \rangle \sim z_i + z_0$ in eq.(\ref{prop})
\cite{koide05}.

What about neutrinos?
On the first sight, because of  weaker mass hierarchy eq.(\ref{r-delta}),  the  
neutrino masses
do not satisfy the Koide relation. Depending on the unknown
absolute mass scale one finds
$Q_{\nu} = 0.33 - 0.60$ \cite{esposito,bo05,xing05},
where the lower bound corresponds to
the degenerate spectrum and the upper one to $m_1 = 0$. 
However, it was noticed recently, that the relation can 
be fulfilled provided that $\sqrt{m_1} < 0$ and two others are positive \cite{brannen}. 
So that for neutrinos the relation is 
\be 
Q_\nu \equiv \frac{m_1 +
m_2 + m_3} {(- |\sqrt{m_1}| + \sqrt{m_2} +
\sqrt{m_3})^2} = \frac{2}{3}.  
\label{koi-nu}
\ee
Of course, the question is still why only one lepton 
$\nu_1$ has negative root squared of  mass.  
The relation (\ref{koi-nu}) together with the 
best fit values of the mass squared differences 
implies strongly hierarchical mass spectrum 
with prediction for the lightest mass \cite{brannen} (see also \cite{Koide6})
\be
m_1 = 3.9 \cdot 10^{-4}~~ {\rm eV}.  
\label{mass}
\ee
Two other masses $m_2 = 9.0  \cdot 10^{-3}$  eV and  
$m_3 = 5.1 \cdot 10^{-2}$  eV are given essentially by the 
measured mass squared differences. 

Other proposals include the following: 
The universality can be restored if one uses the pseudo-masses \cite{gerard}.
Notice that since $U_L^l = I$ for charged leptons,
for neutrinos we have $U_L^{\nu} = U_{PMNS}$.
Then from the condition $Q_{\nu} = 2/3$ one finds
for the allowed region of neutrino oscillation parameters:
$m_1 \sim (3 \pm 1) 10^{-2}$ eV, $\theta_{12} > 35^{\circ}$ and
$\theta_{23} > 50^{\circ}$. All neutrino masses are of the same order,
and large lepton mixing is related to the absence of mass hierarchy
{\it a la} the  GST relation.

Another proposal is to modify the Koide relation for the
upper quarks and neutrinos without introduction of mixing \cite{bo05}.
Observing that $Q_{\nu} < 2/3$ but $Q_{u} > 2/3$
one can assume a kind of mass complementarity
$Q_l + Q_d = Q_\nu + Q_u$. That would lead to
the lightest neutrino mass $m_1 \approx 10^{-5}$ eV.

Notice that in these considerations smallness of neutrino mass
and its possible Majorana nature have not been taken into account.
Apparently, the presence of the Majorana mass matrix of the RH neutrinos
in the context of seesaw mechanism can influence
the implications of the Koide relations for neutrinos.
Alternatively one can imagine that mechanism responcible
for  smallness of the neutrino masses does not influence
ratios of masses.

The question: ``real or accidental'' is still open;
and the lesson is that just one very precise
prediction confirmed  by very precise measurements may
not be  enough to verify theory.

\section{ Seesaw: theory and applications}


As noted above, the see-saw mechanism  is one of the simplest
ways to understand the small neutrino masses.  It has important
implications  and connections to a number of fundamental issues
which we discuss in this section.
\begin{itemize}

\item What is the scale of $M_R$ and what determines it?

\item  Is there a natural reason for the existence of the right
handed neutrinos, the main element of the see-saw?

\item  Is the see-saw mechanism alone  enough to explain all
aspects of neutrino masses and mixings?

\item On a phenomenological level, what is the flavor structure of
the right handed neutrino sector? Can we determine it from, for example,
purely low-energy neutrino observations?

\end{itemize}

\subsection{Right-handed neutrino masses and scale of seesaw}

The scale of the seesaw (type I) is related to the scale of right handed 
neutrino masses. Some idea about $M_R$ can be obtained from the
naive estimation of masses for the third generation: \be M_R \sim
k(M_R) \frac{m_D^2}{m_3} = k(M_R) \frac{m_t^2}{\sqrt{\Delta
m_{23}^2}}  \approx 5 \cdot 10^{14}~ {\rm GeV}, \label{ssscale}
\ee where $m_t$ is the top quark mass,   $ k(M_R)$ is the
renormalization group factor of the $D = 5$ operator.  (Here we assume
normal mass hierarchy.) It is this large scale that  indicates
that neutrino mass is related
 to new physics beyond that implied by the charged fermion masses.
The scale in eq.(\ref{ssscale}) is rather close to the GUT scale and
in fact can be immediately related to the GUT scale. In this sense
the smallness of the neutrino mass is the direct indication of
GUT.

The situation is more complicated if one considers all three neutrinos
and takes into account mixing among them. 
For instance, for the hierarchical mass spectrum of light
neutrinos  the heaviest RH neutrino mass is determines by 
the smallest neutrino mass $m_1$ and not $m_3$:
$M_3 \sim m_t^2/ m_1$. If $m_1 < 10^{-3}$ eV,
the mass  $M_3$ can be at the GUT scale or even bigger.
There are  a number of
uncertainties and ambiguities in determining  $M_R$: (i)  we
do not yet know the  scale of light
 neutrino masses (which can change the estimation
by approximately one  order of magnitude); (ii) the Dirac masses of
neutrinos are not known and must be assumed; 
(iii) mixing can strongly influence the masses of right-handed neutrinos;
(iv) it is not clear yet that seesaw type I gives the main
contribution to the neutrino mass.  If it is subdominant, the
masses of the right-handed neutrinos can be larger; (v) more than three 
singlet fermions (RH neutrinos) can be involved in the generation of the 
light
neutrino masses. In this case (as we  see below) the scale in 
eq.(\ref{ssscale}) may turn out to be a phantom scale that 
does not correspond to any physical reality.

The  assumption that $m_D \propto m_q$ leads typically to a rather
strong hierarchy of the RH neutrino masses: $M_{i} \propto
m_{qi}$ or even stronger. Their values can cover the interval
$(10^5 - 10^{16})$ GeV, although  in some particular cases two masses
can be quasi-degenerate. 
To get small masses for the  usual active neutrinos, it is 
sufficient  to have
only two RH neutrinos, which means that the third one can be
arbitrarily heavy: {\it e.g.},  at the GUT or even Planck-mass
scale.





\subsection{Seesaw, B - L and Left-Right symmetries}

What is the physics content of this new scale? The seesaw scale can be
identified as the scale of violation of certain symmetries. The
fact that $M_R$ can be  much smaller than the Planck scale is an
indication in favor of this. It is therefore appropriate at this
point to discuss possible origins of the RH neutrino masses.

To answer this question, it is important to note the changes
that occur in the standard model with the addition of one right
handed neutrino per generation: The most obvious change is that it
restores the quark-lepton symmetry. On a more fundamental
level, it turns out that in the presence of three right-handed neutrinos, 
the symmetry $B-L$ which was a global symmetry in the standard model
becomes a gaugeable symmetry because  the condition
$Tr(B-L)^3=0$ is satisfied,  implying that gauge anomalies cancel. The
gauge group of weak interactions expands to become the left-right
symmetric group $SU(2)_L\times SU(2)_R\times U(1)_{B-L}$
\cite{lrs} which is a subgroup of the $SU(2)_L\times SU(2)_R\times
SU(4)_c$ group introduced by Pati and Salam \cite{pati}. This leads
to a picture of the weak interactions that is fundamentally different
from that envisaged in the standard model in that the weak
interactions,  like the strong and gravitational interactions, 
become parity conserving. Furthermore, in this theory, the electric
charge formula becomes~\cite{marshak}:
\begin{eqnarray}
Q~=~I_{3L}~+~I_{3R}~+~\frac{B-L}{2},
\end{eqnarray}
where each term has a physical meaning unlike the case of the
standard model. When only the gauge symmetry $SU(2)_R\times
U(1)_{B-L}$ is broken down, one finds the relation $\Delta
I_{3R}~=-\Delta\left(\frac{B-L}{2}\right)$. This connects $B-L$
breaking, {\it i.e.} $\Delta(B-L)\neq 0$, to the breakdown of
parity symmetry,  $\Delta I_{3R}\neq 0$. It also reveals the true
meaning of the standard model hypercharge as
$\frac{Y}{2}~=~I_{3,R}~+\frac{B-L}{2}$.

\begin{figure}[!tbp]
\begin{center}
\epsfxsize11cm\epsffile{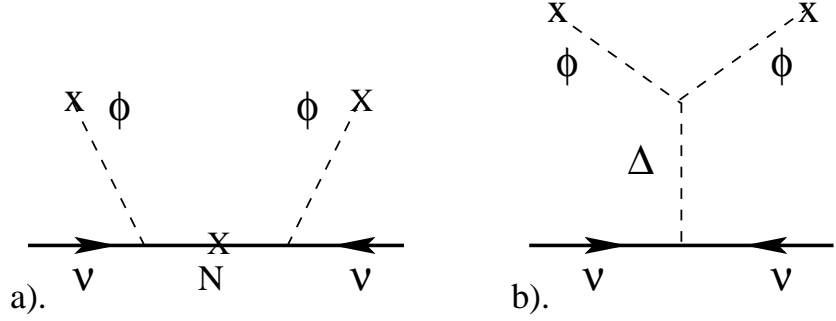}
\end{center}
 \caption{Feynman diagrams for a). type I and b). type II  seesaw
mechanisms.}
\label{see2}
\end{figure}

To discuss the implications of these observations for the see-saw
mechanism, note that in the first stage, the gauge symmetry is broken by
the Higgs multiplets $\Delta_L(3,1,2)\oplus \Delta_R(1,3,2)$ to
the standard model and in the second stage -  by the bi-doublet 
$\phi(2,2,0)$.
In the first stage, the right handed neutrino picks up a mass of
order $f<\Delta^0_R>\equiv fv_R$, then $\phi$ produces the Dirac
mass term. The presence of the coupling of the triplets with the 
bi-doublet
$\lambda \Delta_L \Delta^{\dagger}_R \phi \phi$ leads to a shift
of the minimum of the potential from $\Delta_L = 0$, so that this triplet
acquires the so-called induced VEV 
of the size 
\be 
v_L = \langle \Delta_L^0 \rangle
= \frac{\lambda v^2_{wk}}{v_R} 
\ee 
from the diagram in fig.\ref{see2}b.  
As a consequence the mass
matrix (\ref{mgeneral}) is generated with the components 
\be 
m_L =
f v_L,~~~ m_D =  Y v_{wk}, ~~~ M_R = f v_R. 
\ee 
The light neutrino
mass matrix can  then be written as 
\be 
{\cal M}_{\nu} =
\frac{v^2_{wk}}{v_R} (\lambda f - Y^T f^{-1} Y). \label{ssLR} 
\ee
Note that $v_L$ and the see-saw type II term
are suppressed by the same factor as the see-saw  type I
contribution, so that the overall see-saw suppression remains
\cite{seesaw2}. As a consequence of the left-right symmetry, the two
contributions are partly correlated: Both depend on the same
matrix $f$.
An important problem is to reconstruct $f$ from the low-energy data.
In this connection, an interesting property of the formula in 
eq.(\ref{ssLR}) in the lowest order approximation (before RGE corrections)
is the see-saw duality \cite{duality}: For any
solution $f$, a dual solution $\tilde{f} = {\cal M}_{\nu}/v_L - f$
exists.
In the limit of very large right handed neutrino masses, the general
seesaw formula for neutrino masses reduces to the triplet seesaw
(type II) formula.

Note that the scale of the right handed
neutrino masses (or the seesaw scale) is now the scale of B-L
breaking, which shows that it is not the Planck scale as would be
the case if the left-right symmetry group were not considered. We 
see below that in the context of SO(10) GUTs 
which embed the left-right model, the see-saw
scale can indeed be the GUT scale, removing one arbitrariness in
the description of neutrino masses.


\subsection{Seesaw and the  RGE effects}

As we saw in sec. 6.1  the masses of  RH neutrinos are in general
substantially smaller than the GUT scale. Furthermore, typically
they have an extremely large spread (often related to the
quadratic mass hierarchy of quarks): from $10^6$ to  $10^{15}$
GeV. That determines important features of the RGE effects if the
flavor physics (structure of the Yukawa couplings) is fixed at
$M_f \geq M_{GUT}$. In this case there are three different energy
regions with different RGE behavior:

1). Region below the seesaw scales,
$\mu < M_{1}$ where  $M_{1}$ is the mass of the
lightest RH neutrino.
The RGE effects in this region have been studied in sec. 3.6.

2).  Region between the seesaw scales:  $M_{1} < \mu <  M_{3}$, where
$M_{3}$ is the heaviest RH neutrino.

3). Region above the seesaw scales:  $\mu >  M_{3}$.

In the regions 2) and 3)  the key new feature
is that some or all RH neutrinos are
not decoupled and therefore the neutrino
Yukawa couplings, $Y_{\nu}$, contribute to the running in addition to  $Y_e$.
The couplings  $Y_{\nu}$ now run.
The term  $C_{\nu} Y_{\nu}^{\dagger} Y_{\nu}$,
where $C_{\nu} = 0.5$ in SM and  $C_{\nu} = 1$ in MSSM,
should be added to the RGE eq.(\ref{rge-eq}).   Also $\alpha$ should be
modified.

The couplings $Y_{\nu}$ can be large -
of the order 1  both in SM and
MSSM independently of $\tan \beta$.
Therefore  in general the RGE effects due to
$Y_{\nu}$
are large.

Another important feature is that  the matrices  $Y_{\nu}$ and  $Y_{e}$
can not be made both diagonal, {\it e.g.},  $Y_{\nu}$ is nondiagonal in the
flavor basis where $Y_{e}$ is diagonal. This means that RGE running
due to  $Y_{\nu}$ generates the  flavor transitions and therefore
leads to rotation of the flavor basis.
This running can produce flavor mixing even if initial (at boundary) mixing
matrix is proportional to unity \cite{manfth}.

In the region  above the seesaw scales the running has similar
features to those in the range (1). In particular, similar
enhancement factors appear in the case of degenerate or partially
degenerate spectra. Also CP-violating phases influence running
substantially leading in certain cases to damping of the
enhancement. The 1-2 angle can undergo the strongest
renormalization.

Let us note some interesting possibilities. For the degenerate
spectrum and certain values of phases, running  above the seesaw
scales, $(10^{14} - 10^{16})$ GeV can reduce $\theta_{12}$ from
$45^{\circ}$ down to
 $\sim 30^{\circ}$, thus explaining deviation of the lepton mixing from
bi-maximal.
It can correct the QLC-1 relation reducing  $\theta_{12}$ at low energies.
Running  of other two angles is substantially weaker.
Renormalization effects in two other region can be small,
{\it e.g.},  due to small  $\tan\beta$.

For the hierarchical mass spectrum ($m_1 < 0.01$ eV) the RGE  induced changes of the
mixing parameters are relatively small: $\Delta \theta_{12} < (1 - 2)^{\circ}$
(though it may be relevant for the QLC relation),
$\Delta \sin^2 \theta_{13} < 3\cdot  10^{-5}$ and  $\Delta \sin^2
\theta_{23} < 0.02$.
Mass squared difference $\Delta m^2_{12}$ can decrease  due to running
between and above seesaw scales by
factor of 2 for partially degenerate spectrum,
{\it etc.} \cite{manfth}.

In the region between seesaw scales  one or two  RH neutrinos
decouple. The effective neutrino mass matrix has two different
contributions: ${\cal M}_{\nu} =    {\cal M}_{run} + {\cal
M}_{dec}$ -- the  d=5 type term from the decoupled states, ${\cal
M}_{dec}$,  and the running term, ${\cal M}_{run} =
\tilde{Y}_{\nu}^T(\mu) M_R^{-1}(\mu) \tilde{Y}_{\nu}(\mu)$, where
where $\tilde{Y}_{\nu}$ is the submatrix ($3\times2$ or
$3\times1$) of the Yukawa couplings for ``undecoupled'' RH
neutrinos. In non-supersymmetric models these two contributions
renormalize differently due to vertex  corrections (in the SM case)
to the d=5  operators. 
For the  undecoupled RH neutrinos that corresponds to the  converging box diagram with 
the RH neutrino propagator. The vertex corrections can change
substantially the observables, {\it e.g.}, leading to  $\Delta
\theta_{12} \sim 10^{\circ}$ even for the hierarchical spectrum
\cite{manfth}.

\subsection{Other realizations of seesaw}

 If it turns out that the scale of $B-L$ symmetry breaking, $M_{B-L}$,  
is in the TeV
range, as, for example, in a class of string models discussed
recently \cite{langacker}, the small neutrino masses can be understood by
a double seesaw mechanism \cite{valle1}  where, in addition to the
right handed neutrino, $N$, one postulates the existence of a
singlet neutrino $S$. The symmetries of the model are assumed to
be such that the Majorana mass of $N$ as well as the coupling of
$S$ to the lepton doublet are forbidden. We then have
a neutrino mass matrix in the basis $(\nu, N, S)$ of the form:
\begin{eqnarray}
{ M}~=~\left(\begin{array}{ccc} 0 & m_D^T & 0 \\
m_D & 0 & M \\
0 & M^T & \mu\end{array}\right). \label{dss1}
\end{eqnarray}
For the case $\mu \ll M \approx M_{B-L}$, 
this matrix has one light and two heavy
quasi-degenerate states for each generation. The mass matrix of
light neutrinos  is given by
\be {\cal M}_\nu \sim  m_D^T M^{T
-1}\mu M^{-1} m_D.
\label{fdss}
\ee
There is a double suppression
by the heavy mass compared with the usual seesaw mechanism, hence
the name double seesaw.
One important point here is that to keep $\mu\sim m_D$, one also
needs some additional gauge symmetries, which often are a part of
the string models.

Another possibility which is motivated by the fact that the
required masses of the RH neutrinos are at a somewhat smaller scale
than the GUT scale, is that the RH neutrinos themselves get 
mass via a  see-saw mechanism generated by $N$ and $S$. That would
correspond to $\mu \gg M$ in  eq.(\ref{dss1}),  so that
\be
M_R =
- M \mu^{-1} M^{T}. \label{cascade}
\ee
For $\mu \sim M_{Pl}$ and
$M \sim M_{GUT}$, that gives the required masses of the RH
neutrinos. In particular, eq.(\ref{cascade})  can produce a strong
hierarchy of masses. The formula for the light masses is  the
same as in eq.(\ref{fdss}).

If there is parity
symmetry in models that implement the double  seesaw mechanism,
then the $13$ and $31$ entries of the above neutrino mass matrix
(\ref{dss1}) get filled by small see-saw suppressed masses~\cite{barr}. 
This leads to
\begin{eqnarray}
{ M}~=~\left(\begin{array}{ccc} 0 & m_D^T & M^T\epsilon' \\
m_D & 0 & M \\
M\epsilon' & M^T & \mu
\end{array}\right), \label{dss3}
\end{eqnarray}
where $\epsilon' \simeq v_{wk}/V_0$ and $V_0$ is of the order the
mass $M$ in eq.(\ref{dss3}). The left handed neutrino mass in this case is
given by  
\be {\cal M}_\nu~=~m_D^T M^{T-1}\mu M^{-1}
m_D-(m_D+m^T_D)\epsilon'. 
\ee 
The last contribution, linear in the
Dirac masses, is termed the seesaw type III. There have been a few
applications of this mechanism to model-building \cite{barr1}.

There are also other variations on the seesaw theme for instance
having two right handed neutrinos rather than three. Two RH
neutrinos is the minimum number that will give a realistic
spectrum for neutrinos after the see-saw mechanism. There are schemes
in which new symmetries beyond the standard model  can realize  such
a possibility~\cite{kuchi,frampton}. For instance, if we
supplement the standard model by a local $SU(2)_H$ horizontal symmetry
that acts on the first two generations, then global anomaly
freedom requires that there be only two right handed neutrinos
transforming as a doublet under $SU(2)_H$. This model leads to a
$3\times 2$ seesaw and has features similar to the two RH
dominance models~\cite{king}.

\subsection{Seesaw and large lepton mixing}





\subsubsection{See-Saw enhancement of mixing} 

Can the same mechanism that  explains the smallness of the neutrino mass,
i.e.,  the see-saw also explain the large lepton mixing, so that
eventually large mixing originates from zero neutrino charges and
Majorana nature? 
The idea is that  due to the (approximate) quark-lepton symmetry,
or grand unification, the Dirac mass matrices of the quarks and leptons 
all have 
the same (or similar) structure:  $m_D \sim m_{up} \sim m_l \sim
m_{down}$ leading to zero (small) mixings in the first
approximation. Owing to the  non-diagonal mass matrix of RH neutrinos,
$M_R$, which has no analogue in the quark sector, the seesaw
mechanism produces  non-zero lepton mixing already in the lowest
order.

The problem with this scenario is the strong hierarchy of the
quark and charged lepton masses. Indeed, taking the neutrino Dirac
masses as $m_D = diag(m_u, m_c, m_t)$ in the spirit 
of grand unification, we find
that for a generic $M_R$ the see-saw type I formula produces
strongly hierarchical mass matrix of light neutrinos with small
mixings. The mixing becomes large only for special
structure of $M_R$ which compensates for the  strong hierarchy in
$m_D$.

Two different possibilities are~\cite{senhan,senhan1}

\begin{itemize}

\item strong (nearly quadratic) hierarchy of the RH neutrino
masses: $M_{iR} \sim (m_{i up})^2$ which can be 
reproduced naturally by the double seesaw; and

\item strongly off-diagonal (pseudo-Dirac) structure of $M_R$ 
such as
\begin{equation}
M_R = \left(
\begin{tabular}{lll}
A & 0 & 0\\
0 & 0 & B\\
0 & B & 0 \\
\end{tabular}
\right) \label{eq:}
\end{equation}
which implies certain symmetry. Alternatively, the 12, 21 and
33 elements can be non-zero. An interesting consequence of these
structures is that the pair of  RH neutrinos turns out to be
nearly degenerate, which can lead to the resonant leptogenesis.

\end{itemize}

In the three-neutrino context both possibilities can be realized
simultaneously, so that the pseudo-Dirac structure leads to
maximal 2-3 mixing, whereas the strong hierarchy
$A \ll B$ enhances the 1-2 mixing \cite{AFS}.

There are several  alternatives to the seesaw enhancement.

\subsubsection{Large mixing from type II see-saw} 

In general, the
structure of the neutrino mass matrix generated  by the type II
(triplet) see-saw is not related to the structures of the matrices of
other fermions and it can produce large mixing (see sect. 7). In
some particular cases, however, the relations can appear leading
to interesting consequences.


\subsubsection{Single RH neutrino dominance} 

The large neutrino
mixing and relatively strong mass hierarchy implied by the solar
and atmospheric neutrino data can be reconciled if only one RH
neutrino gives the dominant contribution to the see-saw~\cite{sdom}.  
(This leads to the 2-3 submatrix of $m_{\nu}$ with nearly zero
determinant.) There are two different realizations of this
possibility. In one case the large mixing originates  from the
large mixing in  the Dirac neutrino mass matrix $m_D$: two left-handed 
neutrinos have nearly equal couplings to the  dominating right-handed
component. Suppose that $(m_D)_{23} \approx (m_D)_{33} = m$,
$(m_D)_{13} = \lambda m$ ($\lambda \approx 0.2$) and all other
elements of $m_D$ are much smaller. Then if only $(M^{-1})_{33}$
is large in the inverted matrix, the see-saw will lead to a mass
matrix with the dominant $\mu - \tau$ block. 
The mechanism can also be extended to enhance 1-2 mixing. It
requires the so-called  sequential dominance related to the second
RH neutrino \cite{secdom}.

In another version, the  dominance is realized when two RH
neutrinos are much heavier than the third (dominating) one and no
large mixing in $m_D$ appears. This is equivalent to the strong
mass hierarchy case of the see-saw enhancement mechanism. A
realization requires  $(m_D)_{22} \approx (m_D)_{23} \ll
(m_D)_{33}$, and  dominance of the $(M^{-1})_{22}$ element.

It may happen that the enhancement of the mixing is not related to
the seesaw mechanism at all being, {\it e.g.},  of the radiative origin.
Let us consider the  following
possibility.

\subsubsection{Lopsided models} 

Large lepton mixing in these models follows from the charged lepton
mass matrix in the symmetry basis which  should be left-right
non-symmetric~\cite{lops}. 
This does not contradict the grand unification
since in GUT models such as SU(5), the left-handed components of leptons 
are unified with the RH components of quarks: $5 = (d^c, d^c, d^c, l,
\nu)$. Therefore, large mixing of the left-handed leptonic components is  
accompanied by large mixing of the RH $d$-quarks which is
unobservable. By introducing a Dirac mass matrix of the charged
leptons in which  the only large elements are $(m_l)_{33} \sim (m_l)_{23}$
in the basis where neutrino mass matrix is nearly diagonal, one
obtains the large 2-3 lepton mixing. This scenario can also  be
realized in $SO(10)$, if the symmetry is broken via $SU(5)$. A
double lopsided matrix for both large mixings (solar and
atmospheric) is also possible.



\subsection{Screening of Dirac structure}

The quark - lepton symmetry manifests itself  as a certain relation
(similarity) between  the Dirac mass matrices of quarks and leptons, and
this is the origin of  problems in explaning the  strong
difference of mixings and possible existence of neutrino
symmetries. However, in the context of  double
seesaw mechanism the Dirac structure in the lepton sector can be
completely eliminated - ``screened''\cite{senhan,lind,Kim} thus opening
new possibilities.

Indeed, the double  seesaw mechanism leads to the light neutrino
mass matrix given in eq.(\ref{fdss}). Suppose  that due to a certain
family symmetry or grand unification (which includes also new
singlets $S$) the two Dirac mass matrices are proportional to each
other: 
\be M_D = K^{-1} m_D, ~~~~ K \equiv  v_{EW}/V_{GU}. \ee In
this case the Dirac matrices  cancel in (\ref{fdss}) and we obtain
\be 
{\cal M}_{\nu} = K^2 \mu. 
\ee 
That is, the structure of the light
neutrino mass matrix is determined directly by $\mu$ and does not
depend on the Dirac mass matrix. Here the seesaw mechanism
provides the scale of neutrino masses but not the flavor structure
of the mass matrix. It can be shown that at least in the SUSY version, 
the radiative corrections do not destroy screening \cite{lind}.

The structure of the light neutrino mass matrix is given (up to small
corrections) by  $\mu$ which can be related  to some new physics
at, {\it e.g.}, the  Planck scale. In particular,

1). $\mu$ can be the origin of neutrino symmetry;

2). $\mu \propto I$  leads to a quasi-degenerate spectrum of
light neutrinos;

3). $\mu$ can be the origin of bi-maximal or maximal mixing thus
leading to the QLC relation \cite{qlc-fm} if the charged lepton
mass matrix generates the CKM rotation (QLC-1).
In general,  screening allows one to ``automatically''
reconcile the quark-lepton symmetry with the strong difference of mixings 
of leptons and quarks.

\subsection{Seesaw: tests and applications}

A major problem in neutrino physics is to find ways to test the
proposed mechanisms and scenarios of neutrino mass generation. The
seesaw scenarios are related to physics at very high energy scales
which can not be achieved by the direct studies. Furthermore, it
is practically impossible to reconstruct the right handed neutrino
mass matrix from the low energy observables \cite{davidson}
without additional assumptions like involvement of only two RH
neutrinos \cite{ibarra}, {\it etc.}.

The situation  can change if  the seesaw mechanism is embedded in
into bigger picture so that one will be able to test the whole
context. This will allow one to connect neutrinos with other
phenomena and observables. Moreover, some parameters of the seesaw
mechanism can be determined (see sect. 7). 
The hope is that in future on the basis of certain models we will
be able to make predictions with {\it very small uncertainties}
which can be tested in {\it precision} measurements. This will
then provide a direct test of the model.

In what follow we will describe briefly some connections of the seesaw
with other phenomena which can help to check the mechanism.

\subsubsection{Testing see-saw in colliders}

See-saw mechanism combined with supersymmetry can provide another
testable signature in colliders~\cite{baer} by affecting the mass
spectrum of sleptons at the weak scale. In simple MSSM scenarios, one
assumes a universal scalar mass for all superpartners at the susy
breaking scale (say $M_{GUT}$ or $M_{P\ell}$ ). The weak scale
masses are derived from this by renormalization group extrapolations,
which are sensitive to the interactions at various scales. Above the
see-saw scale, the neutrino Dirac couplings $Y_\nu$  affect the
evolution of both the left-handed slepton and sneutrino masses, as has
been pointed out in \cite{baer}. They find effects which can be
observable at the LHC. Specifically, combination of slepton masses
\be
2\Delta_{\tau} \equiv
(m^2_{\tilde{e_L}} + m^2_{\tilde{e_R}} -
m^2_{\tilde{\nu_{eL}}} )-
(m^2_{\tilde{\tau_L}}+m^2_{\tilde{\tau_R}}-m^2_{\tilde{\nu_{\tau L}}})
+2  m^2_\tau
\ee
can be as large as $0.6\times 10^{3}$ GeV$^2$. This
combination would vanish in the absence of the see-saw scale. There
are also similar other mass combinations that have similar effects.

\subsubsection{Origin of matter.}

A very interesting aspect of the seesaw mechanism is the
possibility that the heavy right handed neutrino decays and  CP
violation in the lepton sector may provide a way to understand the
origin of matter - baryon asymmetry of the Universe~\cite{fuku}.

The original scenario consists of out-of-equilibrium CP violating
decays of the RH neutrinos $ N \rightarrow l + H $ which lead the to
production of a leptonic asymmetry. This asymmetry  is partly
transformed to baryon asymmetry by sphaleron processes (which
conserve B - L but violate B+L).

One of the goals of this discussion is to learn about the
right handed neutrinos and the nature of leptonic CP violation
from the condition of successful leptogenesis \cite{leptog}. 
Although 
there are many possible ways to achieve successful leptogenesis,
{\it e.g.}, resonant leptogenesis, non-thermal leptogenesis,
{\it etc.},  we 
restrict our discussion to the simplest case of a  hierarchical
pattern for right handed neutrino masses, {\it i.e.} $M_1 <
M_{2,3}$ and thermal leptogenesis, and outline the consequences
for neutrinos.

The first implication of leptogenesis for the right handed neutrino
spectrum comes immediately from the out-of-equilibrium condition for their
decay:
\begin{eqnarray}
\Gamma_i\leq H(M_i) \simeq \sqrt{g_*}\frac{M^2_i}{M_{P\ell}},
\end{eqnarray}
where $\Gamma_1$ is the decay rate
\be
\Gamma_i \sim \frac{(Y_\nu Y^{\dagger}_\nu)_{ii} M_i}{8\pi}
\label{decayr}
\ee
and  $M_i$ is the mass of the $i$th RH neutrino $H(M_i)$ is the
expansion rate of the Universe  in the epoch with temperature $T
\sim M_i$, and $g_*$ is the number of relativistic degrees of freedom in
the epoch $T$. This condition leads to a lower bound on the mass
of the RH neutrino
 \begin{eqnarray}
M_i\geq \frac{M_{P\ell} |Y_{\nu,ik}|^2}{8\pi\sqrt{g_*}}.
\label{outof}
\end{eqnarray}


One can get some idea about the required values of the
masses $M_i$, {\it e.g.} assuming  the up quark
Yukawa couplings as a guideline for the Dirac neutrino couplings:
$Y_{\nu,ik} \sim m_{u_i}/v_{wk}$. Then  the out-of-equilibrium
conditions in eq.(\ref{outof}) would imply that $ M_1\geq 10^7$ GeV,
$M_2\geq 10^{12}$ GeV and $M_3\geq  10^{16}$ GeV.
So, the lepton asymmetry is assumed to be produced by the decay of
the lightest $N_1$.

In the leptogenesis scenario the baryon asymmetry,
$\eta_B \equiv n_B/s$, where $n_B$ is the number density of baryons
and $s$ is the entropy density,  can be written as~\cite{leptog}
\be
\eta_B = \frac{8}{23}\frac{n_1}{s} \epsilon_1 \kappa_1.
\label{bau}
\ee
Here $n_1$ is the number density of the RH neutrinos,
$\epsilon_1$ is the lepton asymmetry produced in the decay of
$N_1$ and $\kappa_1$ is the wash out factor which
describes the degree of out-of-equilibrium condition;
the factor 8/23 is the fraction of the $L-$ ($B - L$)
asymmetry which is converted
to the baryon asymmetry by sphalerons.
The quantities $n_1$, $\epsilon_1$ and $\kappa_1$ are all the functions
of the RH neutrino masses and the Dirac type Yukawa couplings.
So,  the bounds on the neutrino parameters
can be obtained from  simultaneous analysis
these quantities \cite{leptog,Buchmuller}.
Indeed,  some more information can be gained by analyzing the magnitude
of the lepton asymmetry $\epsilon_1$ in terms of the Yukawa couplings
$Y_\nu$:
\begin{eqnarray}
\epsilon_1 = \frac{-3}{16\pi(Y_\nu Y^{\dagger}_\nu)_{11}}
\sum_{k\neq 1} Im\left[(Y_\nu Y^{\dagger}_\nu)^2_{1k}\right]
f\left(\frac{M^2_k}{M^2_1}\right),
\label{epsilon1}
\end{eqnarray}
where in the case of hierarchical mass spectrum of the RH neutrinos,
$x \equiv M_k^2/M_1^2 \gg 1$,  and the function $f(x)$ can be approximated as
$f(x)\simeq -3/(2\sqrt{x})$ simplifying
the above expression.  Because  $\eta_B =  6.3 \cdot 10^{-10}$,
and  $\kappa_1$ is roughly of order
$10^{-3}$  (although it  depends strongly on parameters in the
model), we must have $\epsilon_1 \geq 10^{-6}$, which puts
according to (\ref{epsilon1}) a
constraint on the flavor structure of $Y_\nu$.

\begin{figure}[!tbp]
\begin{center}
\epsfxsize9cm\epsffile{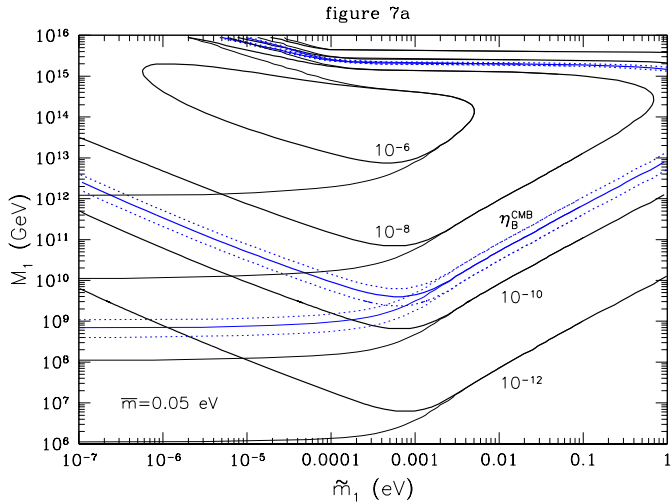}
\end{center}
 \caption{Contour plot of the
 baryon to photon ratio produced in thermal leptogenesis,
 as a function of  $M_{1}$ and $  \tilde{m}_1$, from \cite{Buchmuller}.
The decay asymmetry
 $\epsilon_1$
 was taken to be $10^{-6}$. The three (blue) close-together
 lines are the observed asymmetry. The horizontal contours, for
 small $\tilde{m}_1$ assume a thermal $N_1$ abundance as the initial
 condition.}
\label{lepto}
\end{figure}

Another constraint on  $Y_\nu$ follows from consideration  of the
decay rate of $N_1$, which can be rewritten as (\ref{decayr})
\be
\Gamma_1 =
\frac{\tilde{m}_1M^2_1}{8\pi v^2_{wk}},
~~~~~~ \tilde{m}_1 \equiv
\frac{(Y_\nu Y^{\dagger}_\nu)_{11} v_{wk}^2}{M_1}.
\label{mtilde}
\ee
This rate  controls the initial abundance of $N_1$ and also the
out-of-equilibrium condition. Successful leptogenesis restricts
the values of $\tilde{m}_1$ and $M_1$ as shown in
Fig.~\ref{lepto}.
According to results of fig. \ref{lepto} this standard scenario  implies a
lower bound on the lightest
RH neutrino mass $M_1 \geq 10^8$ GeV and correspondingly gives an  upper
bound on the light neutrino masses thereby essentially excluding the
degenerate spectrum for the case of a type I see-saw.
These constraints can be avoided/weakened if one assumes type II
seesaw \cite{hambye} and/or some specific flavor
structures of the Yukawa couplings  \cite{vives}. The bound can be also
weakened in the case of strong degeneracy of RH neutrino masses
$M_1 \approx M_2$ which leads to enhancement of the asymmetry
$\epsilon_1$ (resonance leptogenesis \cite{reslep})
and therefore allows for
smaller  $k_1$. Consequently,  the washout (out of equilibrium) conditions
relaxes the bound on $M_1$.\\

It was proposed recently that  cosmological density perturbations can be
generated by the inhomogeneous decay of right-handed neutrinos
\cite{Boubekeur}. That requires coupling of the RH neutrinos  with a
scalar field whose fluctuations are created
during inflation.

\subsubsection{Lepton flavor violation as tests of seesaw.}

Once one includes  the right handed neutrinos $N$ in the
standard model so that neutrinos acquire masses and mixings, 
lepton flavor changing  effects such as $\mu\rightarrow e+\gamma$,
$\tau \rightarrow e \gamma$, $\tau \rightarrow  \mu \gamma$ appear.  
However, a
simple estimate of the one-loop contribution to such effects leads
to an unobservable branching ratio (of order $\sim 10^{-40}$).

The situation changes drastically as soon as the seesaw
mechanism  is embedded into the supersymmetric models. Flavor
changing effects arise from the mixings among sleptons
(superpartners of leptons) of different flavors caused by the
renormalization group corrections which via loop diagrams lead to
lepton flavor-violating (LFV) effects at low energies~\cite{lfv}.

The way this happens is as follows. In the simplest $N = 1$ supergravity
models, the supersymmetry breaking terms at the Planck scale are
taken to have only few parameters: a universal scalar mass $m_0$,
universal $A$ terms, and one gaugino mass $m_{1/2}$ for all three
types of gauginos. Clearly, a universal scalar mass implies that
at Planck scale, there is no flavor violation anywhere except in
the Yukawa couplings. However,  as we extrapolate this theory to
the weak scale, the flavor mixings in the Yukawa interactions
induce flavor-violating scalar mass terms. In the absence of
neutrino masses, the Yukawa matrices for leptons can be
diagonalized so that there is no flavor violation in the lepton
sector even after extrapolation down to the weak scale. 
However, when neutrino mixings are present, there is no basis
where all leptonic flavor mixings can be made to disappear. In
fact, in the most general case, of the three matrices:  $Y_{l}$ 
-  the charged lepton coupling matrix, $Y_{\nu}$ - the  RH neutrino
Yukawa coupling and $M_{R}$ -  the matrix characterizing the
heavy RH neutrino mixing, only one can be diagonalized by an
appropriate choice of basis and the flavor mixing in the other two
remains. In a somewhat restricted case in which  the right handed
neutrinos do not have any interaction other than the Yukawa
one and an interaction that generates the Majorana mass
for the right handed neutrino, one can only diagonalize two out of
the three matrices ({\it i.e.}, $Y_\nu, Y_\ell$ and $M_R$). Thus,
there will always be lepton flavor violating terms in the basic
Lagrangian, no matter what basis is chosen. These LFV terms can
then induce mixings between the sleptons of different flavor and
lead to LFV processes.
The smallness of neutrino mass does not matter here.

In the flavor basis, searches for LFV processes such as
$\tau\rightarrow \mu +\gamma$ and/or $\mu\rightarrow e +\gamma$
can throw light on the RH neutrino mixings and/or family mixings in
$m_D$, as has already been discussed.

In the absence of CP violation  there are at least six
mixing angles (nine if $m_D$ is not symmetric) in the seesaw
formula,  only three of which are observable in neutrino oscillation. 
To get useful information on the fundamental high scale theory from
LFV processes, it is often assumed that $M_{R}$ is diagonal so
that one has a direct correlation between the observed neutrino
mixings and the fundamental high-scale parameters of the theory.
The important point is that the flavor mixings in $Y_{\nu}$ then
reflect themselves in the slepton mixings that lead to the LFV
processes via the RGEs.

To give a typical estimate of the magnitude of lepton flavor
violation in seesaw models, we can make a simple ansatz of equal
RH neutrino masses and assume CP conservation. The slepton mixing
defined by $\Delta_{LL,ij} \equiv  \frac{\delta m^2_{ij}}{m^2_0}$
can be estimated from the renormalization group equation to be
\begin{eqnarray}
\Delta_{LL, ij}~=~\frac{3}{8\pi^2}[Y^{\dagger}_{\nu}Y_{\nu}]_{ij}
ln \frac{M^2}{v^2_{wk}}\simeq \frac{1}{4\pi^2}\frac{M
(m_\nu)_{ij}}{v^2_{wk}}ln \frac{M_{P\ell}}{M},
\end{eqnarray}
where $M$ is the seesaw scale. Using $M\sim 10^{13}$ GeV or so,
one finds that the branching ratios for $\mu\rightarrow e+\gamma$
and $\tau\rightarrow \mu+\gamma$ depend on the slepton masses as
$\tilde{m}^{-4}$ and go down as slepton masses increase as can be seen
from
\begin{eqnarray}
R(\ell_j\rightarrow \ell_i +\gamma)  \equiv {B(\ell_j\rightarrow
\ell_i +\gamma) \over B(\ell_j \rightarrow \ell_i +\nu_j +
\bar{\nu}_i)} 
\simeq 
\frac{3\alpha_{em}(c_1g^2_1 + g^2_2)^2}{32\pi G^2_F \tilde{m}^4} 
\left({\Delta _{LL,ij}}\right)^2\tan^2\beta~.
\end{eqnarray}
For the masses   $\tilde{m}$ in the (200 - 500) GeV range, 
the $R(\mu\rightarrow e+\gamma)$ can be above $10^{-14}$, a value that 
can be probed by the MEG experiment in progress \cite{MEG}. Similarly for 
the same slepton masses,  $R(\tau\rightarrow \mu +\gamma)$ can be in the
range of $10^{-9}-10^{-8}$ or so.

\section{Neutrino mass and New physics: ``top-down''}

\subsection{Neutrino mass and Grand Unification.}

One of the major ideas for physics beyond the Standard Model is
supersymmetric grand unification (SUSY GUT) \cite{raby}. It is
stimulated by a number of observations that are in accord with the
general expectations from SUSY GUT's: (i) A solution to the gauge
hierarchy problem, {\it i.e.} why $v_{\rm wk}\ll M_{\rm Pl}$; (ii)
unification of electro-weak, {\it i.e.},  \ $SU(2)_L\times U(1)_Y$
and strong $SU(3)_c$ gauge couplings assuming supersymmetry
breaking masses are in the TeV range, as would be required by the
solution to the gauge hierarchy problem; and (iii) a natural way to
understand the origin of electroweak symmetry breaking.

As noted above, the closeness of the gauge coupling unification
scale of approximately $10^{16}$ GeV to  the  estimate of the seesaw scale
from atmospheric neutrino data of $M_{3}\sim 10^{15}$ GeV
 suggests that the see-saw scale could be the GUT scale itself. Thus 
the smallness of neutrino mass goes quite well with the idea of
supersymmetric grand unification. However, in contrast with 
items (i) through (iii) listed above, the abundance of information
for neutrinos makes seeing whether
the neutrino mixings indeed fit well into simple SUSY GUTs  
a nontrivial exercise.

The simplest GUT group is SU(5). Because the basic matter
representations of SU(5), ${\bf \bar{5}}\oplus {\bf 10}$,  do not
contain the right handed neutrino, one must extend the model by
adding three right-handed neutrinos, one per generation. The
problem then is that the Majorana mass of the gauge singlet right
handed neutrino is unconstrained and can be the same as the Planck
mass which will make it difficult to accommodate the neutrino data.
The right handed neutrino mass fine-tuning question, {\it i.e.},
of why $M_R\ll M_{Pl}$ arises again. However, if one includes the
{\bf 15}-dimensional Higgs boson, then the Yukawa interaction ${\bf
\bar{5}}_m {\bf \bar{5}}_m{\bf 15}_H$ in the superpotential leads
to the coupling $LL\Delta_L$ where $\Delta_L$ is the $SU(2)_L$
triplet in the {\bf 15}-Higgs. The $\Delta^0_L$ field acquires a
VEV of order $v^2_{wk}/\lambda M_U$ \cite{sarkar}, where $\lambda$
is a typical coupling parameter for the Higgs fields among
themselves and $M_U$ is the scale of grand unification. This
leads to neutrino masses (as in type II seesaw) of the right order 
of magnitude to explain the data.

However, if one considers the SO(10) group~\cite{so10},
then its basic spinor representation automatically 
contains the right-handed neutrino 
along with the other fifteen fermions of
the Standard Model (for each family). In order to give a mass to
the right handed neutrino, one must therefore break SO(10)
symmetry (or more precisely, the B-L subgroup of SO(10)). This
naturally solves the right handed neutrino mass fine-tuning
problem. Thus,
 one could argue that small neutrino masses have already
chosen $SO(10)$ GUT as the most natural way to proceed beyond the
Standard Model. Therefore $SO(10)$ has rightly been the focus of
many attempts to understand neutrino mixings.


The $SO(10)$ SUSY GUT models  can be broadly classified into two
classes. One class of models that employ the {\bf 16}-Higgs
representation to give mass to the right handed neutrinos and
another that employs a {\bf 126} Higgs. We outline below their major
features and differences.

As noted, one of the features that distinguishes $SO(10)$ from
SU(5) is the presence of local $B-L$ symmetry as a subgroup, and
the two classes of the $SO(10)$ models mentioned above differ in
the way the $B-L$ symmetry is broken: Breaking by ${\bf 16}_H$ Higgs field
gives $\Delta(B-L)=1$,  whereas ${\bf 126}_H$ leads to
$\Delta(B-L)=2$. In the first case the right-handed neutrino mass
necessarily arises out of a nonrenormalizable coupling, whereas in the
second case it arises from a renormalizable coupling. In addition, the
breaking of $B-L$ by {\bf 16 } Higgs 
leads necessarily  to low
energy MSSM with R-parity breaking so that the model cannot have
cold dark matter without additional assumptions such as matter
parity which forbids specific couplings such as
$({\bf 16}_m)^3 {\bf 16}_H$,
where ${\bf 16}_m$ stands for the matter spinor.

However, ${\bf 126}_H$ breaking of $B-L$ preserves R-parity
at low energies, so that the low-energy MSSM that derives from
such an $SO(10)$ has a natural dark matter candidate, {\it i.e.}
the lightest SUSY particle.

Because $SO(10)$ contains the left-right symmetric group as a subgroup, it
can have either a
type II or type I seesaw formula for neutrino masses depending on
the details of symmetry breaking and parameter ranges of the
theory. For instance, in the ${\bf 16}_H$ based models, the type II
seesaw term is negligible and therefore the neutrino masses are
dictated by type I seesaw formula. In contrast,  in {\bf 126}
Higgs models, the neutrino mass can be given by either the first
term or the second term in the general seesaw formula, or both.

\subsubsection{A minimal {\bf 126}-based ${\bf{SO(10)}}$ model.}

Because ${\bf 16}\otimes {\bf 16}={\bf 10}\oplus {\bf 120} \oplus
{\bf 126}$, the natural minimal model considers all the three
Higgs fields and couples them to the matter {\bf 16}. A
simpler model contains only ${\bf 10}\oplus {\bf 126}$ in which
case there are only two Yukawa coupling matrices: (i) $h$ for the
{\bf 10} Higgs and (ii) $f$ for the {\bf 126} Higgs \cite{babu}.
SO(10)  has the property that 
the Yukawa couplings involving the {\bf 10} and
{\bf 126} Higgs representations are symmetric. Therefore, 
to further reduce the number of free parameters, one may assume  that
Yukawa couplings are CP-conserving, and CP-violation arises from
other sectors of the theory ({\it e.g.} squark masses). In a certain  
basis, 
one of these two sets of Yukawa coupling matrices is diagonal, and
the Yukawa sector has only nine parameters. Noting 
that the (2,2,15) submultiplet of ${\bf 126}_H$ as well as the (2,2,1) 
submultiplet of 
${\bf 10}_H$ each have a pair of standard
model doublets that contribute to charged-fermion masses, one can
write the quark and lepton mass matrices as follows~\cite{babu}:
\begin{eqnarray}
M_u~=~ h \kappa_u + f v_u,  ~~~~~ M_d~=~ h \kappa_d + f v_d, ~~~~\\
M_l~=~ h \kappa_d -3 f v_d, ~~~~~m_D ~=~ h \kappa_u -3 f v_u , \\
\nonumber \label{sumrule}
\end{eqnarray}
where $\kappa_{u,d}$ are the VEV's of the up and down standard
model type Higgs fields in the ${\bf 10}_H$ multiplet and $v_{u,d}$
are the corresponding VEV's for the same doublets in ${\bf 126}_H$.
Note that there are 13 parameters (nine
parameters in the Yukawa couplings noted above and four VEV's 
for the four MSSM doublets in the {\bf 10}, and
{\bf 126} Higgs fields) in the equations above, and  13 inputs (six quark 
masses 
three lepton masses and three quark mixing angles and the weak scale).
Thus, all parameters of the model that go into fermion masses are
determined.

To generate the light neutrino masses, we use the seesaw formula
in eq.(\ref{ssLR}), where $f$ is the same 
${\bf 126}_H$ Yukawa coupling as above. Thus all parameters that give
neutrino mixings except an overall scale are determined~\cite{last}.

To see how large mixings arise in this model let us assume that
the seesaw type II gives the dominant contribution to the neutrino
mass,  so that 
\be 
{\cal M}_\nu =  f v_{\nu}, 
\label{calmass}
\ee 
where $v_{\nu}$ is the VEV of EW triplet from ${\bf 126}_H$. 
Then,  using eq.(\ref{sumrule}), we obtain
\begin{eqnarray}
{\cal M }_\nu~=~ \frac{v_{\nu}}{4 v_d} (M_d - M_l). 
\label{key}
\end{eqnarray}
Furthermore, note 
that the minimality of the Higgs content leads to the following
sum rule among the mass matrices:
\begin{eqnarray}
k{M}_{l}~=~r{ M}_d +{ M}_u. \label{5}
\end{eqnarray}
We then find using the known mass and mixing pattern for quarks, 
that
\begin{eqnarray}
M_{d, l}~\approx ~m_{b,\tau}\pmatrix{\lambda^3 & \lambda^3
&\lambda^3\cr \lambda^3 & \lambda^2& \lambda^2 \cr \lambda^3 &
\lambda^2 & 1},
\label{matr-dl}
\end{eqnarray}
where $\lambda \sim \sin \theta_C =  0.22$ and the matrix elements
are supposed to give only the approximate order of magnitude. An
important consequence of the relation between the charged lepton
and the quark mass matrices in eq.(\ref{matr-dl})  is that the charged
lepton contribution to the neutrino mixing matrix, is 
$U_l \simeq {\bf 1} + O(\lambda)$, close to the identity matrix.
As a result large neutrino mixings must arise predominantly from
the neutrino mass matrix given by the type II seesaw formula. In the
actual calculations of course the charged lepton mixings are also
taken into account. The phenomenological fact that $m_b-m_\tau
\approx m_{\tau}\lambda^2$ for a wide range of values of
tan$\beta$ and Eqs. (\ref{key}), (\ref{matr-dl}) 
imply that the neutrino mass matrix takes roughly the form
\begin{eqnarray}
{\cal M}_{\nu}~=
\frac{v_{\nu}}{4 v_d} (M_d-M_l)\approx ~m_0\pmatrix{\lambda^3 & 
\lambda^3
&\lambda^3\cr \lambda^3 & \lambda^2 & \lambda^2 \cr \lambda^3 &
\lambda^2 & \lambda^2},\label{lambda}
\end{eqnarray}
where, except for numbers of order one, the entire neutrino mass
matrix is characterized by the Cabibbo angle alone. It is easy to
see that both $\theta_{12}$ (the solar angle) and $\theta_{23}$
(the atmospheric angle) are now large \cite{btau,3gen}.

The main point illustrated by this model is that the large
neutrino mixings need not be a consequence of symmetries but
rather could arise  dynamically out of $b-\tau$ unification at
high scale. Note that this requires the choice of small $f_{33}$
which was however made to fit the quark sector and not to ``fix''
the neutrino mixings. Of course, one must understand the flavor
structure of the $h$ and $f$ Yukawa couplings ({\it e.g.} why $f_{33}$
is so small) from a higher scale theory.

There are various ways to incorporate CP violation into these models
(see  \cite{stef,mimura}). One could simply assume that the Yukawa
couplings are complex. However, in this case the simple connection
between $b-\tau$ unification and large neutrino mixing is lost. 
In the pure type II seesaw SO(10) models \cite{btau}, \cite{3gen} 
the fermion mass fits give the CKM phase in the second quadrant 
whereas in the  mixed (type I plus type II) seesaw
generalization, one can get a good fit to CP violation with correct CKM
phase \cite{stef}.

Coming to the rest of the Higgs sector, one can define the
minimal SO(10) model \cite{kuo,melfo} as the one with the complete Higgs
set ${\bf 10} \oplus {\bf 126} \oplus {\bf {\bar{126}}} \oplus {\bf 210}$
for all symmetry breakings and fermion masses. In this model it is hard to
get the type II seesaw to dominate, a problem easily cured by adding a
{\bf 54} Higgs without affecting the fermion mass discussion~\cite{nasri}.
The parameter range of the minimal model that fits fermion masses also
has trouble giving rise to coupling unification\cite{bert}. Thus, it
seems that the non-fermionic part of the Higgs sector
of the {\bf 126} type models has to
be extended to keep fermion mass fits.

An alternative way to incorporate CP violation is to consider the Higgs
set {\bf 10}, {\bf 126} and {\bf 120}~\cite{mimura,other120}, with all 
couplings
constrained by CP invariance. Despite five extra parameters over the
minimal model, one can retain the understanding of large
atmospheric mixing as a consequence of $b-\tau$ unification and a
prediction for $\theta_{13}$ as well as the Dirac
phase of neutrinos in the observable range.

\subsubsection{{\bf 16} Higgs-based SO(10) models.}
The other class of SO(10) models for neutrinos that has been
 widely discussed in the literature includes just ${\bf 10}_H$, ${\bf 16}_H$,
$\overline{\bf 16}_H$ and ${\bf 45}_H$ 
\cite{Babu:1998wi,Albright:1998vf,Blazek:1999hz,li}. An advantage
of these models is that they use low dimensional Higgs multiplets.
However, because  the only renormalizable term in these models  is 
$ {\bf 16}_m {\bf 16}_m {\bf 10}_H$, they can neither explain the
observed quark and lepton masses nor can they  explain the neutrino
masses.  One has to therefore include higher-dimensional operators
in the Yukawa coupling such as
${\bf 16}_m {\bf 16}_m {\bf 16}_H {\bf 16}_H$,~
${\bf 16}_m {\bf 16}_m \overline{\bf 16}_H \overline{\bf 16}_H$,~
${\bf 16}_m {\bf 16}_H {\bf 16}_m {\bf 16}_H$,~
${\bf 16}_m  \overline{\bf 16}_H {\bf 16}_m
\overline{\bf 16}_H~~$, ${\bf 16}_m {\bf 16}_m {\bf 10}_H {\bf 45}_H$,
where ${\bf 16}_m$ stands for various fermion generations. Of these, the
first
two give symmetric Yukawa couplings, the next two have no
symmetry property, and the last one can be both symmetric as well
as antisymmetric. Because each coupling is a $3\times 3$ matrix,
there are many more free parameters in such models than
observables. A strategy employed is to impose additional discrete
symmetries to reduce the number of parameters. This and the fact
that one can have large R-parity violation are drawbacks for these
models.

However, these models have certain advantages: (i) it is
possible to implement the doublet-triplet splitting in a simple
way such that the low energy theory below the GUT scale is the
MSSM  and (ii) the threshold corrections to the gauge couplings
are not excessive, so that no particular constraint on symmetry
breaking is necessary for the gauge couplings to remain
perturbative. Another distinction from the ${\bf 126}_H$-based models
is that the type II seesaw contribution to neutrino masses is
small in this model. The MSSM Higgs doublet fields, {\it i.e.} $H_{u,d}$,  
are linear combinations of the doublets in ${\bf 10}_H$,
${\bf 16}_H$
and $\overline{\bf 16}_H$. The right handed neutrino mass arises
from the ${\bf 16}_m {\bf 16}_m \overline{\bf 16}_H \overline{\bf 16}_H$
couplings when $\tilde{\nu}^c$ component of $\overline{\bf 16}_H$
acquires a VEV. Typically, one uses the lopsided
mechanism~\cite{lops} to generate large atmospheric neutrino
mixing. 
Most {\bf 16}-Higgs  based SO(10) models lead to a small value for
$\theta_{13}$,  although it is possible to have variations of the
model that  lead to a bigger value~\cite{li}.

\subsection{Grand unification and flavor symmetry}

Although the hypothesis of grand unification goes naturally with the
seesaw scale, the detailed flavor pattern, {\it i.e.}, the  hierarchical
mass and mixing among quarks and large mixings for leptons, is
perhaps suggestive of some kind of flavor symmetry connecting
different generations. A possible symmetry group for such models
that unify quark and lepton flavor textures while 
implementing the seesaw mechanism could be, {\it e.g.},
SO(10)$\otimes G_{family}$, where $G_{family}$  can be either
SO(3), SU(3)  \cite{ross}, or SU(2)~\cite{chen} or $U(1)$~\cite{lavi}
group, or a discrete group such as $S_4$~\cite{leemo}, $Z_2$ or
$A_4$, all of which have been attempted. The groups such as
$SU(3), SO(3)$ as well as $S_4$ and $A_4$ have an advantage over
the $U(1)$ and $Z_2$ groups since they have {\bf 3} dimensional
representations into which the three families can fit unlike the
other groups.

The main feature of these models is that in the case of abelian
discrete group one can reproduce the flavor structure selecting
the Yukawa couplings,  whereas in the case of non-abelian ones, the
problem shifts to VEV alignment and particular form of the scalar
potential. This generally requires large number of Higgs fields
with specific couplings. However, this appears to be a
straightforward and promising direction for both quark-lepton and
flavor unification and better models must be pursued.

It is also worth noting that if simpler models such as the minimal
SO(10) model with ${\bf 126}_H$ discussed above are experimentally
favored, we must find a natural way based on some higher symmetry
 to generate the necessary form of the ${\bf 126}_H$ Yukawa coupling $f$
as in eq.(\ref{lambda}).

Research along this line are mostly at an exploratory stage but it
is probably fair to conjecture that such a unified approach is
more likely to succeed if the neutrino mass hierarchy is
established to be normal rather than inverted since the
unification group connects all fermion textures. Furthermore, a
strong signal of an underlying symmetry would be a degenerate
spectrum. Examples of such symmetries which in conjunction
with type II seesaw lead to degenerate spectrum have been
discussed in the literature \cite{leemo,io}.

\subsection{Grand unification and proton decay}

Since grand unified theories connect quarks and leptons, most such
theories predict an unstable proton and therefore one could use
proton decay as a signal of the specific nature of the grand
unified theory. In supersymmetric theories since the dominant
contribution to proton decay arises from dimension five operators
which involve Yukawa couplings responsible for flavor structure of
fermions, one may also hope to learn about the fermion textures
from the proton decay modes.

In the context of SO(10) models, the predictions for proton decay
have been studied in both ${\bf 16}_H$ based \cite{bpw} as well as 
${\bf 126}_H$ based models \cite{mimura1}. Both cases have typical
predictions for distinguishing models, {\it e.g.} the cannonical
$p\rightarrow \bar{\nu}K^+$ in the case of Ref.~\cite{bpw} and
$n\rightarrow \pi^0\bar{\nu}$ in the ${\bf 126}_H$ case at an observable
level.

It must however be stressed that a true test of grand unification
would be the discovery of the gauge mediated proton decay mode
$p\rightarrow e^+\pi^0$ which is completely model independent. The 
present lower limit on the partial lifetime for this mode is $5\times 
10^{33}$ years~\cite{SKP}. 
For supersymmetric theories however they are expected to be at the
level of $10^{36}$ years or longer and are therefore beyond the
reach of experiments with conventional technology. The situation
is more hopeful for nonsupersymmetric theories. For example in two step 
non-SUSY SO(10) theories where $SO(10)\rightarrow SU(2)_L\times 
SU(2)_R\times SU(4)_c\rightarrow SM$, the lifetime is predicted to 
be
$1.44\times 10^{32.1\pm 0.7\pm 1.0\pm 1.9}$ years 
\cite{dglee} (where the 
uncertainties include threshold effects and gauge coupling uncertainties 
{\it etc.}). The intermediate scale in these theories is around $10^{13.6}$ GeV
which is of the right order to explain the neutrino masses via the seesaw 
mechanism. Other non-SUSY examples are minimal SU(5) models with a {\bf 
15}-plet to give neutrino masses by type II seesaw~\cite{pavi}. In these 
theories predicted proton life time can be close to the present limit but 
in any case has an upper limit of $1.4\times 10^{36}$ yrs.

Clearly, proton decay search will be a crucial part of 
the search for grand unification that explain neutrino masses.

\subsection{Neutrinos and extra dimensions}

One of the important predictions of string theories is the
existence of more than three space dimensions. For a long time, 
these extra dimensions were believed to be small and 
therefore practically inconsequential as far as low energy physics
is concerned. However, recent progress in the understanding of the
nonperturbative aspects of string theories has opened up the
possibility that some of these extra dimensions could be large
without contradicting observations. In particular, models in which 
some of the extra dimensions have sizes as large as a
sub-millimeter and  the string scale $M_*$ is  in the few TeV
range have attracted a great deal of phenomenological
attention~\cite{nima}. The basic assumption of these models,
inspired by the D-branes in string theories, is that 
space-time has a brane-bulk structure, where the brane is the
familiar (3+1)-dimensional space-time, with the standard model
particles and forces residing in it, and the bulk consists of all
space dimensions where gravity and other possible gauge singlet
particles live. One could of course envision (3+d+1)-dimensional
D-branes where d space dimensions have miniscule ($\leq {\rm
TeV}^{-1}$) size. The main interest in these models has been due
to the fact that the low string scale provides an opportunity to
test the models  using existing collider facilities.

In general the extra dimensional theories can be divided
into three broad classes: (i) very small size flat extra
dimensions ($r\sim M^{-1}_U$ or so); (ii) large flat extra
dimensions (i.e. $r\sim$ millimeter and (iii) warped extra
dimensions of Randall-Sundrum type. 

In models with $M^{-1}_U$ sized extra dimensions, one can
implement the seesaw mechanism to generate neutrino masses. These
models fit in very well with the conventional grand unified
theories. These models have become popular as a way to providing an 
alternative resolution of the doublet-triplet
splitting problem of grand unified theories via orbifold
compactification~\cite{kawa}. As far as the flavor problem goes,
if all the flavors are in the same brane, the presence of extra dimension
does not distinguish between them and therefore does not throw any 
light on this issue. There are however models where different fermion 
generations are put in different locations in extra dimensions~\cite{feru} 
which then leads to nontrivial flavor structure and a possible way to 
approach the flavor problem. Usually however extra assumptions such as 
symmetries are needed to get realistic models.

Coming to models with large extra D models (case (ii)), a major
challenge to them comes from the neutrino sector. There are
several problems: (i) how to understand the small neutrino masses
in a natural manner since the see-saw mechanism does not work here
due to lack of a high scale; (ii) second problem is that if one
considers only the standard model group in the brane, operators
such as $LH LH/M_*$ could be induced by string theory in the low
energy effective Lagrangian. For TeV scale strings this would
obviously lead to unacceptable neutrino masses.

One mechanism suggested in Ref.~\cite{dienes} is to postulate the
existence of  gauge singlet neutrinos, $\nu_B$, in the
bulk that couple to the lepton doublets in the brane and
additionally demand the theory to be invariant under the $B-L$
symmetry so that the higher-dimensional operator $LH LH/M_*$ is
absent. In four dimensions the Yukawa couplings and consequently
Dirac masses  turn out to be  suppressed by the ratio
$M_*/M_{P\ell}$, where $M_{P\ell}$ is the Planck mass. The latter
is now an effective parameter related to the volume  of the extra
dimensions, $V_d = (2\pi)^d R_1 ... R_d$, and the fundamental
scale as 
\be 
M_{Pl}^2 = M_*^{2 + d} V_d. 
\label{fundeff} 
\ee 
This suppression  is sufficient to explain small neutrino masses and
owes its origin to the large bulk volume in comparison with width
of the brane $(1/M_*)^d$. The volume suppresses the effective
Yukawa couplings of the Kaluza-Klein (KK) modes of the bulk
neutrino to the brane fields.

The appearance of the suppression can be shown using one extra
dimension with coordinate $y$ and radius $R$. The full action
involving the $\nu_B (x,y)$ can be written as
\begin{eqnarray}
{\cal S} = \int d^4xdy
[i\bar{\nu}_B\gamma_{\mu}\partial^{\mu}\nu_B +
i \bar{\nu}_{BL}(x,y)\partial_y \nu_{BR}(x,y)  + 
\frac{h}{\sqrt{M_*}}\delta(y)  \bar{L} H
\nu_{BR}(x, y) +  h.c.],
\label{l1}
\end{eqnarray}
where $\mu = 0,1,2,3$ and  $H$ denotes the standard model Higgs
doublet. By expanding the bulk field in the Fourier series we obtain
\be
\nu_R(x,y)= \sum_k \frac{1}{\sqrt{2\pi R}} \nu^{(k)}_R
\cos\frac{ky}{R},
\label{l222}
\ee
where $\nu^{(k)}_R$ is the $k$th KK mode and
the prefactor follows from  normalization of the wave function.
Then,  according to eq.(\ref{l1}) the effective 4-dimensional Dirac
coupling of the neutrino $\nu^{(k)}_R$  equals   $\kappa =
h/\sqrt{2 \pi R M_*}$. Generalization for the case of $d$ extra
dimensions is straightforward: $2\pi R M_*  \rightarrow V_d
M_*^d$. Now using the  relation between the four and $4 + d$ -
dimensional Planck masses in eq.(\ref{fundeff})
we get $\kappa= h {M_*\over M_{P\ell}}$, independent of the
number and configuration of the extra dimensions. After standard model
gauge symmetry breaking, this generates a Dirac mass for the
neutrino \cite{dienes} given by
\begin{eqnarray}
m = \frac{h v_{wk} M_*}{M_{P\ell}}. \label{mvsmstar}
\end{eqnarray}
For $M_*\sim 10-100$ TeV, eq.(\ref{mvsmstar}) leads to $m \simeq
(10^{-3}-10^{-2}) h$ eV. Because $h$ is a five-dimensional coupling,
its value could perhaps be chosen to be $\sim 10$ in which case we get
neutrino mass in the range of interest in the discussion of
neutrino oscillations.  Furthermore, the usual left-handed neutrino is 
mixed
with all the KK modes of the bulk neutrino, with the same  mixing
mass $\sim \sqrt{2} m$. Because the $k$th KK mode has a mass $m_k =
kR^{-1}$, the mixing angle is given by $\sqrt{2} m R/k$. 
Note that
for $R\sim 0.1$ mm, this mixing angle could be relevant for 
the subdominant effects in solar and supernova neutrinos.

The above discussion can be extended in a very straightforward
manner to the case of three generations. The simplest thing to do
is to add three bulk neutrinos ascribing  the generation label to
all fermion fields. Now $\kappa$ becomes a $3\times 3$ matrix. One
can first diagonalize this by rotating both the bulk and the active
neutrinos. The mixing matrix then becomes the neutrino mixing
matrix $U$ discussed in the text. After this
diagonalization one can perform the KK expansion, which leads to
mixing of 
the active neutrinos and the bulk towers. There are now three
mixing parameters, one for each mass eigenstate denoted by
$\xi_i\equiv \sqrt{2} m_i R$, and the  mixing angle for each mass
eigenstate to the $k$th KK mode of the corresponding bulk
neutrino is given by $\xi/k$.

In four  dimensions the KK modes of the  RH neutrinos  show up as
sterile neutrinos. The main feature is that there is an infinite number
of such neutrinos with increasing mass and decreasing mixing. This
can lead to peculiar effects in  neutrino oscillations. Untill  now,
however,  no effects are found which leads to the upper limits on
$\xi_i$ and hence on the radius of the extra dimension $R$,  given a
value of the neutrino mass $m_i$ (or the coupling $h$).
According to detailed
analysis performed in~\cite{5dothers} one has  $R^{-1} \geq 0.02$ eV
for a hierarchical, $\geq 0.22$ eV for an inverted, and $\geq 4.1$ eV
for a degenerate neutrino spectrum. In generical, for all three cases
the most stringent bound comes from the solar neutrino data 

Coming to the third type of extra D models {\it i.e.} the
Randall-Sundrum scenario, where one invokes a warped extra space
dimension, understanding small neutrino masses is less
straightforward  and has not yet reached a level where its
detailed phenomenological implications can be discussed although
some interesting attempts have been made~\cite{rsnu}.

Essentially,  theories of extra dimensions provide us with
qualitatively new mechanism of generation of a  small {\it Dirac}
neutrino mass. There are different scenarios; however, their common
feature can be termed the overlap suppression: the overlap of wave
functions of the left-handed, $\nu_L(y)$, and right-handed, $\nu_R(y)$,  
components in extra dimensions (coordinate $y$). The suppression
occurs owing  to different localizations of the $\nu_L(y)$ and
$\nu_R(y)$ in the extra space. The effective Yukawa coupling is
proportional to the overlap. Thus, in the large flat extra dimensional
scenario described above 
$\nu_L$ is localized in the brane which has  volume $1/(M_*)^d$ in
extra space, whereas $\nu_R$ propagates in the whole extra space
volume  $V_n$. Thus, the overlap equals the ratio of the two:
$(1/M_*^d)/(V_d)$ which is precisely the factor we have discussed
above. In the Randall-Sundrum scenario, $\nu_L$ and $\nu_R$ are
localized into two different branes and the overlap of their wave
functions is exponentially suppressed.
In addition, extra dimensions can be the origin of light
sterile neutrinos.

\section{Beyond three neutrinos: sterile neutrinos and new physics}

An important part of our understanding of physics beyond the
standard model involves a knowledge of whether there are only
three light neutrinos $\nu_{e,\mu,\tau}$ or there are others.
Known low-energy particle physics as well as cosmology constrain
the number and properties of any extra neutrino. The fact that
the measurement of the invisible Z-width at LEP is accounted for
by three known neutrinos to a very high degree of accuracy
\cite{LEP} implies that any extra light neutrino must not couple
to the Z-boson and hence not the W-boson either. Extra neutrinos
are therefore termed sterile neutrinos ($\nu_s$).

Sterile neutrinos can communicate with the usual active particles via
Yukawa interactions. Non-zero VEVs of the corresponding scalar
bosons generate the Dirac-type mass terms which lead to mixing of
active and sterile neutrinos. In turn this mixing may have
important theoretical and phenomenological consequences.

\subsection{Phenomenology of sterile neutrinos}

The possible existence
of sterile neutrinos  and their mixing have interesting
consequences in particle physics, astrophysics and cosmology.
Most of the studies however give bounds on masses and mixing
of these neutrinos (see Ref. \cite{CSV} for a recent review).
In particular, if mixing of sterile neutrinos  with  active
neutrinos is strong enough,   they can come into equilibrium in the
early universe and affect the  big bang nucleosynthesis. Present
data on primordial $^4$He, $^2$D and $^7$Li abundances impose a 
constraint $N_{eff} < 1.5$ \cite{olive} on the effective
number of sterile neutrinos,  which were in equilibrium in the
epoch of nucleosynthesis. This in turn  leads to the bound on the
active-sterile mixing $\theta_S$ as a function of the mass $m_S$.

Strong bounds on parameters of sterile neutrinos also come from 
structure formation in the Universe, from solar and supernova
neutrinos, and from  studies of the electromagnetic radiation in the
Universe (because  sterile neutrinos have a radiative decay mode).

One may ask whether there is any need to introduce light sterile
neutrinos. There are several reasons that are very suggestive:

(i) Interpretation of the excess of $e^+ n$ events observed in
the LSND experiment~\cite{LSND} in terms of
$\bar{\nu}_\mu-\bar{\nu}_e$ oscillations imply the existence of one or
more  extra sterile neutrinos with mass $1 - 5$ eV~
\cite{LSND3,LSND12}. 
Such an interpretation has its own problems.
Furthermore it contradicts result of analysis of  the large scale
structure (LSS) in the Universe \cite{dodelson}. Another possibility
is decay of a  relatively heavy sterile neutrino with mass $\sim 0.01 -
0.1$ MeV \cite{lsndste}. The MiniBooNE experiment~\cite{miniboone} is
testing the LSND result.

(ii) Spherically asymmetric emission of  sterile neutrinos with
mass in the keV range during  supernova collapses may explain the
phenomenon of pulsar kicks \cite{kusenko}.

(iii) Sterile neutrinos with mass $m_S \sim 1 - 3$ keV were
proposed to be the warm component of the dark matter in the
universe \cite{fuller,warm}. However recent analysis of  the LSS
data is not compatible with this proposal \cite{uros}.

(iv) Oscillations of sterile neutrinos in the early Universe can be
the origin of the lepton asymmetry in the Universe \cite{slept,warm}.

(v)  Weak (statistically insignificant)  indications of the
presence of sterile states come from solar neutrino data: the low
Homestake rate and  absence of the upturn of the energy spectrum
at low energies \cite{ssol}.


If for some  reason  the existence of sterile neutrinos
is confirmed, it will be a major revolution in the landscape of
neutrino physics. We discuss some physics implications of this 
below, focusing mainly on the question of how to
understand the  lightness of sterile neutrinos 
in the context of  extensions of the
standard model.

\subsection{Sterile neutrinos and properties of active neutrinos}

Sterile neutrinos may have very small mixings for a
given mass and therefore their  astrophysical and cosmological
effects may be  unobservable. Despite of this, they can strongly
influence the mass matrix of active neutrinos and therefore affect
the implications of the established experimental results for 
fundamental theory
\cite{astaup}.

Suppose the active neutrinos acquire ({\it e.g.}, via the see-saw) the
Majorana mass matrix $m_a$. Consider one sterile neutrino, $S$,
with Majorana mass $m_S$ and mixing masses with active neutrinos
$m_{aS}^T = (m_{eS}, m_{\mu S}, m_{\tau S})$ . If $m_S \gg
m_{aS}$, then after
decoupling of $S$ the mass matrix of active neutrinos becomes
\be
{\cal M}_{\nu} = m_a -  \frac{m_{aS} m_{aS}^T}{m_S},
\ee where the last
term is the matrix induced by $S$. Let us consider some possible
effects.

The active-sterile mixing (induced matrix) can be the origin of
large lepton mixing. Indeed, $m_a$ may have the usual hierarchical
structure with small mixing. The mixing parameters $m_{aS}$ can be
chosen in such a way that the resulting matrix 
${\cal M}_{\nu}$ leads to large or
maximal mixing of active neutrinos \cite{abdel}.

The induced matrix can be the origin of particular neutrino
symmetries. Consider universal  couplings of a singlet
field $S$ with active neutrinos: $ m_{aS}^T = m_0 (1,
1, 1) = m_2/\sqrt{3}$. Then the induced matrix has the form 
\be
\delta m_S = \frac{m_2}{3} D, 
\ee 
where $D$ is the democratic
matrix with all elements equal to one. Suppose that the original
active neutrino mass matrix has the structure 
\be m_a =
\frac{m_3}{\sqrt{2}} \left(\begin{array}{ccc}
0 & 0 & 0\\
0 & 1 & - 1\\
0 & - 1 & 1
\end{array}
\right). \label{ddd} 
\ee 
Then the sum $m_{\nu} = m_a + \delta m_S$
reproduces the mass matrix for the tri-bimaximal mixing
\cite{astaup}.
Second sterile neutrino  can generate matrix 
eq.(\ref{ddd}).
Clearly this changes the implications of the neutrino results, 
which would require existence of sterile neutrinos, flavor blindness of
their couplings, {\it etc.}. Because $S$ is
outside the SM structure (with RH neutrinos) it may be easier
to realize some particular symmetries for the induced matrix.



\subsection{On the origin of sterile neutrinos. }

Understanding the origin of light sterile neutrinos is a  challenge.
Their masses are not protected by the EW symmetry and some new
physics (symmetries, dynamics) should exist to explain why they
are light.

\subsubsection{Mirror model for the sterile neutrino.}
An interesting scenario for physics beyond the standard model 
in which there is an identical copy of both the forces and matter
present side by side with known forces and matter 
has been discussed in literature. This new
copy is called the mirror sector of the familiar universe. The
mirror sector communicates with the familiar one only via
gravitational interactions. 
This idea was originally proposed by

Lee and Yang originally proposed 
this idea to maintain an exact parity
symmetry in the full universe containing the mirror sector even
though in each sector parity is violated in its weak
interactions~\cite{mirror}. Such scenarios have emerged recently 
in the context of string theories, where one has $E_8\times E_8$
symmetry of matter and forces, with each $E_8$ acting on one 10-
dimensional brane world, and where under mirror parity one brane goes
into another. They are completely consistent with what is
known about the low energy particles and forces as well as the
standard big bang model of the universe if one assumes that in the
process of evolution of the universe, the reheat temperature of
the mirror sector is somewhat lower than that of the visible sector.  Many
interesting phenomenological consequences can follow in generic
versions of such a theory at low energies such as neutrino
oscillations, the dark matter of the universe {\it etc.}.

The mirror model  was applied to the description of neutrino oscillation 
physics
in \cite{bere}, where it was noted that if sterile neutrinos
indicated by the LSND results are confirmed,
one of the ways to explain their lightness is to postulate the
existence of the mirror sector of the universe,  in which case the
mirror neutrinos can play the role of the sterile neutrinos and
their lightness will follow from arguments similar to the familiar
neutrinos, {\it e.g.}, via mirror seesaw. Electroweak symmetry can
generate their mixings via operators of the form $LHL'H'/M$, where
$M$ could be the Planck mass realizing the possibility that the
two sectors mix via gravitational interactions. In general, 
$M$ could represent the mass of any standard model singlet
particle. This model also has the potential to lead to sterile 
neutrinos in the keV range that mix with known neutrinos.

\subsubsection{Other possibilities. }
Other models for the sterile neutrino include the possibility that
it may be a modulino - one of the standard model singlet fields present in
string models~\cite{smir}, one of the extra singlet fermions in
the $E_6$ models~\cite{e6},  or one of the seesaw right handed
neutrinos that becomes massless owing to leptonic
symmetries~\cite{me} such as $L_e-L_\mu-L_\tau$ or $\mu-\tau$
exchange symmetry. A general feature of these models is that in
the symmetry limit one of the SM singlet fermions remains massless
and  can be identified with the sterile neutrino. Its small mass
and mixing with  active neutrinos
are  generated via the terms that break the symmetry.

\section{Conclusion}

Recent discoveries in neutrino physics have opened up a new vista
of physics beyond the standard model. In this review we have
attempted to provide a glimpse of what we have learned from the 
discoveries  and
what future experiments hold in terms how far this
understanding can go. A broad theme is the appearance of new
lepton flavor physics that was absent in the standard model with
massless neutrinos  and possibly important ramifications for the
flavor physics of quarks. The main areas we focused on were 
(i) understanding small neutrino masses; (ii) understanding the
flavor structure of leptons that leads to large mixings and
possible new symmetries implied by it, and (iii) some possible
implications of the existence of new types of neutrinos. Of the
several scenarios for understanding the small neutrino masses, the
seesaw mechanism seems to have an advantage over others in many
respects: (i) it provides a bridge to quark physics via grand
unified theories; (ii) it gives a simple mechanism for
understanding the origin of matter in the universe; and (iii) it  has
interesting low energy tests in the arena of lepton flavor
violation and  electric dipole moment of leptons. As far as 
the lepton flavor puzzle is concerned, although there are many
interesting proposals, the final answer is far from clear and the
next generation of experiments will  very likely  shed light
on this issue. This process will quite possibly  reveal new
symmetries for leptons, which, in the broad framework of
quark-lepton unification, may throw new light on the quark flavor
structure. We summarized different ways to understand lepton
mixings with and without the use of symmetries and discussed
possible tests. Evidence for new neutrino species
mixing with known neutrinos  will be a new surprise in addition to 
the large mixing surprise and will be another revolution. It could
raise questions such as: Are there new quark species corresponding
to the new neutrinos? What role do extra neutrinos  play in the
evolution of the universe, {\it e.g.}, is there a mirror sector to
the universe or are there extra dimensions?

We emphasize  that the field of neutrino
physics is at an important crossroad in its evolution at the moment, 
and further advances will depend on how we answer the
questions raised in this review. Some of the answers will very likely come
from the proposed experiments that will test issues such as, is
the neutrino its own antiparticle, how are the neutrino masses
ordered, and what is the absolute scale of neutrino mass? Further
precision measurements of neutrino parameters, as well as searches
for new (sterile) neutrinos, are of fundamental
importance.\\

The work of R. N. M. is supported by the National Science
Foundation grant number PHY-0354401.



\begin{thebibliography}{99}


\bibitem{threv}
H.~Fritzsch and Z.~z.~Xing, Prog.\ Part.\ Nucl.\ Phys.\  {\bf 45},
1 (2000); V.~Barger, D.~Marfatia and K.~Whisnant,
  Int.\ J.\ Mod.\ Phys.\ E {\bf 12}, 569 (2003); G. Altarelli and F. 
Feruglio, New J.\ Phys.\  {\bf 6},
106 (2004); S. F. King, Rept.\ Prog.\ Phys.\  {\bf 67}, 107
(2004); Z.~z.~Xing, Int.\ J.\ Mod.\ Phys.\ A {\bf 19}, 1 (2004);
R. N. Mohapatra, New J. Phys. {\bf 6}, 82 (2004); A. Yu. Smirnov,
Int.\ J.\ Mod.\ Phys.\ A {\bf 19}, 1180 (2004); R.~N.~Mohapatra
{\it et al.}, {\it ``Theory of neutrinos: A white paper,''}
hep-ph/0510213.

\bibitem{eff1}
S. Weinberg,  Phys. Rev. Lett. {\bf 43}, 1566 (1979).

\bibitem{eff} R. Barbieri, J. Ellis and M. K. Gaillard, Phys.
Lett. {\bf B90}, 249 (1980); E. Akhmedov, Z. Berezhiani and G.
Senjanovic, Phys. Rev. Lett. {\bf 69}, 3013 (1992).

\bibitem{BerV}  F.~Vissani, M.~Narayan and V.~Berezinsky,
  Phys.\ Lett.\ B {\bf 571}, 209 (2003).


\bibitem{seesaw} P. Minkowski, Phys. Lett. {\bf B 67}, 421 (1977),
 M.~Gell-Mann, P.~Ramond, and R.~Slansky, \emph{Supergravity}
(P.~van Nieuwenhuizen et al. eds.), North Holland, Amsterdam,
1980, p.~315;
 T.~Yanagida, in \emph{Proceedings of the
Workshop on the Unified Theory and the Baryon Number in the
Universe} (O.~Sawada and A.~Sugamoto, eds.), KEK, Tsukuba, Japan,
1979, p.~95; S.~L. Glashow, \emph{The future of elementary
particle physics}, in
  \emph{Proceedings of the 1979 Carg{\`e}se Summer Institute on Quarks and
  Leptons} (M.~L{\'e}vy et al. eds.), Plenum Press, New York, 1980, pp.~687--71
 R.~N. Mohapatra and G.~Senjanovi\'c, Phys. Rev. Lett.
 \textbf{44},912 (1980).

\bibitem{seesaw2}  G. Lazarides, Q. Shafi and C. Wetterich,
Nucl.Phys.{\bf B181}, 287 (1981); R. N. Mohapatra and G.
Senjanovi\'c, Phys. Rev. {\bf D 23}, 165 (1981);

\bibitem{valle} For an analysis of mixed seesaw models without the
high scale suppression, see J. Schecter and J. W. F. Valle, Phys.
Rev. {\bf D22}, 2227  (1980).
%

\bibitem{pseudo} L. Wolfenstein, Phys. Lett. {\bf B 107}, 77 (1981);
J. Schecter and J. W. F. Valle, Phys. Rev. {\bf D 24}, 1883
(1981).



\bibitem{pontosc} B. Pontecorvo, Zh. Eksp. Theor. Fiz. {\bf 33} (1957) 
549; {\it ibidem}  {\bf 34} (1958)  247.

\bibitem{mns} Z. Maki, M. Nakagawa and S. Sakata, Prog. Theor. Phys. {\bf
28} (1962) 870.


\bibitem{homestake}
  B.~T.~Cleveland {\it et al.},
Astrophys.\ J.\  {\bf 496} (1998) 505; J.~N.~Abdurashitov {\it et
al.}  [SAGE Collaboration],
  J.\ Exp.\ Theor.\ Phys.\  {\bf 95} (2002) 181
  [Zh.\ Eksp.\ Teor.\ Fiz.\  {\bf 122} (2002) 211];
  W.~Hampel {\it et al.}  [GALLEX Collaboration],
  Phys.\ Lett.\ B {\bf 447} (1999) 127;
J.~Hosaka {\it et al.}  [Super-Kamkiokande Collaboration],
hep-ex/0508053; M.~Altmann {\it et al.}  [GNO COLLABORATION
Collaboration], Phys.\ Lett.\ B {\bf 616} (2005) 174; SNO
Collaboration (B. Aharmim et al.). {\it Phys. Rev.} C {\bf 72}
(2005) 055502; Phys.\ Rev.\ D {\bf 72}, 052010 (2005).


\bibitem{kamland}
 T.~Araki {\it et al.}  [KamLAND Collaboration],
 Phys.\ Rev.\ Lett.\  {\bf 94} (2005) 081801.

\bibitem{atm} Super-Kamiokande Collaboration (Y. Ashie et al.),
Phys. Rev. D {\bf 71}  (2005) 112005; Y.~Ashie {\it et al.}
[Super-Kamiokande Collaboration], Phys.\ Rev.\ Lett.\  {\bf 93}
(2004) 101801; M.~Ambrosio {\it et al.}  [MACRO Collaboration],
Eur.\ Phys.\ J.\ C {\bf 36} (2004) 323; M.~C.~Sanchez {\it et al.}
[Soudan 2 Collaboration],
 Phys.\ Rev.\ D {\bf 68} (2003) 113004;
[MINOS Collaboration], hep-ex/0512036.


\bibitem{k2k}
E.~Aliu {\it et al.}  [K2K Collaboration], Phys.\ Rev.\ Lett.\
{\bf 94}  (2005) 081802.

\bibitem{CHOOZ}
  M.~Apollonio {\it et al.},
nuclear
  Eur.\ Phys.\ J.\ C {\bf 27} (2003) 331.


\bibitem{pontprob} B. Pontecorvo, ZETF, {\bf 53}, 1771 (1967) [Sov. Phys.
JETP, {\bf 26}, 984 (1968)]; V. N. Gribov and B. Pontecorvo, Phys.
Lett. {\bf 28B} (1969)  493.


\bibitem{w1} L. Wolfenstein, Phys. Rev. D{\bf 17} (1978) 2369;
in {\it ``Neutrino -78"}, Purdue Univ. C3, (1978), Phys. Rev. {\bf
D20} (1979) 2634.

\bibitem{ms1} S. P. Mikheyev and A. Yu. Smirnov, Sov. J. Nucl. Phys.
{\bf 42} (1985)  913;  Nuovo Cim. {\bf C9} (1986) 17; S.P. Mikheev
and  A.Yu. Smirnov, Sov. Phys. JETP {\bf 64} (1986) 4.


\bibitem{sv}
A. Strumia, F. Vissani, {\it Nucl. Phys.} B {\bf 726} (2005) 294.



\bibitem{bari}
G. L. Fogli et al,  hep-ph/0506083.

\bibitem{T2K} Y.~Itow {\it et al.}, hep-ex/0106019.

\bibitem{DC} F.~Ardellier {\it et al.},
{\it ``Letter of intent for double-CHOOZ''},  hep-ex/0405032.

\bibitem{concha}
M. C. Gonzalez-Garcia, M. Maltoni, A. Yu. Smirnov, Phys.
Rev. D {\bf 70},   093005 (2004).

\bibitem{orl}
O. L. G. Peres, A. Yu. Smirnov,  Phys. Lett. B {\bf 456},
204 (1999);  {\it Nucl. Phys.} B {\bf 680}, 479 (2004).

\bibitem{HM-neg}
H.V. Klapdor-Kleingrothaus {\it et al.}, {\it Eur. Phys. J.} A
{\bf 12}, 147 (2001); A. M. Bakalyarov {\it et al.,} talk given at
{\it the 4th International Conference on Non-accelerator New
Physics} (NANP 03), Dubna,  Russia, 23-28 Jun. 2003,
hep-ex/0309016.

\bibitem{HM-pos}
H.V. Klapdor-Kleingrothaus  {\it et al.,} {\it Mod. Phys. Lett.} A
{\bf 16} (2001)  2409; H.V. Klapdor-Kleingrothaus, et al, {\it
Phys. Lett.} B {\bf 586} (2004) 198.

\bibitem{deg}  D. Caldwell and R. N. Mohapatra, Phys. Rev. {\bf D 48 },
3259 (1993); A. Joshipura, Phys. Rev. {\bf D51}, 1321 (1995).

\bibitem{cos}
U. Seljak et al.. {\it Phys. Rev.} D {\bf 71} (2005) 103515.

\bibitem{goobar} A.~Goobar, S.~Hannestad, E.~Mortsell and H.~Tu,
astro-ph/0602155.

\bibitem{Seljak06}
U.~Seljak, A.~Slosar and P.~McDonald,
arXiv:astro-ph/0604335.

\bibitem{troitsk}
  V.~M.~Lobashev {\it et al.},
Nucl.\ Phys.\ Proc.\ Suppl.\  {\bf 91} (2001) 280; C.~Kraus {\it
et al.}, Eur.\ Phys.\ J.\ C {\bf 40} (2005) 447.

\bibitem{katrin}
  A.~Osipowicz {\it et al.}  [KATRIN Collaboration],
  arXiv:hep-ex/0109033.

\bibitem{scalar} M. Kawasaki, H. Murayama and T. Yanagida, {\it Mod. Phys. Lett.} A
{\bf 7}, 563 (1992); G. J. Stephenson, T. Goldman, B.H.J.McKellar,
{\it Int. J. Mod. Phys.} A {\bf 13}, 2765 (1998); {\it Mod. Phys.
Lett.} A {\bf 12}, 2391 (1997); R.~N.~Mohapatra and S.~Nussinov,
Phys.\ Lett.\ B {\bf 395}, 63 (1997).

\bibitem{mavan} P. Q. Hung, hep-ph/0010126; Peihong Gu, Xiulian Wang, 
Xinmin Zhang, Phys.Rev. {\bf D68}, 087301 (2003);
R. Fardon, A. E. Nelson, N.
Weiner, {\it  JCAP} {\bf 0410} 005 (2004); hep-ph/0507235;  D. B.
Kaplan, A. E. Nelson, N. Weiner, {\it Phys. Rev. Lett.}  {\bf 93},
091801 (2004); P.~Q.~Hung and H.~Pas, Mod.\ Phys.\ Lett.\ A {\bf
20}, 1209 (2005);  A. Brookfield, C. van de Bruck, D.F. Mota, D. 
Tocchini-Valentini, Phys. Rev. Lett. {\bf 96}, 061301 ( 2006).



\bibitem{DS}
A. S. Dighe and  A. Yu. Smirnov, Phys. Rev. D{\bf 62},  033007
(2000); C. Lunardini and   A. Yu. Smirnov,  JCAP {\bf 0306}, 009
(2003).


\bibitem{bim}F. Vissani, hep-ph/9708483;
V.~Barger, S.~Pakvasa, T.~Weiler and K.~Whisnant, Phys.\ Lett.\ B
{\bf 437}, 107 (1998); A.~Baltz, A.S.~Goldhaber and M.~Goldhaber,
Phys.\ Rev.\ Lett. {\bf 81} 5730 (1998); G.~Altarelli and
F.~Feruglio, Phys.\ Lett.\ B {\bf 439}, 112 (1998); M.~Jezabek and
Y.~Sumino, Phys.\ Lett.\ B {\bf 440}, 327 (1998); D. V. Ahluwalia,
Mod. Phys. Lett. {\bf A13}, 2249 (1998).


\bibitem{tbm} L. Wolfenstein, Phys. Rev. {\bf D 18}, 958 (1978); P. F. 
Harrison, D. H. Perkins and W. G. Scott, Phys. Lett. {\bf B
530}, 167 (2002); P. F. Harrison and W. G. Scott, Phys. Lett. {\bf
B 535},  163 (2002);  Z.~z.~Xing, Phys.\ Lett.\ B {\bf 533}, 85 (2002); 


\bibitem{xing} F.~Plentinger and
W.~Rodejohann, Phys.\ Lett.\ B {\bf 625}, 264 (2005); I.~de
Medeiros Varzielas, S.~F.~King and G.~G.~Ross,
hep-ph/0512313.

\bibitem{Frige}
  M.~Frigerio and A.~Y.~Smirnov,
Nucl.\ Phys.\ B {\bf 640}, 233 (2002); Phys.\ Rev.\ D {\bf 67},
013007 (2003).


\bibitem{mutau} T. Fukuyama and H. Nishiura,
hep-ph/9702253; R.~N.~Mohapatra and S.~Nussinov,
Phys.\ Rev.\ D {\bf 60}, 013002 (1999);
C.~S.~Lam, Phys.\ Lett.\ B {\bf 507}, 214 (2001);
P.~F.~Harrison and W.~G.~Scott,
  Phys.\ Lett.\ B {\bf 547}, 219 (2002);
T. Kitabayashi and M. Yasue, Phys.Rev. {\bf D67} 015006 (2003);
W. Grimus and L. Lavoura, Phys.\ Lett.\ B {\bf 572}, 189 (2003);
J.\ Phys.\ G {\bf 30}, 73 (2004);
A. Ghosal,hep-ph/0304090;
Mod.\ Phys.\ Lett.\ A {\bf 19} (2004) 2579;
Y.~Koide, Phys.\ Rev.\ D {\bf 69}, 093001 (2004).


\bibitem{mutaubr}  R. N. Mohapatra, SLAC Summer
Inst. lecture; http://www-conf.slac.stanford.edu/ssi/2004;
hep-ph/0408187; JHEP, {\bf 0410}, 027 (2004); A. de Gouvea, Phys.
Rev. {\bf D 69}, 093007 (2004);
  W. Grimus, A. S.Joshipura, S. Kaneko, L.
Lavoura, H. Sawanaka, M. Tanimoto, Nucl.\ Phys.\ B {\bf 713}, 151 (2005);
R. N. Mohapatra and W. Rodejohann, Phys. Rev. {\bf D72}, 053001
 (2005).

\bibitem{emutau} S.~T.~Petcov,
Phys.\ Lett.\ B {\bf 110}, 245 (1982);
an incomplete list of more recent studies is: R.~Barbieri {\it et
al.}, JHEP {\bf 9812}, 017 (1998);
A.~S.~Joshipura and S.~D.~Rindani, Eur.\ Phys.\ J.\ C {\bf 14}, 85
(2000);
R.~N.~Mohapatra, A.~Perez-Lorenzana and C.~A.~de Sousa Pires,
Phys.\ Lett.\ B {\bf 474}, 355 (2000);
Q.~Shafi and Z.~Tavartkiladze, Phys.\ Lett.\ B {\bf 482}, 145
(2000).
L.~Lavoura, Phys.\ Rev.\ D {\bf 62}, 093011 (2000); T.~Kitabayashi
and M.~Yasue, Phys.\ Rev.\ D {\bf 63}, 095002 (2001);
K.~S.~Babu and R.~N.~Mohapatra, Phys.\ Lett.\ B {\bf 532}, 77
(2002);
H.~J.~He, D.~A.~Dicus and J.~N.~Ng, Phys.\ Lett.\ B {\bf 536}, 83
(2002)
H.~S.~Goh, R.~N.~Mohapatra and S.~P.~Ng, Phys.\ Lett.\ B {\bf
542}, 116 (2002);
G.~K.~Leontaris, J.~Rizos and A.~Psallidas, Phys.\ Lett.\ B {\bf
597}, 182 (2004);
 W.~Grimus and L.~Lavoura,
  J.\ Phys.\ G {\bf 31}, 683 (2005).

\bibitem{s4} D. G. Lee and R. N. Mohapatra, Phys. Lett.
{\bf B 329}, 463 (1994).

\bibitem{so3} P. Bamert and C. Burgess, Phys. Lett. {\bf B
329}, 109 (1994); A. Ionissian and J. W. F. Valle, Phys. Lett.
{\bf B 332}, 93 (1994); Y. L. Wu,
Phys.\ Rev.\ D {\bf 60}, 073010 (1999);
Int.\ J.\ Mod.\ Phys.\ A {\bf 14}, 4313 (1999).



\bibitem{a4}
E. Ma, Mod. Phys. Lett. A {\bf 17}, 2361 (2002) E. Ma and G.
Rajasekaran, Phys. Rev. {\bf D 64}, 113012 (2001).

\bibitem{zee} A. Zee, Phys. Lett. {\bf 93B}, 389 (1980); L. Wolfenstein,
Nucl. Phys. {\bf B175}, 93 (1980); for recent discussion, Y.
Koide, Phys. Rev.{\bf D 64}, 077301 (2001).

\bibitem{frampton} P. Frampton, S. L. Glashow and D. Marfatia,
Phys.\ Lett.\ B {\bf 536}, 79 (2002); A. Kageyama, S. Kaneko, N. Shimoyama
and M. Tanimoto, Phys. Lett. {\bf B 538}, 96 (2002); B. Desai, D. P. Roy
and A. Vaucher, Mod.\ Phys.\ Lett.\ A {\bf 18}, 1355 (2003).


\bibitem{anarchy}  L.~J.~Hall, H.~Murayama and N.~Weiner,
  Phys.\ Rev.\ Lett.\  {\bf 84}, 2572 (2000);
A.~de Gouvea and H.~Murayama,
  Phys.\ Lett.\ B {\bf 573}, 94 (2003);
  G.~Altarelli, F.~Feruglio and I.~Masina,
  JHEP {\bf 0301}, 035 (2003);
  Y. E. Antebi, Y. Nir, T.  Volansky, hep-ph/0512211.



\bibitem{rad} K. S. Babu, C. N. Leung and J. Pantaleone,
Phys. Lett. B {\bf 319}, 191 (1993); P. Chankowski and Z.
Plucieniek, Phys. Lett. {\bf 316}, 312 (1993).


\bibitem{manfD}
  M.~Lindner, M.~Ratz and M.~A.~Schmidt,
  JHEP {\bf 0509}, 081 (2005).

\bibitem{manfth}
M. Tanimoto, Phys. Lett. B{\bf 360}, 41 (1995);
S.~F.~King and N.~N.~Singh, Nucl.\ Phys.\ B {\bf 591}, 3 (2000);
  S.~Antusch, J.~Kersten, M.~Lindner, M.~Ratz and M.~A.~Schmidt,
  JHEP {\bf 0503}, 024 (2005);

\bibitem{rge-eq}
J. A. Casas {\it et al.,} {\it Nucl. Phys.} B {\bf 573}, 652 (2000);
S. Antusch, M. Drees, J. Kersten, M. Lindner, M. Ratz, Phys.
Lett. B {\bf 519}, 238 (2001);  P.~H.~Chankowski and S.~Pokorski,
  Int.\ J.\ Mod.\ Phys.\ A {\bf 17}, 575 (2002);
  P.~H.~Chankowski, W.~Krolikowski and S.~Pokorski,
  Phys.\ Lett.\ B {\bf 473}, 109 (2000)


\bibitem{massmatrrg}
M.~K.~Parida, C.~R.~Das and G.~Rajasekaran,
  Pramana {\bf 62}, 647 (2004);
C.~Hagedorn, J.~Kersten and M.~Lindner,
  Phys.\ Lett.\ B {\bf 597}, 63 (2004);
  T.~Miura, E.~Takasugi and M.~Yoshimura,
  Prog.\ Theor.\ Phys.\  {\bf 104}, 1173 (2000)



\bibitem{frigren}
M.~Frigerio and A.~Y.~Smirnov,  JHEP {\bf 0302}, 004 (2003).


\bibitem{degen}
J. A. Casas {\it et al.,} {\it Nucl. Phys.} B {\bf 556}, 3 (1999),
{\it ibidem},  {\bf 569}, 82 (2000), {\it ibidem} {\bf 573}, 659
(2000);  J.~R.~Ellis and S.~Lola, Phys.\ Lett.\ B {\bf 458}, 310 (1999)
P.~H.~Chankowski, A.~Ioannisian, S.~Pokorski and J.~W.~F.~Valle,
  Phys.\ Rev.\ Lett.\  {\bf 86}, 3488 (2001);
M.~C.~Chen and K.~T.~Mahanthappa,
Int.\ J.\ Mod.\ Phys.\ A {\bf 16}, 3923 (2001).
\bibitem{rad1} K. R. S. Balaji, A. Dighe, R. N. Mohapatra and M. K.
Parida, Phys. Rev. Lett. {\bf 84}, 5034 (2000); Phys. Lett. {\bf B
481}, 33 (2000); S.~Antusch and M.~Ratz,  JHEP {\bf 0211}, 010
(2002); R. N. Mohapatra, G. Rajasekaran and M. K. Parida, Phys.
Rev. {\bf D 69}, 053007 (2004);


\bibitem{small} E.~J.~Chun,  Phys.\ Lett.\ B {\bf 505}, 155 (2001);
G. Bhattacharyya, A. Raichoudhuri and A. Sil,
Phys.Rev. {\bf D67}, 073004 (2003);
A.~S.~Joshipura, S.~D.~Rindani and N.~N.~Singh,
  Nucl.\ Phys.\ B {\bf 660}, 362 (2003).

\bibitem{manf13}
  S.~Antusch, J.~Kersten, M.~Lindner and M.~Ratz,
  Nucl.\ Phys.\ B {\bf 674}, 401 (2003);
J.~w.~Mei and Z.~z.~Xing,
  Phys.\ Rev.\ D {\bf 70}, 053002 (2004);
  S.~Antusch, P.~Huber, J.~Kersten, T.~Schwetz and W.~Winter,
  Phys.\ Rev.\ D {\bf 70}, 097302 (2004).


\bibitem{renphases}
  N.~Haba, Y.~Matsui and N.~Okamura,
  Eur.\ Phys.\ J.\ C {\bf 17}, 513 (2000)






\bibitem{bmrad}
  S.~Antusch, J.~Kersten, M.~Lindner and M.~Ratz,
  Phys.\ Lett.\ B {\bf 544}, 1 (2002);
 T.~Miura, T.~Shindou and E.~Takasugi,
  Phys.\ Rev.\ D {\bf 68}, 093009 (2003);
T.~Shindou and E.~Takasugi,
  Phys.\ Rev.\ D {\bf 70}, 013005 (2004)




\bibitem{grimus} W. Grimus and L. Lavoura, Phys.\ Lett.\ B {\bf 572}, 189
(2003); J.\ Phys.\ G {\bf 30}, 73 (2004).

\bibitem{yu} R. N. Mohapatra, S. Nasri and H. Yu, 
Phys. Lett. B {\bf 636}, 114 (2006)
hep-ph/0603020.

\bibitem{joship} A.~S.~Joshipura, hep-ph/0512252.

\bibitem{kubo}J. Kubo {\it et al.}, {\it Prog. Theor. Phys.}
{\bf 109} (2003) 795.

\bibitem{Z4} E.~Ma and G.~Rajasekaran,
  Phys.\ Rev.\ D {\bf 68}, 071302 (2003).


\bibitem{D4}
W.~Grimus and L.~Lavoura,
  Phys.\ Lett.\ B {\bf 572}, 189 (2003).


\bibitem{a4a}
K. S. Babu, E. Ma and J. W. F. Valle, Phys. Lett. {\bf 552}, 207
(2003).

\bibitem{a4b} For some recent publications see:
 W.~Grimus and L.~Lavoura, JHEP {\bf 0508}, 013 (2005);
E.~Ma, Mod.\ Phys.\ Lett. A {\bf 20}, 2601 (2005).

\bibitem{AF06} G.~Altarelli and F.~Feruglio,
hep-ph/0512103.


\bibitem{babu06}
K.~S.~Babu and X.~G.~He, hep-ph/0507217.

\bibitem{ma06}
E.~Ma, Mod.\ Phys.\ Lett.\ A {\bf 20}, 2601 (2005).



\bibitem{qlc} A. Yu. Smirnov, hep-ph/0402264.

\bibitem{raidal}M. Raidal,  Phys.\ Rev.\ Lett.\  {\bf 93} (2004) 161801.

\bibitem{qlc-ms} H. Minakata and A. Y. Smirnov; Phys.\ Rev.\ D {\bf 70},
073009 (2004).




\bibitem{pati}J. C. Pati and A. Salam, Phys. Rev. D {\bf 10}, 275
(1974).

\bibitem{so10}H. Georgi,
{\it In Coral Gables 1979 Proceeding, Theory and experiment
in high energy physics}, New York 1975, 329 and H. Fritzsch and P.
Minkowski, Annals Phys. {\bf 93} 193 (1975).


\bibitem{dors}I. Dorsner, A.Yu. Smirnov,
Nucl. Phys. B {\bf 698} (2004) 386.



\bibitem{sing}E.~K.~Akhmedov, et al.,
Phys.\ Lett.\ B {\bf 498}, 237 (2001); R. Dermisek,
Phys. Rev. D {\bf 70}, 033007 (2004).

\bibitem{FN}
C. D. Froggatt and H. B. Nielsen, Nucl. Phys. B {\bf
147}, 277 (1979).

\bibitem{JS05}A.~S.~Joshipura and A.~Y.~Smirnov,
hep-ph/0512024.


\bibitem{jarlskog}
C.~Jarlskog, Phys.\ Lett.\ B {\bf 625}, 63 (2005).

\bibitem{petcov}
S.~T.~Petcov and A.~Y.~Smirnov,
Phys.\ Lett.\ B {\bf 322}, 109 (1994).


\bibitem{parametr}
M.~Jezabek and Y.~Sumino,
Phys.\ Lett.\ B {\bf 457}, 139 (1999);
C.~Giunti and M.~Tanimoto, Phys.\ Rev.\ D {\bf 66}, 053013 (2002);
W.~Rodejohann, Phys.\ Rev.\ D {\bf 69}, 033005 (2004);
P.~H.~Frampton, S.~T.~Petcov and W.~Rodejohann,
Nucl.\ Phys.\ B {\bf 687}, 31 (2004);

\bibitem{qlc-fm} P. Frampton and R. N. Mohapatra, JHEP {\bf 0501}, 025
(2005).

\bibitem{qlc-cp} T.~Ohlsson and G.~Seidl,  Nucl.\ Phys.\ B {\bf 643}, 247
(2002);
S.~Antusch and S.~F.~King, Phys.\ Lett.\ B {\bf 631}, 42 (2005);
I.~Masina,  Phys.\ Lett.\ B {\bf 633}, 134 (2006); J.~Harada,
hep-ph/0512294.


\bibitem{shift}
J. Ferrandis and S. Pakvasa, Phys.\ Rev.\ D {\bf 71}, 033004 (2005);
D.~Falcone, hep-ph/0509028; A.~Ghosal and D.~Majumdar, hep-ph/0505173.

\bibitem{falcone} See D.~Falcone in (\cite{shift}).


\bibitem{qlc-km}
S. Antusch, S. F. King and R. N. Mohapatra,
Phys.\ Lett.\ B {\bf 618}, 150 (2005).


\bibitem{qlc-ren}
Sin Kyu Kang, C. S. Kim, Jake Lee,
Phys.\ Lett.\ B {\bf 619}, 129 (2005);
K.~Cheung, S.~K.~Kang, C.~S.~Kim and J.~Lee,
Phys.\ Rev.\ D {\bf 72}, 036003 (2005);
A.~Dighe, S.~Goswami and P.~Roy, hep-ph/0602062.


\bibitem{parametr2}
N.~Li and B.~Q.~Ma, Phys.\ Lett.\ B {\bf 600}, 248 (2004);
N.~Li and B.~Q.~Ma, Eur.\ Phys.\ J.\ C {\bf 42}, 17 (2005).
N.~Li and B.~Q.~Ma,
Phys.\ Rev.\ D {\bf 71}, 097301 (2005);
Z.~z.~Xing, Phys.\ Lett.\ B {\bf 618}, 141 (2005).


\bibitem{par-gen}
A.~Datta, L.~Everett  and P.~Ramond, Phys.\ Lett.\ B {\bf 620}, 42 (2005);
T.~Ohlsson,  Phys.\ Lett.\ B {\bf 622}, 159 (2005);
L.~L.~Everett, Phys.\ Rev.\ D {\bf 73}, 013011 (2006).


\bibitem{GST}R. Gatto, G. Sartori and M. Tonin, Phys. Lett. B{\bf 28},
128 (1968).

\bibitem{koide} Y. Koide, Lett. Nuovo Cim. {\bf 34}, 201 (1982).

\bibitem{koide1} Y. Koide, Phys. Rev. D{\bf 28}, 252 (1983).

\bibitem{koide2}Y. Koide, Mod. Phys. Lett. {\bf A5}, 2319 (1990).

\bibitem{koide05} Y. Koide, hep-ph/0506247.

\bibitem{foot} R. Foot, hep-ph/9402242.

\bibitem{esposito}S. Esposito and P. Santorelli, Mod. Phys. Lett.
A{\bf 10}, 3077 (1995).

\bibitem{bo05}
N. Li and B.-Q. Ma, Phys. Lett. B {\bf 609}, 309 (2005);
hep-ph/0601031.

\bibitem{xing05} Z. Z. Xing and He Zhang, hep-ph/0602134.

\bibitem{gerard} J. M. G\'erard, F. Goffinet and M. Herquet,
Phys. Lett. B {\bf 633}, 563 (2006).


\bibitem{koidess} Y. Koide and H. Fusaoka, Z. Phys. C{\bf 71}, 459 (1996);
Y. Koide, Phys. Rev. D{\bf 60}, 077301 (1999).

\bibitem{koidedem}Y. Koide, Mod. Phys. Lett. A{\bf 8}, 2071 (1993).


\bibitem{brannen}C. A. Brannen, http://brannenworks.com/MASSES2.pdf.

\bibitem{Koide6}  Y.~Koide, arXiv:hep-ph/0605074.



\bibitem{lrs} J. C. Pati and A. Salam, \cite{pati};
R. N. Mohapatra and J. C. Pati, Phys. Rev. {\bf D 11}, 566, 2558
(1975); G. Senjanovi\'c and R. N. Mohapatra, Phys. Rev. {\bf D
12}, 1502 (1975).

\bibitem{marshak} R. N. Mohapatra and R. E. Marshak, Phys. Lett. {\bf B
91}, 222 (1980); A. Davidson, Phys. Rev. {\bf D20}, 776 (1979).



\bibitem{duality}
  E.~K.~Akhmedov and M.~Frigerio, Phys.\ Rev.\ Lett.\  {\bf 96}, 061802 
(2006); For further analysis of this approach, see P.~Hosteins, 
S.~Lavignac and C.~A.~Savoy,  arXiv:hep-ph/0606078; E.~K.~Akhmedov, 
M.~Blennow, T.~Hallgren, T.~Konstandin and T.~Ohlsson,
  arXiv:hep-ph/0612194.
 
 

\bibitem{langacker} J. Kang, P. Langacker, and T. Li, UPR-1010T


\bibitem{valle1} R.~N.~Mohapatra,
Phys.\ Rev.\ Lett.\  {\bf 56}, 561 (1986); R. N. Mohapatra and J.
W. F. Valle, Phys. Rev. {\bf D 34},  1642 (1986).


\bibitem{barr}
E.~Akhmedov, M.~Lindner, E.~Schnapka and J.~W.~F.~Valle,
Phys.\ Rev.\ D {\bf 53}, 2752 (1996);
S.~M.~Barr, Phys.\ Rev.\ Lett.\  {\bf 92}, 101601
(2004);  C.~H.~Albright and S.~M.~Barr, Phys.\ Rev.\ D {\bf 69},
073010 (2004).

\bibitem{barr1}
S.~M.~Barr and I.~Dorsner,   Phys.\ Lett.\ B {\bf 632}, 527
(2006); M.~Malinsky, J.~C.~Romao and J.~W.~F.~Valle,
  Phys.\ Rev.\ Lett.\  {\bf 95}, 161801 (2005).

\bibitem{kuchi}  R. Kuchimanchi and R. N. Mohapatra,
Phys. Rev. {\bf D 66}, 051301 (2002); and hep-ph/0107373.

\bibitem{king} S. F. King, Phys. Lett. {\bf B 439}, 350 (1998);
Nucl. Phys. {\bf B 562}, 57 (1999).

\bibitem{senhan}
A.~Yu.~Smirnov, Phys.\ Rev.\ D {\bf 48}, 3264 (1993).

\bibitem{senhan1} M. Tanimoto, Phys. Lett. B {\bf 345} (1995) 477;
T.K. Kuo, Guo-Hong Wu, Sadek W. Mansour, Phys. Rev. D {\bf 61},
111301 (2000); G. Altarelli F. Feruglio and I. Masina, Phys. Lett.
B {\bf 472}, 382 (2000); S. Lavignac, I. Masina, C. A. Savoy,
Nucl. Phys. B {\bf 633}, 139 (2002). A. Datta, F. S. Ling and P.
Ramond,  Nucl.\ Phys.\ B {\bf 671} (2003) 383; M. Bando, {\it et
al.,} Phys.\ Lett.\ B {\bf 580} (2004) 229.


\bibitem{AFS}
E.~K.~Akhmedov, M.~Frigerio and A.~Y.~Smirnov, JHEP {\bf 0309}, 021 (2003).

\bibitem{sdom} S. F. King, Phys. Lett. {\bf B 439}, 350 (1998).


\bibitem{secdom}
  S.~Antusch and S.~F.~King,
  New J.\ Phys.\  {\bf 6}, 110 (2004).


\bibitem{lops} C.~H.~Albright, K.~S.~Babu and S.~M.~Barr,
  Phys.\ Rev.\ Lett.\  {\bf 81}, 1167 (1998).



\bibitem{lind} M. Lindner, A. Y. Smirnov and M. Schmidt,
{\it JHEP} {\bf 0507}, 048 (2005).

\bibitem{Kim} J.~E.~Kim and J.~C.~Park, hep-ph/0512130.

\bibitem{davidson} S.~Davidson and A.~Ibarra,
  JHEP {\bf 0109}, 013 (2001);
  S.~Davidson, hep-ph/0409339.

\bibitem{ibarra}
  A.~Ibarra, JHEP {\bf 0601}, 064 (2006)


\bibitem{baer} H. Baer, C. Balazs, J. Mizukoshi and X. Tata, Phys. Rev.
{\bf D 63}, 055011 (2001); G. Blair, W. Porod, P.M. Zerwas, Eur. Phys. J.
{\bf C 27}, 263 (2003); M.~R.~Buckley and H.~Murayama, arXiv:hep-ph/0606088.


\bibitem{fuku} M. Fukugita and T. Yanagida, Phys. Lett. {\bf 74 B}, 45
(1986);  V. Kuzmin, V. Rubakov and M. Shaposnikov. Phys. Lett.
{\bf 185B}, 36 (1985).

\bibitem{leptog} M.~Flanz, E.~A.~Paschos and U.~Sarkar,
Phys.\ Lett.\ B {\bf 345} (1995) 248 [Erratum-ibid.\ B {\bf 382}
(1996) 447]; L.~Covi, E.~Roulet and F.~Vissani, Phys.\ Lett.\ B
{\bf 384} (1996) 169; W.~Buchm\"uller and M.~Pl\"umacher, Int.\
J.\ Mod.\ Phys.\ A {\bf 15} (2000) 5047 A. Pilaftsis, Phys.\ Rev.\
D {\bf 56} (1997) 5431; Nucl.\ Phys.\ B {\bf 504} (1997) 61;
A.~Pilaftsis and T.~E.~J.~Underwood, Nucl.\ Phys.\ B {\bf 692},
303 (2004); G.~F.~Giudice {\em et al.}, Nucl.\ Phys.\ B {\bf 685},
89 (2004);   S.~Davidson and A.~Ibarra, Phys.\
Lett.\ B {\bf 535} (2002) 25;
T.~Hambye, Y.~Lin, A.~Notari, M.~Papucci and A.~Strumia,
Nucl.\ Phys.\ B {\bf 695}, 169 (2004);
S.~Davidson, JHEP {\bf 0303} (2003) 037; G.~C.~Branco,
T.~Morozumi, B.~M.~Nobre and M.~N.~Rebelo, Nucl.\ Phys.\ B {\bf
617} (2001) 475; see also M.~N.~Rebelo,; Phys.\
Rev.\ D {\bf 67}, 013008 (2003).


\bibitem{Buchmuller}
W.~Buchm\"{u}ller, P.~Di Bari and M.~Pl\"{u}macher,
Annals Phys.\  {\bf 315}, 305 (2005);
Nucl.\ Phys.\ B {\bf
643} (2002) 367; Nucl.\ Phys.\ B {\bf 665} (2003) 445.

\bibitem{hambye} T.~Hambye and G.~Senjanovic,
  Phys.\ Lett.\ B {\bf 582}, 73 (2004);  S.~Antusch and S.~F.~King,
  Phys.\ Lett.\ B {\bf 597}, 199 (2004).
 

\bibitem{vives} O.~Vives, hep-ph/0512160.

\bibitem{reslep} See A. Pilaftsis, in \cite{leptog}.

\bibitem{Boubekeur}
  L.~Boubekeur and P.~Creminelli, hep-ph/0602052.



\bibitem{lfv} F. Borzumati and A. Masiero, Phys. Rev. Lett. {\bf 57}, 961
(1986); J. Hisano, T. Moroi, K. Tobe and M. Yamaguchi, Phys. Rev.
{\bf D 53}, 2442 (1996); J. Hisano and D. Nomura, Phys. Rev. {\bf
D 59}, 116005 (1999); S. F. King and M. Oliviera, Phys. Rev. {\bf
D 60}, 035003 (1999);
K. S. Babu, B. Dutta and R. N. Mohapatra, Phys. Lett. {\bf 458},
93 (1999); S. Lavgnac, I. Masina and C. Savoy, Phys. Lett. {\bf B520}, 269
(2001);
D. Chang, A. Masiero and H. Murayama, Phys.\ Rev.\ D {\bf
67}, 075013 (2003);
A. Rossi, Phys.\ Rev.\ D {\bf 66}, 075003 (2002);
K. S. Babu and C. Kolda, Phys.\ Rev.\ Lett.\  {\bf 89}, 241802 (2002);
J. Ellis, J. Hisano, S. Lola and M. Raidal, Nucl. Phys. {\bf B621}, 208
(2002);
F.~Deppisch, H.~Pas, A.~Redelbach, R.~Ruckl and Y.~Shimizu,
Eur.\ Phys.\ J.\ C {\bf 28}, 365 (2003);
A. Masiero, S. Vempati and O. Vives, Nucl.\ Phys.\ B {\bf 649}, 189
(2003);
S.~T.~Petcov, W.~Rodejohann, T.~Shindou and Y.~Takanishi,
arXiv:hep-ph/0510404.


\bibitem{MEG} T. Mori, et al. [MEG collaboration], 
exp PSI-R-99-05,  http://meg.web.psi.ch.

\bibitem{raby} For two recent reviews, see S. Raby, Phys. Lett. {\bf B
592}, 1 (2004); R.~N.~Mohapatra, hep-ph/9911272.



\bibitem{sarkar}  R. N. Mohapatra and P. B. Pal, {\it Massive Neutrinos
in Physics and Astrophysics, First Edition, 1991}, p. 127 and 128;
Eq. 7.19; E.~Ma and U.~Sarkar, Phys.\ Rev.\ Lett.\  {\bf 80}, 5716
(1998).


\bibitem{babu} K. S. Babu and R. N. Mohapatra,
Phys. Rev. Lett. {\bf 70}, 2845 (1993).

\bibitem{last} D. G. Lee and R. N. Mohapatra, Phys. Rev. {\bf D 51}, 1353
(1995); L. Lavoura, Phys. Rev. {\bf D 48}, 5440 (1993); B.
Brahmachari and R. N. Mohapatra, Phys. Rev. {\bf D58}, 015001
(1998); K. Oda, E. Takasugi, M. Tanaka and M. Yoshimura, Phys.
Rev. {\bf D 58}, 055001 (1999); K. Matsuda, Y. Koide, T. Fukuyama
and N. Okada, Phys. Rev. {\bf D 65}, 033008 (2002); T. Fukuyama
and N. Okada, JHEP {\bf 0211}, 011 (2002); N. Oshimo, Nucl.\
Phys.\ B {\bf 668}, 258 (2003); see M. C. Chen and K. T.
  Mahanthappa, Int. J. Mod. Phys. {\bf A 18}, 5819
(2003) for a review of SO(10) models.


\bibitem{btau}  B.~Bajc, G.~Senjanovic and F.~Vissani,
Phys.\ Rev.\ Lett.\  {\bf 90}, 051802 (2003).


\bibitem{3gen}  H.~S.~Goh, R.~N.~Mohapatra and S.~P.~Ng,
Phys.\ Lett.\ B {\bf 570}, 215 (2003);
  Phys.\ Rev.\ D {\bf 68}, 115008 (2003).



\bibitem{stef} H.~S.~Goh, R.~N.~Mohapatra and S.~P.~Ng,
Phys.\ Rev.\ D {\bf 68}, 115008 (2003);
 S.~Bertolini and M.~Malinsky,
  Phys.\ Rev.\ D {\bf 72}, 055021 (2005); K.~S.~Babu and C.~Macesanu,
hep-ph/0505200;  S.~Bertolini, T.~Schwetz and M.~Malinsky,
  Phys.\ Rev.\ D {\bf 73}, 115012 (2006).

\bibitem{mimura} B. Dutta, Y. Mimura and R. N. Mohapatra,
Phys.\ Lett.\ B {\bf 603}, 35 (2004);  Phys.\ Rev.\ Lett.\  {\bf
94}, 091804 (2005).

\bibitem{other120}   W.~Grimus and H.~Kuhbock,
  arXiv:hep-ph/0607197; arXiv:hep-ph/0612132;
  C.~S.~Aulakh, arXiv:hep-ph/0607252.


\bibitem{kuo} T.~E.~Clark, T.~K.~Kuo and N.~Nakagawa,  Phys.\ Lett.\ B
{\bf 115}, 26 (1982);  C.~S.~Aulakh and  R.~N.~Mohapatra, Phys.\ Rev.\ D
{\bf 28}, 217 (1983).

\bibitem{melfo} C.~S.~Aulakh, B.~Bajc, A.~Melfo, G.~Senjanovic and
F.~Vissani,  Phys.\ Lett.\ B {\bf 588}, 196 (2004);  B.~Bajc, A.~Melfo,
G.~Senjanovic and F.~Vissani,  Phys.\ Rev.\ D {\bf 70}, 035007 (2004);
T.~Fukuyama, A.~Ilakovac, T.~Kikuchi, S.~Meljanac and N.~Okada,
  Phys.\ Rev.\ D {\bf 72}, 051701 (2005).

\bibitem{nasri} H.~S.~Goh, R.~N.~Mohapatra and S.~Nasri,
  Phys.\ Rev.\ D {\bf 70}, 075022 (2004).

\bibitem{bert}S.~Bertolini, T.~Schwetz and M.~Malinsky,
  arXiv:hep-ph/0605006.

 
 






%








\bibitem{Babu:1998wi}
  K.~S.~Babu, J.~C.~Pati and F.~Wilczek,
  Nucl.\ Phys.\ B {\bf 566}, 33 (2000).

\bibitem{Albright:1998vf}
C.~H.~Albright and S.~M.~Barr,
  Phys.\ Rev.\ D {\bf 58}, 013002 (1998); C.~H.~Albright and S.~M.~Barr,
  Phys.\ Rev.\ D {\bf 62}, 093008 (2000).

\bibitem{Blazek:1999hz}
  T.~Blazek, S.~Raby and K.~Tobe,
  Phys.\ Rev.\ D {\bf 62}, 055001 (2000);
  Z.~Berezhiani and A.~Rossi,
  Nucl.\ Phys.\ B {\bf 594}, 113 (2001).

\bibitem{li} X. Ji, Y. Li and R. N. Mohapatra, Phys. Lett. {\bf B 633}, 
755 (2006).

\bibitem{ross} See for examples S. F. King and G. G. Ross, Phys.
Lett. {\bf B 520}, 243 (2001);  Z.~Berezhiani and A.~Rossi,
  Nucl.\ Phys.\ B {\bf 594}, 113 (2001); I.~de Medeiros Varzielas and
G.~G.~Ross,  Nucl.\ Phys.\ B {\bf 733}, 31 (2006)

\bibitem{lavi}  N.~Irges, S.~Lavignac and P.~Ramond,
  Phys.\ Rev.\ D {\bf 58}, 035003 (1998).

\bibitem{chen} M.~C.~Chen and K.~T.~Mahanthappa,
  Phys.\ Rev.\ D {\bf 62}, 113007 (2000).

\bibitem{leemo} D.~G.~Lee and R.~N.~Mohapatra,
  Phys.\ Lett.\ B {\bf 329}, 463 (1994);  C.~Hagedorn, M.~Lindner and R.~N.~Mohapatra,
  arXiv:hep-ph/0602244.

\bibitem{io}  A.~Ioannisian and J.~W.~F.~Valle,
  Phys.\ Lett.\ B {\bf 332}, 93 (1994).

\bibitem{bpw}  K.~S.~Babu, J.~C.~Pati and F.~Wilczek,
  Nucl.\ Phys.\ B {\bf 566}, 33 (2000).

\bibitem{mimura1} H.~S.~Goh, R.~N.~Mohapatra, S.~Nasri and S.~P.~Ng,
  Phys.\ Lett.\ B {\bf 587}, 105 (2004);  B.~Dutta, Y.~Mimura and 
R.~N.~Mohapatra, Phys.\ Rev.\ Lett.\  {\bf 94}, 091804 (2005);  
T.~Fukuyama, A.~Ilakovac, T.~Kikuchi, S.~Meljanac and N.~Okada,
  JHEP {\bf 0409}, 052 (2004).


\bibitem{SKP} Super-Kamiokande collaboration (2005).

\bibitem{dglee}  D.~G.~Lee, R.~N.~Mohapatra, M.~K.~Parida and M.~Rani,
  Phys.\ Rev.\ D {\bf 51}, 229 (1995)

\bibitem{pavi} I.~Dorsner and P.~F.~Perez,
  arXiv:hep-ph/0606062;  P.~Nath and P.~F.~Perez,
  arXiv:hep-ph/0601023.

 

\bibitem{nima} N. Arkani-Hamed, S. Dimopoulos and G. Dvali, Phys.
Lett. {\bf B429}, 263 (1998); I. Antoniadis et al. Nucl. Phys.
{\bf B 516}, 70 (1998).

\bibitem{kawa}  Y. Kawamura, hep-ph/0012125; G.~Altarelli, F.~Feruglio and I.~Masina,
  JHEP {\bf 0011}, 040 (2000); G.~Altarelli and F.~Feruglio,
  Phys.\ Lett.\ B {\bf 511}, 257 (2001).

\bibitem{feru} T.~Asaka, W.~Buchmuller and L.~Covi,  Phys.\ Lett.\ B {\bf 
563}, 209 (2003);  R.~Dermisek and A.~Mafi, Phys.\ Rev.\ D {\bf 65}, 
055002 (2002); M.~L.~Alciati, F.~Feruglio, Y.~Lin and A.~Varagnolo,
  arXiv:hep-ph/0603086;


\bibitem{dienes}
K.R. Dienes, E. Dudas and  T. Gherghetta, Nucl. Phys. {\bf B557},
25 (1999); N. Arkani-Hamed, S. Dimopoulos, G. Dvali and J.
March-Russell, hep-ph/9811448, Phys.\ Rev.\ D {\bf 65}, 024032 (2002); G.
Dvali and A. Yu. Smirnov, Nucl. Phys. {\bf B563}, 63 (1999).

\bibitem{5dothers}
A. E. Faraggi, M. Pospelov, 
Phys. Lett. {\bf B458},  237  (1999); 
 R. Barbieri, P. Creminelli 
and A. Strumia, Nucl. Phys. {\bf B585}, 28
(2000); D. Caldwell, R. N. Mohapatra and S. Yellin, Phys. Rev.
Lett. {\bf 87}, 041061 (2001); Phys. Rev. {\bf D 64}, 073001
(2002);
 A. Lukas, P. Ramond, A. Romanino and G. Ross,
Phys. Lett. {\bf B495}, 136 (2000); R. N. Mohapatra and A.
Perez-Lorenzana, Nucl. Phys. {\bf B593}, 451 (2001); H.
Davoudiasl, P. Langacker and S. Perelstein, Phys. Rev. {\bf D 65},
105015 (2002); Q.~H.~Cao, S.~Gopalakrishna and C.~P.~Yuan,
  Phys.\ Rev.\ D {\bf 69}, 115003 (2004); J.~L.~Hewett, P.~Roy and S.~Roy,
  Phys.\ Rev.\ D {\bf 70}, 051903 (2004).


\bibitem{rsnu} Y. Grossman and M. Neubert, Phys. Lett. {\bf B
474}, 361 (2000);
S. J. Huber and Q. Shafi, Phys.\ Lett.\ B {\bf 583}, 293 (2004);
T.~Gherghetta, Phys.\ Rev.\ Lett.\ {\bf 92}, 161601 (2004).


\bibitem{LEP} For a summary see Particle Data Group, S. Eidelman et 
al. Phys. Lett. {\bf B
592}, 1 (2004).


\bibitem{CSV}
  M.~Cirelli, G.~Marandella, A.~Strumia and F.~Vissani,
  Nucl.\ Phys.\ B {\bf 708}, 215 (2005)

\bibitem{olive} R. Cyburt, B. D. Fields, K. Olive and E. Skillman,
Astropart. Phys. {\bf 23}, 313 (2005).

\bibitem{LSND}
A. Aguilar {\it et al.}, (LSND Collaboration) {\it Phys. Rev.} D
{\bf 64} (2001)  112007.

\bibitem{LSND3} D. Caldwell and R. N. Mohapatra, Phys. Rev. {\bf D 46},
3259 (1993); J. Peltoniemi and J. W. F. Valle, Nucl. Phys. {\bf B
406}, 409 (1993); S. Bilenky, W. Grimus, C. Giunti and T. Schwetz,
hep-ph/9904316; Phys.\ Rev.\ D {\bf 60}, 073007 (1999);  V. Barger, B.
Kayser, J. Learned, T. Weiler and
K. Whisnant, Phys. Lett. {\bf B 489}, 345 (2000);  C.~Giunti and 
M.~Laveder, JHEP {\bf 0102}, 001 (2001).

\bibitem{LSND12} O. L. G. Peres, A.Yu. Smirnov,
{\it Nucl. Phys.} B {\bf 599} (2001) 3; M. Sorel, J. Conrad, M.
Shaevitz, Phys.\ Rev.\ D {\bf 70} (2004) 073004.

\bibitem{dodelson}
  S.~Dodelson, A.~Melchiorri and A.~Slosar, astro-ph/0511500.

\bibitem{lsndste}
  S.~Palomares-Ruiz, S.~Pascoli and T.~Schwetz,
  JHEP {\bf 0509} (2005) 048.

\bibitem{miniboone}
  M.~H.~Shaevitz  [MiniBooNE Collaboration],
  Nucl.\ Phys.\ Proc.\ Suppl.\  {\bf 137} (2004) 46.


\bibitem{kusenko}  A.~Kusenko and G.~Segre, 
  Phys.\ Lett.\ B {\bf 396}, 197 (1997); 
  G.~M.~Fuller, A.~Kusenko, I.~Mocioiu and S.~Pascoli,
  Phys.\ Rev.\ D {\bf 68}, 103002 (2003); 
  A.~Kusenko,
  Int.\ J.\ Mod.\ Phys.\ D {\bf 13}, 2065 (2004).

\bibitem{fuller}  K.~Abazajian, G.~M.~Fuller and M.~Patel,
  Phys.\ Rev.\ D {\bf 64}, 023501 (2001);

\bibitem{warm}
T. Asaka, S. Blanchet, M. Shaposhnikov,
Phys.\ Lett.\ B {\bf 631} 151 (2005).

\bibitem{uros}
  U.~Seljak, A.~Makarov, P.~McDonald and H.~Trac, astro-ph/0602430.


\bibitem{slept}
  E.~K.~Akhmedov, V.~A.~Rubakov and A.~Y.~Smirnov,
  Phys.\ Rev.\ Lett.\  {\bf 81}, 1359 (1998);
T.~Asaka and M.~Shaposhnikov,
  Phys.\ Lett.\ B {\bf 620}, 17 (2005).


\bibitem{ssol}P.~C.~de Holanda and A.~Y.~Smirnov,
Phys.\ Rev.\ D {\bf 69}, 113002 (2004).

\bibitem{astaup}
  A.~Y.~Smirnov, hep-ph/0512303.


\bibitem{abdel}K.R.S. Balaji, A. Perez-Lorenzana, A.Yu. Smirnov,
{\it Phys. Lett.} B {\bf 509}, 111 (2001).


\bibitem{mirror} T. D. Lee and C. N. Yang, Phys. Rev. {\bf 104}, 254
(1956); K. Nishijima, private communication; Y. Kobzarev, L. Okun
and I. Ya Pomeranchuk, Yad. Fiz. {\bf 3}, 1154 (1966);  M. Pavsic,
Int. J. T. P. {\bf 9}, 229 (1974); S. I. Blinnikov and M. Y.
Khlopov, Astro. Zh. {\bf 60}, 632 (1983); R. Foot, H. Lew and R.
R. Volkas, Phys. Lett. B272, 67 (1991); R. Foot, H. Lew and R. R.
Volkas, Mod. Phys. Lett. A7, 2567 (1992).
\bibitem{bere} R. Foot and R.
Volkas, Phys. Rev. {\bf D 52}, 6595 (1995); Z. Berezhiani and R.
N. Mohapatra, Phys. Rev. {\bf D 52}, 6607 (1995); Z. Berezhiani,
A. Dolgov and R. N. Mohapatra, Phys. Lett. {\bf B 375}, 26 (1996).


\bibitem{smir} K.~Benakli and A.~Y.~Smirnov,
  Phys.\ Rev.\ Lett.\  {\bf 79}, 4314 (1997).
\bibitem{e6}  Z.~Chacko and R.~N.~Mohapatra,
  Phys.\ Rev.\ D {\bf 61}, 053002 (2000); M.~Frank, I.~Turan and M.~Sher,
  Phys.\ Rev.\ D {\bf 71}, 113001 (2005).

\bibitem{me} R.~N.~Mohapatra,
  Phys.\ Rev.\ D {\bf 64}, 091301 (2001);  R.~N.~Mohapatra, S.~Nasri and
H.~B. Yu~Phys.\ Rev.\ D {\bf 72}, 033007 (2005); A.~G.~Dias, C.~A.~de
S.Pires and P.~S.~Rodrigues da Silva,
  Phys.\ Lett.\ B {\bf 628}, 85 (2005); K.~S.~Babu and G.~Seidl,
  Phys.\ Rev.\ D {\bf 70}, 113014 (2004).





\end{thebibliography}
\end{document}